# Molecular Transport Junctions: Vibrational Effects

Michael Galperin and Mark A. Ratner
Department of Chemistry, Northwestern University, Evanston, IL 60208
Abraham Nitzan
School of Chemistry, Tel Aviv University, Tel Aviv 69978, Israel

# 1. Introduction

Molecular transport junctions (MTJs), the simplest components of molecular electronics, are structures in which a molecule is inserted between two electrodes, and subjected to applied voltage. Monitoring MTJ current as a function of applied voltage can be viewed as a kind of spectroscopy.[1-19] This spectroscopy is characterized by several factors. First, of course, is the identity of the molecule and the geometry that the molecule adopts within the junction. Second are the parameters of the Hamiltonian that describe the system and determine the band structure of the electrodes, the electronic structure of the molecule and the electronic coupling between the electrodes and the molecule. The latter includes electronic correlations such as the image effect that is often disregarded in theoretical studies. Finally, effects of the underlying nuclear configuration as well as dynamic coupling between transmitted electrons and molecular vibrations can strongly affect the electron transmission process. Figure 1 sketches, in a light way, a two terminal junction and indicates the couplings that are important in understanding MTJ's.

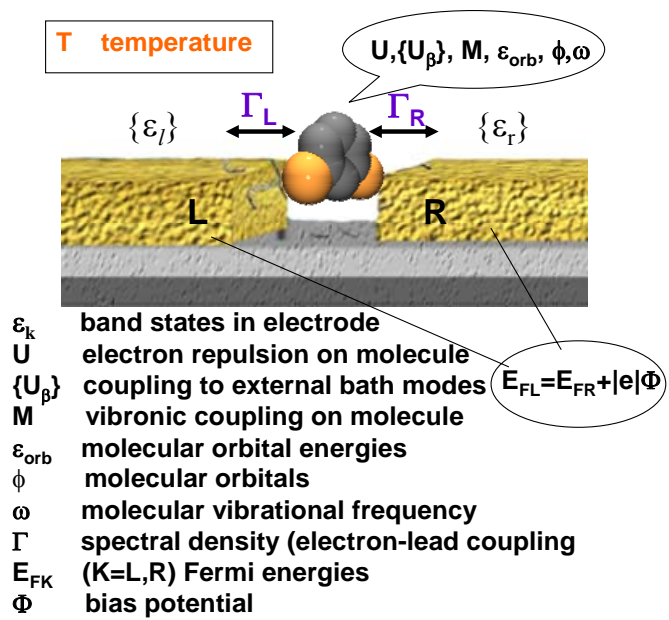

$\varepsilon_k$ band states in electrode
$U$ electron repulsion on molecule
$\{U_\beta\}$ coupling to external bath modes
$M$ vibronic coupling on molecule
$\varepsilon_{orb}$ molecular orbital energies
$\phi$ molecular orbitals
$\omega$ molecular vibrational frequency
$\Gamma$ spectral density (electron-lead coupling
$E_{FK}$ (K=L,R) Fermi energies
$\Phi$ bias potential

Figure 1: Cartoon view of some of the parameters important for molecular junction transport. The molecule is shown schematically in a metal junction. The parameters are the relevant energies that determine the nature and the mechanisms of junction transport.

This review deals with the effect of electron-phonon[1] interactions in molecular conduction junctions. The interplay between electronic and nuclear dynamics in molecular systems is a significant factor in molecular energetics and dynamics with important, sometimes critical, implications for molecular structure, spectroscopy, electron transfer and chemical reactions. For example, electron transfer in condensed phases, a process akin to molecular conduction, would not take place without the active participation of nuclear motions. This statement cannot be made about molecular conduction, still electron-phonon interactions are associated with some key junction properties and can strongly affect their operation.

Consider condensed phase electron transfer between a donor and an acceptor center in a molecular system. As pointed out above, a strong interaction of the electronic process with the nuclear environment is a critical component of this process. Indeed, this rate process is driven by the polaron-like localization of the transferred electron at the donor and acceptor sites. This is expressed explicitly by the Marcus expression for the non-adiabatic electron transfer rate[20-22]

$$k_{et} = \frac{2\pi}{\hbar} |V_{DA}|^2 \, \mathcal{F} \tag{1}$$

where $V_{DA}$ is the coupling between the donor (D) and acceptor (A) electronic states and where

$$\mathcal{F} = \mathcal{F}(E_{AD}) = \sum_{v_D}\sum_{v_A} P_{th}\left(\varepsilon_D(v_D)\right) \left|\langle v_D | v_A \rangle\right|^2 \delta\left(\varepsilon_A(v_A) - \varepsilon_D(v_D) + E_{AD}\right) \tag{2}$$

is the thermally averaged and Franck Condon (FC) weighted density of nuclear states. In Eq. (2) $v_D$ and $v_A$ denote donor and acceptor nuclear states, $P_{th}$ is the Boltzmann distribution over donor states, $\varepsilon_D(v_D)$ and $\varepsilon_A(v_A)$ are nuclear energies above the corresponding electronic origin and $E_{AD} = E_A - E_D$ is the electronic energy gap between the donor and acceptor states. In the classical limit $\mathcal{F}$ is given by

$$\mathcal{F}(E_{AD}) = \frac{e^{-(\lambda + E_{AD})^2 / 4\lambda k_B T}}{\sqrt{4\pi\lambda k_B T}} \tag{3}$$

where $k_B$ is the Boltzmann constant and $T$ is the temperature, and where $\lambda$ is the reorganization energy, a measure of the nuclear energy that would be dissipated after a sudden jump from the electronic state describing an electron on the donor to that associated with an electron on the acceptor. A simple approximate expression for the relationship between electron transfer rate $k_{D \to A}$ across a given molecular species and the low bias conduction $g$ of the same species in the coherent transport regime is[23]

$$g \approx \frac{8e^2}{\pi^2 \Gamma_D^{(L)} \Gamma_A^{(R)} \mathcal{F}} k_{D \to A} \qquad (4)$$

where $\Gamma_D^{(L)}$ is the rate of electron transfer from the donor (assumed to be attached to the left electrode) into the electrode while $\Gamma_A^{(R)}$ is the equivalent rate for the acceptor on the right electrode.[2] Eq. (4) shows that the nuclear processes that dominate the electron transfer rate do not appear in the corresponding conduction, in which the driving force originates from the coupling between electrons on the molecule and the infinite electronic baths provided by the leads. Indeed, inelastic effects in molecular conduction junctions originate from coupling between the transmitted electron(s) and nuclear degrees of freedom on the bridge during the electron passage.[3] Important consequences of this coupling are:

(1) Far from resonance, when the energy gap between the molecular highest occupied molecular orbital (HOMO) or lowest unoccupied molecular orbital (LUMO)[4] and the nearest lead Fermi energy is large relative to the relevant phonon frequencies and corresponding electron-phonon couplings (a normal situation for low bias ungated molecular junctions), this coupling leads to distinct features in the current-voltage response. Indeed, inelastic electron tunneling spectroscopy (IETS, see Section 5b) provides a tool of increasing importance in the study of structure and dynamics of MTJs, and much of the impetus for the current interest in electron-phonon effects in MTJs is derived from these experimental studies.[29] [30] [31-36] [37, 35, 38-40] Such experiments not only confirm the presence of the molecule in the transport junction, but can also be analyzed to show particular normal modes and intensities, help interpret the junction geometry and indicate mechanisms and transport pathways.[37, 38, 41-45]

---

[1] The term "phonons" is used in this review for vibrational modes associated with any nuclear vibrations, including molecular normal modes.

---

[2] Eq. (4) is valid when the molecular D and A electronic levels are not too far (relative to their corresponding widths $\Gamma$) from the metal Fermi energy.

[3] Obviously such effects exist also in electron transfer processes, see, e.g. Refs. They are responsible, for example, for the crossover from coherent tunneling to non-coherent hopping in long range electron transfer.

[4] The HOMO/LUMO language is very commonly used in describing molecular transport junctions, but it can be quite deceptive. For a bulk system, electron affinity and ionization energy are the same in magnitude. This is not true for a molecule, where the one-electron levels change substantially upon charging. This is a significant issue, involving the so-called 'band lineup" problem[27], [28] in descriptions of junction transport.

(2) When conditions for resonance tunneling are satisfied, i.e. when the bias is large enough and/or appropriate gating is applied, and provided that the molecule-lead coupling is not too large, the inelastic tunneling spectrum (resonance inelastic electron tunneling spectroscopy, RIETS, see Sect. 5c) changes qualitatively, displaying features associated with the vibrational structure of the intermediate molecular ion.

(3) When temperature is high enough, thermal activation and dephasing can change the nature of the transport process from coherent tunneling (away from resonance) or coherent band motion (in resonance) to incoherent hopping. This is manifested in the temperature and length dependence of the transport process and the ensuing conduction. (Section 4)

(4) Electron phonon coupling is directly related to the issue of junction heating and consequently junction stability.[46] This in turn raises the important problem of heat conduction by molecular junction. A stable steady state operation of a biased molecular junction depends on the balance between heat generation in the junction and heat dissipation by thermal conduction.(Section 9).

In addition, some of the most important properties of molecular junctions are associated with the strong dependence of the junction transport properties on the bridge nuclear conformation. This static limit corresponds to "frozen" vibrations – that is, configurational modulation of the electronic properties, leading to changes in the conductance spectrum due to displacement of nuclear coordinates[47-56, 43, 57, 58] similar to the well known geometry-dependence of optical spectra. A recent demonstration of a rather extreme situation is a system where conduction can take place only when the bridge is vibrationally excited.[59] This is essentially a breakdown of the Condon approximation for conductance, which assumes that the transport is independent of the geometry. Experimentally this is observed in the so-called "stochastic switching" phenomenon.[47-49] An example of this behavior is shown in Figure 2, demonstrating the rapid changes in the current observed in a MTJ based on a self-assembled monolayer (SAM) using a transporting impurity molecule in an alkane thiol host.[60, 61, 47] The simplest understanding of such stochastic switching is very similar to that involved in the understanding of spectral diffusion in single-molecule spectroscopy,[62] where evolution of the environment results in shifting of the peaks in the absorption spectrum in a random, stochastic fashion. In molecular junctions, changing the geometry in which the transport occurs similarly modulates the electronic Hamiltonian, and therefore changes the conductance. Because these geometric changes occur by random excitations, the process appears to be stochastic. Recent electronic structure studies[43, 57, 63, 58] demonstrate quite clearly that modulation of the geometry at the interfacial atom (still usually gold/thiol) can result in changes in the conductance ranging from factors of several fold for the most common situations to factors as large as 1000 if coupling along the electrode/molecule tunneling direction changes.[64-67] Under some conditions switching can be controlled, indicating the potential for device application.[68-76] Note that switching is sometimes associated with charging (changing the oxidation state) of the molecular bridge, that may in turn induce configurational change as a secondary effect.[77-81]) See also Section 8 for a simple model for this effect.

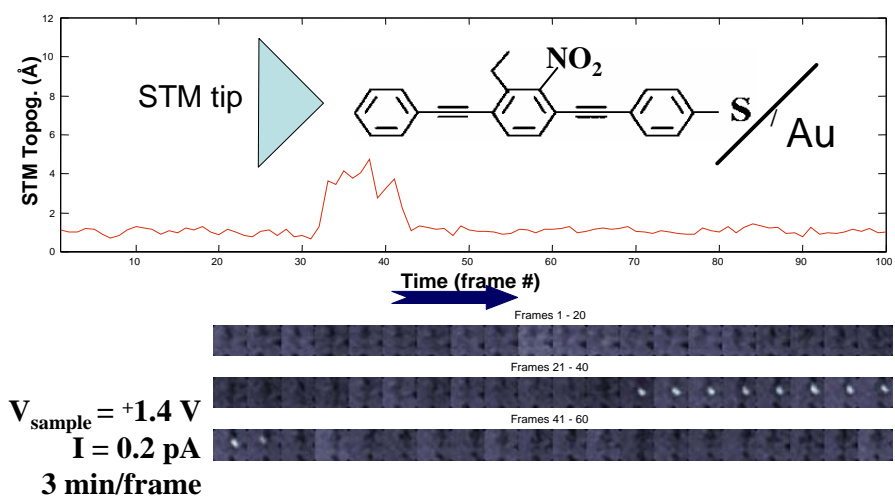

Figure 2: Stochastic switching in a molecular junction. The conjugated phenyleneethynylene oligomer is present on a gold surface, as an impurity molecule in a film of alkane thiol. The current is measured for 180 minutes using an STM tip. Notice the fluctuations in the magnitude, particularly in the area between 30 and 40 time frames. From Ref. [82]

While the discussion in this review focuses on structures in which a molecule or molecules bridge between the source and drain electrodes, a substantial part of the relevant literature focuses on nanodots – mostly small metal or semiconductor particles – as the bridging unit. It should be emphasized that to a large extent the difference is only semantic; in both cases the essential character of the bridge results from its finite size, assumed small enough to show quantum effects in the ensuing dynamics. In some aspects the differences are real even if the borderline is never well defined: molecules can be considered as small and flexible nanodots. The smaller bridge sizes encountered in molecular junctions imply that their energy level spacing is relatively large, larger than $k_B T$ at room temperature, and that charging energies are large so multiple charging is rare (see however Ref. [28]). In addition, molecules are less rigid and more amenable to structural changes that can strongly affect transport behavior. Molecules are therefore expected to show strong consequences of coupling between electronic and nuclear degrees of freedom.

Molecular physics relies on the Born-Oppenheimer (BO) separation between electronic and nuclear dynamics as a crucial theoretical and interpretative tool, and as a starting point for discussing the consequences of electron-vibration interaction. For molecules interacting with metal electrons this separation is not obvious. Still, because for weak molecule-metal interactions the isolated molecule picture is a reasonable starting point, the BO picture and the concept of nuclear molecular potential surfaces are useful for interpreting vibronic effects in conductance spectroscopy. In addition to standard molecular timescale considerations, one should consider the so called tunneling traversal time (see Section 3c). Such considerations suggest that in many experiments vibrational interactions constitute only small perturbations on the electronic transmission process. In gated MTJs that exhibit resonance tunneling transport, the tunneling traversal time can be of the order of or longer than vibrational period, and strong interactions are expected and observed.[83-92, 27]

Electron-phonon models are pervasive in the condensed matter and molecular physics literatures, and most theoretical discussions of effects of electron-phonon interaction in MTJs use extensions of such generic models. In general, treatments of many-body dynamical processes use a convenient separation of the overall system into the system of interest, henceforth referred to as the *system*, possibly subjected to external force(s) (e.g. a radiation field or a deterministic mechanical force), and a bath or baths that characterize the thermal environment(s). The choice of system-bath separation depends on the particular application. When we focus on the electron transmission process it is natural to consider the molecular bridge as the system interacting with several baths: the left and right leads are modeled as free-electron metals, each in its own thermal equilibrium, and the nuclear environment is modeled as a thermal boson bath. In studies of electron-vibration interaction by inelastic tunneling spectroscopy it is convenient to take those vibrational modes that are directly coupled to the tunneling process as part of the system. Such modes are referred to below as 'primary phonons'. They are in turn coupled to the rest of the thermal nuclear environment, which is represented by a boson bath ('secondary phonons'). Finally, we may be interested mainly in the dynamics in the primary vibrational subspace, for example when we focus on heating and/or configuration changes induced by the electronic current. In this case the relevant vibrational modes constitute our system, which is driven by its coupling to the biased electrodes and to the thermal environment.

This overview starts with a brief description of experimental issues (Section 2) and a discussion of the important time and energy scales in the problem (Section 3). Following these introductory sections we focus on the important physical phenomena associated with electron-phonon interactions in molecular junctions: the crossover from coherent to incoherent transport (Section 4), inelastic electron tunneling spectroscopy (section 5) together with manifestations of electron-electron interaction (Section 6) and noise (Section 7) in this spectroscopy. We also discuss non-linear conductance phenomena such as hysteresis and negative differential resistance (NDR) that may result from electron-phonon interaction (Section 8), and processes that center on the phonon subsystem: heating and heat conduction (Section 9) and current induced chemical reactions (Section 10). We end with a brief summary and outlook in Section 11.

## 2. Experimental Background – Test Beds

The experimental realization of MTJs has been strongly associated with the development of both the appropriate chemical methodologies for preparing molecule/electrodes interfaces and the development of nanoscale characterization and preparation techniques, particularly scanning probe microscopy. The interest in MTJ has led to development of test beds, generalized methods for aligning molecules between electrodes and making conductance measurements. There are several different ways in which test beds can be characterized.

(1) The simplest measurements, and some of the most important, were made by placing the molecule on a surface, then inquiring about the nature of that molecule and transport through it using a scanning tunneling microscope tip. This implies that in general there is a large vacuum gap between a non-bonded molecular terminus and the tip edge. This results in effectively all the voltage drop occurring in that vacuum gap, and also in the validity of the Tersoff-Hamann picture,[93-95] which implies that conductance is proportional to the local density of electronic states at the tip position and at the Fermi energy. These STM type measurements have been crucial in understanding many transport properties of molecules,[32, 33, 96-100] however their structure is not that of a typical MTJ because only one of the electrodes is in close contact with the molecule.

(2) Two terminal vs. three terminal junctions. When a molecular adlayer is aligned between two electrodes, in the absence of a third gate electrode, one simply measures the current/voltage spectroscopy with no reference potential. This has been so far the most common measurement experimental observation in MTJs.[101, 102] A third, gate terminal, can be assembled. Because the gate length (source to drain distance) for MTJ's is generally much smaller than in CMOS transistors, large gating voltages are required to modulate the electronic levels of the molecule in the junction. Such setups[39, 83, 84, 86-92, 27] can be used to change the injection gap and consequently the nature of the transport process from coherent tunneling to hopping behavior (see Section 4).

(3) Electrodes – metal or semiconductor. Because of the facility with which thiol/gold structures can be self-assembled, nearly all MTJ's reported to date have used metals, nearly always gold, as either one or both electrodes. Work using semiconductor electrodes including silicon,[103-107] carbon[108-110] and GaAs,[111] have been reported; because transport involving semiconductors is dominated by their band gap, much richer transport behavior can be expected there.[112-114] Moreover the covalent nature of many molecule-semiconductor bonds implies the potential for reducing geometric variability and uncertainty. Nevertheless, the assembly of such structures is more difficult than the thiol/metal structures, and such measurements are still unusual although they are beginning to appear more often.

(4) Single molecules vs. molecular clusters. In the ideal experiment, one would assemble a single molecule between two electrodes, in the presence of a gate electrode, and with well defined geometry. It is in fact difficult to assure that one single molecule is present in the junction. Many measurements have been made using self-assembled monolayers (SAM's) on suitable substrates, with counter electrodes developed using a series of methods ranging from so-called nanopores[115] to metallic flakes[116] to metal dots[117] to indirectly deposited metals.[118, 119] Measurements using SAM structures include so-called crosswire test beds,[120] suspended nanodots test beds, [74] nanopores, [115] mesas[121] and in-wire junctions.[122] The effect of intermolecular interactions is relevant here: while theoretically one might expect measurable dependence on such interactions as well as possible coherence effects over several neighboring tunnel junctions,[123] for the most part the interpretation in terms of simple additivity have been successful.[124, 125]

(5) The presence of solvent can substantially modify transport, both due to effects of solvent polarization (that dominate traditional molecular electron transfer)[18, 126] and because of possible changes in the molecule-lead interaction at the interface due to the presence of a solvent molecule (particularly water).[106] On the other hand, electrochemical break junctions[79] are one of the newest and most effective ways to make statistically significant multiple junction measurements.[89, 80, 127, 51, 128] Electrochemical gating[129] in such structures permits observations of different limits of transport, and the rich statistics obtainable from such measurements increase understanding of the transport spectroscopy.[127] On the other hand, solvated junctions cannot be studied below the solvent freezing point while, as already indicated, low temperatures are generally required for complete characterization of vibrational effects on conductance spectroscopy.

(6) Single junctions or junction networks. Most MTJ measurements are made on one molecular bridging structure (single molecule or SAM) suspended between two electrodes. Recent work in several groups has considered instead a two dimensional network of gold nanoparticles, with molecular entities strung between them, terminated at both ends by thiol groups.[130-132] Transport through such a sheet is more easily measured than through a single junction, and certain sorts of averaging make the interpretation more straightforward. The molecular wires connecting the gold dots have ranged from simple alkane dithiols to more complicated redox wires, in which a transition metal center is located that can undergo oxidation state changes.

While data from all of these test-beds is important for understanding conductance spectroscopy, vibrational effects have generally been studied only with metallic electrodes, in STM junctions and in two terminal or three terminal geometries. Both single molecule break junctions and molecular SAMs have been used as samples, and nearly all vibrationally resolved measurements have been made in the solvent free environment, using a single junction rather than a network.

## 3. Theoretical approaches

### 3a. A microscopic model

We consider a two-terminal junction with leads (left, L and right, R) represented by free electron reservoirs each in its thermal equilibrium, coupled through a bridging molecular system. A third lead, a gate G, capacitively coupled to the bridge, may be present as well (see Fig. 3). This gate provides a potential that changes the energies of molecular states relative to the leads. The molecular bridge, possibly with a few of the lead atoms on both sides, constitutes the "extended molecule" that will be considered as our system. Electron-phonon coupling is assumed to be important only on the extended molecule, and will be disregarded elsewhere. Nuclear motions, within the bridge, on the leads and in the surrounding solvent are described as two groups of harmonic oscillators. The first group includes those local vibrations on the extended molecule that directly couple to the electronic system on the bridge; it is sometimes referred to as 'primary'. The second, 'secondary' group, includes modes of the nuclear thermal environment that are assumed to remain at thermal equilibrium with the given laboratory temperature $T$.[5] Under steady state operation of a biased junction the primary vibrations reach a non-equilibrium steady state driven by the nonequilibrium bridge electronic system on one hand and by their coupling to their electronic and nuclear thermal environments on the other.

Reduced units, $e=1$, $\hbar=1$ and $m_e=1$, are used throughout this review, although at times we write these parameters explicitly for clarity. The Hamiltonian of this model is given by

$$\hat{H} = \hat{H}_M + \sum_{i=1}^{N} \Phi_i^{ext} \hat{d}_i^\dagger \hat{d}_i + \hat{V}_{M-out} + \hat{H}_{out} = \hat{H}_0 + \hat{V} = \hat{\tilde{H}}_0 + \hat{\tilde{V}} \qquad (5)$$

$$\hat{H}_{out} = \sum_{k \in L,R} \varepsilon_k \hat{c}_k^\dagger \hat{c}_k + \sum_{\beta=L,R} \omega_\beta \hat{b}_\beta^\dagger \hat{b}_\beta \qquad (6)$$

$$\hat{H}_M = \sum_{i,j=1}^{N} H_{ij}^M \hat{d}_i^\dagger \hat{d}_j + \sum_{\alpha} \omega_\alpha \hat{a}_\alpha^\dagger \hat{a}_\alpha + \sum_{i,j;\,\alpha} M_{ij}^\alpha \hat{Q}_\alpha^a \hat{d}_i^\dagger \hat{d}_j \qquad (7)$$

$$\hat{V}_{M-out} = \sum_{k \in L,R;\,i} \left( V_{ki} \hat{c}_k^\dagger \hat{d}_i + V_{ik} \hat{d}_i^\dagger \hat{c}_k \right) + \sum_{\alpha;\,\beta=L,R} U_{\alpha\beta} \hat{Q}_\alpha^a \hat{Q}_\beta^b \qquad (8)$$

$$\hat{H}_0 = \sum_{i,j=1}^{N} H_{ij}^M \hat{d}_i^\dagger \hat{d}_j + \sum_{i=1}^{N} \Phi_i^{ext} \hat{d}_i^\dagger \hat{d}_i + \sum_{k \in L,R} \varepsilon_k \hat{c}_k^\dagger \hat{c}_k$$
$$+ \sum_{\alpha} \omega_\alpha \hat{a}_\alpha^\dagger \hat{a}_\alpha + \sum_{\beta=L,R} \omega_\beta \hat{b}_\beta^\dagger \hat{b}_\beta \qquad (9a)$$

$$\hat{V} = \hat{V}_{M-out} + \sum_{i,j;\,\alpha} M_{ij}^\alpha \hat{Q}_\alpha^a \hat{d}_i^\dagger \hat{d}_j \qquad (9b)$$

$$\hat{\tilde{H}}_0 = \hat{H}_0 + \hat{V}_{M-out} \qquad (10a)$$

---

[5] In the discussion of heat conduction, Section 9, the secondary modes include modes of the two leads, represented by thermal baths that may be at different temperatures.

$$\hat{\tilde{V}} = \sum_{i,j;\,\alpha} M_{ij}^{\alpha} \hat{Q}_{\alpha}^{a} \hat{d}_i^{\dagger} \hat{d}_j \qquad (10b)$$

where $\hat{Q}_{\alpha}^{a}$ and $\hat{Q}_{\beta}^{b}$ are vibration coordinate operators

$$\hat{Q}_{\alpha}^{a} = \hat{a}_{\alpha} + \hat{a}_{\alpha}^{\dagger} \qquad \hat{Q}_{\beta}^{b} = \hat{b}_{\beta} + \hat{b}_{\beta}^{\dagger} \qquad (11)$$

and for future reference we also introduce the corresponding momentum operators

$$\hat{P}_{\alpha}^{a} = -i\left(\hat{a}_{\alpha} - \hat{a}_{\alpha}^{\dagger}\right) \qquad \hat{P}_{\beta}^{b} = -i\left(\hat{b}_{\beta} - \hat{b}_{\beta}^{\dagger}\right) \qquad (12)$$

$\hat{a}$ ($\hat{a}^{\dagger}$) and $\hat{b}$ ($\hat{b}^{\dagger}$) are annihilation (creation) operators for system (bridge) and bath phonons while $\hat{d}$ ($\hat{d}^{\dagger}$) and $\hat{c}$ ($\hat{c}^{\dagger}$) are similar operators for the system and leads electrons. In (5) the terms on the right are, respectively, the molecular bridge Hamiltonian, an external controllable potential (e.g. a gate potential, $\Phi^{ext} = \Phi_g$) which is assumed to affect only the bridge, the interaction between the molecular bridge and the external reservoirs (usually the leads) and the Hamiltonian for these external reservoirs. The latter, $\hat{H}_{out}$ (Eq. (6)), contains the free electron Hamiltonians for the right (*R*) and left (*L*) electrodes as well as the Hamiltonians for the external phonon baths that represent nuclear motions in the bridge, leads and surrounding solvent which are not directly coupled to the bridge electronic subsystem. Eq. (7) represents a simple model for the bridge Hamiltonian in which the electronic part is modeled as a 1-particle Hamiltonian using a suitable molecular basis (a set of atomic or molecular orbitals, real space grid points, plane waves or any other convenient basis), the vibrational part is represented by a set of harmonic normal modes and the electron-phonon interaction is taken linear in the phonon coordinate. The interaction terms in Eq. (8) are respectively the molecule-electrodes electron transfer coupling and the coupling between primary and secondary phonons, which is assumed to be bilinear.[6] The sets of system ($\{\alpha\}$) and bath ($\{\beta\}$) phonons constitute respectively the "primary" and "secondary" phonon groups of this model. Eqs. (9-10) represent different separation schemes of the total

---

[6] The bilinear form $\sum U_{\alpha\beta} \hat{Q}_{\alpha}^{a} \hat{Q}_{\beta}^{b}$ is convenient as it yields an exact expression for the self energy of the primary phonons due to their interactions with the secondary ones. This form is however not very realistic for molecular interaction with condensed environments where the Debye cutoff frequency $\omega_c$ is often smaller than that of molecular vibrations. Relaxation of molecular vibrations is then caused by multiphonon processes that result from non-linear interactions. Here we follow the workaround used in Ref.[133] by introducing an effective density $\omega^2 \exp(-\omega/\omega_c)$ of thermal bath modes. Such a bath, coupled biliniarly to the molecule mimics the multiphonon process.

Hamiltonian into "zero order" (exactly soluble) and "perturbation" parts. In particular, with focus on the electron-phonon interaction, we will often use the scheme (10) where the bilinear couplings of the system electrons with the external electron reservoirs and of the primary (system) phonons with the secondary phonon reservoirs are included in the zero order Hamiltonian (implying renormalization of electronic energies and vibrational frequencies by complex additive terms).

This model is characterized by several physical parameters. In addition to the coupling parameters appearing explicitly in (5)-(10), two other groups of parameters are often used:

$$\Gamma_i^{(K)} = 2\pi \sum_{k \in K} |V_{ik}|^2 \delta(\varepsilon_i - \varepsilon_k); \quad K = L, R \qquad (13)$$

represents the molecule-lead coupling by its effect on the lifetime broadening on a molecular level $i$ (see also Eq. (30) below for a more general expression) and

$$\gamma_{\alpha, ph}^{(K)} = 2\pi \sum_{\beta \in K} |U_{\alpha\beta}|^2 \delta(\omega_\alpha - \omega_\beta) \qquad (14)$$

are similarly the lifetime broadenings of the primary phonon α due to its coupling to the bath of secondary phonons (in this case on the lead $K$). Also, in the popular polaron model for electron-phonon coupling, where (7) is replaced by the diagonal form $\hat{\tilde{V}} = \sum_{i,\alpha} M_i^\alpha \hat{Q}_\alpha^a \hat{d}_i^\dagger \hat{d}_i$ we often use the reorganization energy of electronic state $i$, a term borrowed from the theory of electron transfer,

$$E_{ri} = \sum_\alpha \left(M_i^\alpha\right)^2 / \omega_a, \qquad (15)$$

as a measure of the electron-phonon coupling.

Below we will often use a simple version (see Fig. 3) of the model described above, in which the bridge is described by one electronic level of energy $\varepsilon_0$ (representing the molecular orbital relevant for the energy range of interest) coupled to one primary vibrational mode of frequency $\omega_0$. The electron-phonon coupling in (7) then becomes $M\hat{d}_0^\dagger \hat{d}_0 \left(\hat{a}_0^\dagger + \hat{a}_0\right)$. Further issues associated with the electron-phonon coupling are discussed in the next Section.

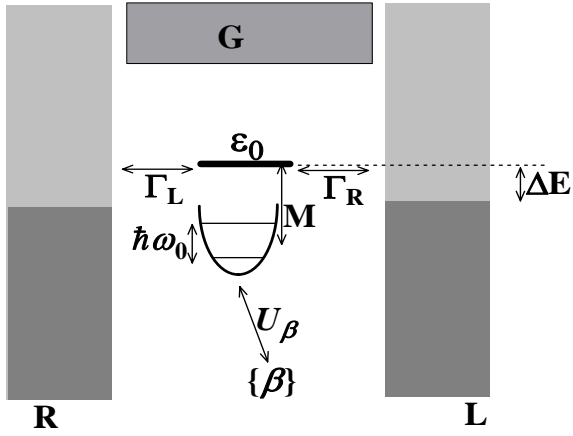

Figure 3. The single bridge level/ single bridge oscillator model. The shaded areas on the right and left denote the continuous manifolds of states of the two leads where dark and light shades correspond to occupied and unoccupied states and the line separating them is the Fermi energy. G denotes a gate electrode. The molecular vibration of frequency $\omega_0$ is coupled linearly to the electronic population in the molecular level of energy $\varepsilon_0$ and to a bath of secondary phonons $\{\beta\}$.

## 3b. The electron-phonon coupling

As in most computational work on inelastic effects in molecular electronic processes, the small amplitude of vibrational motion is usually invoked to expand the electronic Hamiltonian to first order in the deviations of nuclei from their positions in the equilibrium molecular configuration. This leads to the interaction (10b) with coefficients derived from

$$M_{ij}^{\alpha} = \sum_n \sqrt{\frac{\hbar}{2M_n \omega_a}} C_{n\alpha} \langle i | \nabla_{R_n} H_{el}(\mathbf{R}) | j \rangle, \quad (16)$$

where $H_{el}(\mathbf{R})$ is the electronic Hamiltonian for a given nuclear configuration and the sum is over all atomic centers. In (16) $\mathbf{R} = \{R_n\}$ is the vector of nuclear coordinates, $M_n$ are the corresponding masses, $\omega_\alpha$ are the molecular (primary) normal mode frequencies, $C_{n\alpha}$ are coefficients of the transformation between atomic and normal mode coordinates and the derivatives are evaluated in the equilibrium junction configuration. The matrix elements are evaluated as part of the electronic structure calculation that precedes the transport analysis.

Simplified models are often used to study particular issues of the transport process. Often, only diagonal terms $i = j$ are taken in (10b), implying that the electron-phonon coupling is derived from the nuclear coordinate dependence of the energies $\langle i | H_{el} | i \rangle$. The physics of such a model depends on the electronic basis used. When the single electron states $\{i\}$ are molecular orbitals it is hard to justify such simplification. However when $\{i\}$ represents a local (atomic or physically motivated choice of molecular sites) basis, one may argue that a change in the local energy $\langle i | H_{el} | i \rangle$ reflects polarization of the local configuration when the electron occupies the corresponding site, while $\langle i | H_{el} | j \rangle$ is small for $i \neq j$ because of small overlap between functions localized on different sites.

It should be kept in mind however that such an approximation may miss an important part of the tunneling physics. Consider for example two neighboring atoms in the molecular structure. When their relative orientation is such that they lie along the tunneling direction, a motion that modulates the distance between them (and therefore change the overlap between the corresponding atomic orbitals) may strongly affect the tunneling probability. Indeed it is found[44] (see Section 5g) that modes associated with such motions appear prominently in inelastic tunneling spectra. Such motion also dominates shuttle transport in nanojunction[134-141] A different motion with similar or even stronger effect is torsional or rotational motion, that can modulate electronic overlap between neighboring molecular sites.[58] Obviously, disregarding $i \neq j$ terms in Eq. (10b) will miss such effects.

## 3c. Time and energy scales

The measurements discussed throughout this review, and consequently their theoretical consideration, focus on steady-state transport in molecular junctions. Still, time plays a decisive role in the behavior of the system and the nature of the transport process. Indeed, the existence of several important timescales (and associated energies) is what makes the problem rich and interesting as well as complex and sometimes difficult.

Consider first the important energy parameters of the transport problem. $\Delta E$, the injection gap, is the energy difference between the leads Fermi energy and the relevant bridge levels (e.g., the molecular HOMO or LUMO).[4] $\Gamma$ (Eq. (13)), $\gamma_{ph}$ (Eq. (14)) and $E_r$ (Eq. (15)) measure respectively the lifetime broadening of molecular electronic states due to molecule-lead coupling, the broadening of molecular vibrations due to coupling to thermal phonons on the leads and in the environment, and the electron-phonon coupling. In addition, if the molecule is an ordered chain we may consider the bandwidth $V_B$ of its "conduction band" (a single electron property). The charging energy $U$ represents electron-electron interactions on the bridge. Finally the thermal energy $k_B T$ is another important energy parameter.

These energy parameters are directly related to important timescales. $\hbar/\Gamma_K$ measures the lifetime of a bridge electron for escaping into the corresponding lead and $\hbar/\gamma_{ph}$ is the relaxation time for bridge phonons to their thermal phonon environment, with $\Gamma$ and $\gamma$ representing the corresponding rates. For coherent (band) motion on the bridge, $\hbar/V_B$ measures the electron lifetime on a single bridge site.

Some derived time scales (and corresponding rates) that can be obtained from the model Hamiltonian are less obvious. The most important of these is the dephasing time, a measure of the time it takes for the electron to lose its phase due to interactions with its electron and phonon environments. When this time is short enough (see below) the electronic motion becomes incoherent and can be described by successive classical rate processes, sometime referred to as hopping. Another, more elusive and sometimes controversial concept, is the tunneling traversal time, a measure of the time an electron spends in the barrier region as experienced by another degree of freedom, e.g. a vibrational mode, that resides in that region. Using the dynamics of that mode as a clock, the traversal time obtained in the deep tunneling limit for a square barrier of energy height $\Delta E$ and width $D$ is[142] $\tau = D\sqrt{m/2\Delta E}$ where $m$ is the electron mass. If, instead, the bridge is represented by a 1-dimensional lattice of $N$ equivalent sites, $\tau = \hbar N/\Delta E$ .[7] It should be emphasized again that these times correspond to the deep tunneling limit. In the opposite case of resonance tunneling the lifetime of the electron on the bridge is determined by the escape rate $\Gamma$ and/or the bandwidth $V_B$. Sometimes a unified expression

$$\tau \sim \hbar \left(\sqrt{\Delta E^2 + \Gamma^2}\right)^{-1} \tag{17}$$

is used as an estimate of the traversal time per molecular site. For more complex barriers this time can be evaluated numerically.[144]

The relative magnitudes of these energy and time scales determine the physical nature of the transport process. When either $\Delta E$ or $\Gamma$ is large relative to the timescale for electron-phonon interaction dephasing and energy loss can be disregarded and the

---

[7] These two results are limiting cases of a general formula, See Ref. [143].

electron transmission is a coherent quantum process; described as ballistic transmission when $\Delta E \simeq 0$ (resonance). In the opposite case, near resonance transmission in the weak electronic coupling limit, transmission often proceeds by successive hopping. The latter term refers to a process in which the electron hops between the leads and the molecule and between successive sites on the molecule, where at each site complete local thermalization is achieved. The corresponding hopping rates are then calculated from the theory of electron transfer between localized sites[20, 145, 146, 22] or between a molecular site and a metal electrode.[147, 22]

These limits, as well as intermediate situations, can be experimentally monitored by inelastic electron tunneling spectroscopy (Section 5). In the analysis of such experiments we distinguish between weak and strong electron-phonon coupling cases by comparing the magnitudes of the coupling $M$ and the energy parameter $\sqrt{\Delta E^2 + (\Gamma/2)^2}$. For small electron-phonon coupling, $\left| M / \sqrt{\Delta E^2 + (\Gamma/2)^2} \right| \ll 1$, inelastic tunneling can be treated perturbatively with this small parameter using the Migdal-Eliashberg theory.[148] [149] In the opposite limit this treatment breaks down. A transient intermediate molecular ion (essentially a polaron) may form in the junction and its vibrational structure may appear in the inelastic signal as satellite peaks (sidebands) in the conduction/voltage plot near the conduction threshold. For this structure to be resolved another inequality, $\omega_0 > \Gamma/2$, (where $\omega_0$ is the relevant vibrational frequency), has to be satisfied between the system time/energy scales.

These timescale considerations have to be incorporated together with other factors that make the issue more involved. First, temperature plays an important role in determining the dominant transport mechanism. In the framework of the previous paragraph, a finite temperature system is characterized by a distribution of $\Delta E$ values and the ensuing flux is an ensemble average of contributions that can be of different physical nature. Low temperature coherent transmission associated with electron injection with large $|\Delta E|$ can cross over to incoherent transmission at higher temperature where injection at small $|\Delta E|$ dominates. Secondly, molecular chain length is another important factor: incoherent transport becomes more important for longer chains both because dephasing is more effective in such systems and (for off resonance tunneling) because of the exponential falloff of the coherent component. These issues are discussed in Section 4. Finally, the presence of an electron on the molecular bridge can change the physical nature of the bridge itself. Polaron formation is essentially solvation of the electron by the nuclear environment on the bridge (and in the surrounding solvent if present).[8] The reorganization energy (15) is essentially this solvation energy, and the corresponding solvation, or polaron formation, time, is another important and non-trivial time parameter in the problem. When this time is short relative to the time (17) for the electron to remain on the undistorted bridge a transient polaron may form. It is important to note that this process is characterized by an intrinsic feedback mechanism, because as the polaron formation proceeds, the bridge continuously distorts. This changes the relevant energy parameters and consequently the relative timescales. Furthermore, polaron formation brings into play another significant timescale, the polaron lifetime, which now dominates the electron lifetime on the bridge. This feedback property of molecular junctions, associated with their tendency to respond to charging by structural changes, is what sets them apart from junctions based on semiconductor and metal nanodots. Junctions based on molecular bridges that can support charged states by such polaron formation (so called redox molecules) show interesting non-linear transport properties such as multistability, hysteresis and negative differential conductance (see Section 8), and are currently subjects of active research.

The passage of electronic current through a molecular bridge can be accompanied by heating of nuclear degrees of freedom. Junction stability requires that this heating is balanced by thermal relaxation, which brings up the issue of timescales relevant to these processes. The relaxation rate $\gamma_{ph}$, Eq. (14), is one contribution to this process, however for molecules connected to metal surfaces another route for vibrational relaxation is excitation of electron-hole pairs in the metal. It is found that the electronic component, $\gamma_{el}$, often dominates the total rate $\gamma = \gamma_{ph} + \gamma_{el}$.

To end this discussion we reiterate again the emerging general picture. The two extreme limits of junction transport process are (a) an overall transmission process ("cotunneling") whose efficiency is determined by the tunneling probability between the two metal contacts through the molecular barrier and (b) a sequential process in which the electron is transiently localized on the molecule (or successively at several molecular sites) en route between the two contacts. The first occurs when $\sqrt{\Delta E^2 + (\Gamma/2)^2}$ is large enough, i.e. in off resonant transmission or for strong molecule-metal coupling. The second characterizes resonance transmission in weak molecule-lead coupling situations. In the latter case effects of electron-phonon coupling often change the dynamical character of the transmission process, loss of coherence often accompanied by transient stabilization of localized electron states results in hopping conduction with the possibility of strongly non-linear transport.

---

[8] The term "polaron" usually refers to a charging-induced distortion in a polar environment and is usually discussed with the latter represented in the harmonic approximation. The use of this term here should be understood more generally, as any charging-induced configurational change.

These two limiting cases translate into the factors that affect the conduction process. In the non-resonant regime the magnitude of the observed current is dominated by the metal-to-metal tunneling probability. In the resonant case the most important factor is the electron lifetime on the bridge. Assuming that electrons move singly through the junction, the observed current $I$ provides an upper bound to the electron lifetime on the bridge according to $\tau \leq e/I$. This upper limit, about $10^{-9}$s for $I = 1$ nA, indicates that time on the bridge may be long enough for electron-phonon interaction to take effect.

## 3d. Theoretical methods

Theoretical studies of electron-phonon interaction effects in condensed phase dynamics have a long history,[150] still the observation of their consequence in the current-voltage characteristics of metal-insulator-metal junctions, including STM junctions and other types of MTJs, has raised new points for consideration. Treatments of inelastic tunneling are usually done using models similar to those defined by Eqs. (5)-(8). Several theoretical issues are of particular interest

(a) Evaluation and/or estimation of the electron-phonon coupling parameters.

(b) Effects of the thermal environment, as a source of activation, dissipation and dephasing on the electron transport process.

(c) Effects of the non-equilibrium electronic process on dynamical processes in the primary vibrational subsystem, including heating, change of conformation and dissociation.

(d) Evaluation of vibrational signatures in current-voltage spectroscopies.

(e) Manifestation of strong electron-nuclear coupling in resonance transmission situations, in junctions involving redox molecules and in shuttle conduction mechanisms.

(f) The effect of electron-phonon interaction on the current-noise characteristics of the junction.

Electron-phonon coupling, e.g. the parameters defined by Eq. (16), can be estimated from reorganization energies measured in electron transfer reactions,[151] from the lifetime broadening of infrared lineshapes of molecules adsorbed on metal surfaces,[152] and from simple considerations of electron-ion scattering cross-sections.[153] It can also be evaluated from pseudopotential models that were very useful in studies of hydrated electrons[154] and, using first principle calculations, from the nuclear coordinate dependence of the electronic matrix elements (see, e.g., references [155, 44, 156, 157]). A substantial number of recent papers[158, 156, 159-164, 155, 41, 44, 165-167] use the latter approach in combination with some level of transport theory (see below) to make quantitative interpretations and predictions for inelastic tunneling spectra.

Points (b) and (c) above consider opposite ends of the electron-phonon problem in junction transport. On one end we are interested in the way electron transport is affected by thermal interactions. A conceptually simple approach is to consider the transporting electron(s) as a system interacting with its thermal environment, and to seek a reduced equation of motion in the electronic subspace by projecting out the thermal part. In most applications this results in a generalized master equation for the electronic motion,[168-171, 151, 172-177] that can show crossover from coherent tunneling or band motion to activated diffusive transport. For example, Segal and Nitzan[170, 171, 151] have studied an $N$-site tight binding bridge model coupled at its edges to free electron reservoirs, with a general local coupling to a thermal bath:

$$\hat{H} = \hat{H}_M + \hat{H}_B + \hat{H}_{M,B} + \hat{H}_{LEADS} + \hat{H}_{M,LEADS} \tag{18a}$$

$$\hat{H}_M = \sum_{n=1}^{N} E_n |n\rangle\langle n| + \sum_{n=1}^{N-1} \left( V_{n,n+1} |n\rangle\langle n+1| + V_{n+1,n} |n+1\rangle\langle n| \right) \tag{18b}$$

$$\hat{H}_{LEADS} = \sum_{k \in L,R} E_k |k\rangle\langle k| \tag{18c}$$

$$\hat{H}_{M,LEADS} = \sum_{l} \left( V_{l,1} |l\rangle\langle 1| + V_{1,l} |1\rangle\langle l| \right) + \sum_{r} \left( V_{r,N} |r\rangle\langle N| + V_{N,r} |N\rangle\langle r| \right) \tag{18d}$$

$$\hat{H}_{M,B} = \sum_{n=1}^{N} F_n |n\rangle\langle n| \tag{18e}$$

where the thermal bath and its interaction with the system are characterized by the time correlation function

$$\int_{-\infty}^{\infty} dt e^{i\omega t} \langle F_n(t) F_{n'}(0) \rangle = e^{\beta \hbar \omega} \int_{-\infty}^{\infty} dt e^{i\omega t} \langle F_{n'}(0) F_n(t) \rangle \quad ; \quad \beta = (k_B T)^{-1} \tag{19a}$$

for example, a convenient model choice for some applications is

$$\langle F_n(t)F_{n'}(0)\rangle = \delta_{n,n'}\frac{\kappa}{2\tau_c}\exp(-|t|/\tau_c) \tag{19b}$$

This model within the Redfield approximation[178-180] was used[168, 170, 171] to obtain a quantum master equation for electron transport that includes the effect of phonon-induced relaxation and dephasing on the bridge. This equation was in turn applied to evaluate the differential transmission coefficients, $\mathcal{T}_{L\to R}(E_{out}, E_{in})$ and $\mathcal{T}_{R\to L}(E_{out}, E_{in})$, for an electron entering from the left lead with energy $E_{in}$ and scattered into the right lead with energy $E_{out}$, and same from right to left, in the presence of dissipation and dephasing. These resulting transmission coefficients can be used as input for junction transport calculations (see below). A simpler though cruder description of junction transport in the presence of dephasing can be achieved[181-185] by applying a generalized Buttiker probe[186] technique to affect a distribution of phase breaking processes along the conducting channel. We return to these issues in Section 4.

On the opposite end we are concerned with the dynamics in the subspace of the primary phonons. These vibrations are driven out of equilibrium by their interaction with the current carrying electronic system, and their steady state is determined by this interaction together with their coupling to the dissipative environment of the secondary phonons and the thermal electrons in the leads. These phenomena pertain to the issues of heating described in Section 9 and current-induced reactions discussed in Section 10. Theoretical approaches to this problem focus on the balance between the energy deposited into the primary vibrations by the electronic current and the dissipation caused by coupling to the thermal environment. It is usually assumed that this balance is dominated by incoherent dynamics that can be described by kinetic equations in the primary nuclear subspace[187] or in the combined electronic-primary nuclear subspace,[188, 189] and distinction is made between consecutive single phonon excitation processes and multiphonon pathways induced by the formation of a transient molecular ion.[190] Another way to discuss heating and energy balance in the primary nuclear subspace of an MTJ is within the non-equilibrium Green function (NEGF) methodology,[191-195] presented below and further discussed in Sections 5 and 9. This approach makes it possible to describe electronic and energy currents consistently and simultaneously in the electronic and nuclear subspaces, but its complexity limits its applicability to relatively simple models.

Experimentally, the simplest and most direct consequence of electron-phonon interaction in MTJs phenomenology is inelastic tunneling spectroscopy (see Section 5), where vibrational signatures are observed in the current-voltage characteristic of the junction (issue (d)). As in the Landauer approach to transport in static junctions,[196-199] it makes sense to consider also inelastic transport in nanojunctions using scattering theory. Indeed, inelastic electron scattering and tunneling involving vibrating targets in vacuum can be handled essentially exactly.[200-203] Applications to transport in metal-molecule-metal junctions suffer from the fact that scattering cross-sections or transmission coefficients calculated in vacuum do not properly account for the Fermi statistics of the electronic populations in the metal electrodes (see Section 5d). Nevertheless scattering theory based calculations of inelastic junction transport are abundant,[204-211, 158, 212, 156, 213-219, 155, 41, 44, 58, 220] and despite their questionable theoretical basis (see below), appear to provide a practical working approach in the weak (electron-phonon) coupling limit.

A common heuristic way to accommodate scattering theory input in the description of metal-molecule-metal transport is to use Fermi population factors together with vacuum-based transmission coefficients. The core calculation in these approaches is done for the tunneling transmission probability in a scattering-like configuration, where the incoming and outgoing electron is essentially in vacuum. For example an inelastic transmission coefficient $\mathcal{T}(E_{out}, E_{in})$ associated with a process where an electron enters the junction from one electrode, say $L$ (see Fig. 1), with energy $E_{in}$ and leaves to the other electrode ($R$) with energy $E_{out}$, is multiplied by $f_L(E_{in})(1 - f_R(E_{out}))$. Here $f_K(E); K = L, R$ are the corresponding Fermi distribution functions

$$f_K(E) = \left(\left[\exp\left((E - \mu_K)/k_B T\right)\right] + 1\right)^{-1} \tag{20}$$

Perturbative scattering theories, e.g. the Herzberg-Teller-like analysis of the molecular Green's function or the electron propagator,[155, 41, 44] which provide practical methods for calculations involving realistic molecular models in the weak coupling limit(see Section 5d) rely on a similar approach. Such heuristic correction factors are also used in several master equation descriptions of junction transport, see e.g. References [170, 151, 171, 173-175]. As noted above, this approach should be regarded as an approximation (see also Section 5d below and Ref. [199] section 2.6) that is uncontrolled in the sense that it does not become exact when a small parameter vanishes. In addition, many of these approaches disregard the effect of the non-equilibrium electronic system on the phonon dynamics.

In contrast to such heuristically corrected scattering theory methods, a consistent approach to inelastic junction transport is provided by the non-equilibrium Green function (NEGF) formalism.[221-223, 199, 224] In this approach the objects of interest are the Green functions (GFs) of the electron and the primary vibrations on the Keldysh contour[221, 225, 222, 150, 199]

$$G_{ij}(\tau,\tau') = -i\langle T_c \hat{d}_i(\tau)\hat{d}_j^\dagger(\tau')\rangle \tag{21}$$

$$D_{\alpha,\alpha'}(\tau,\tau') = -i\langle T_c \hat{Q}_\alpha(\tau)\hat{Q}_{\alpha'}^\dagger(\tau')\rangle \tag{22}$$

(where the Keldysh time $\tau$ starts and ends at $-\infty$ and where $T_c$ is a contour time ordering operator), their projections $G^r, G^a, G^>, G^<$ (and similarly for $D$) onto the real time axis, and the corresponding self energies (SEs) $\Sigma$ and $\Pi$ and their projections. These functions satisfy the Dyson equations

$$G^r = G_0^r + G^r \Sigma^r G_0^r \ ; \qquad D^r = D_0^r + D^r \Pi^r D_0^r \tag{23}$$

(and similar equations for $G^a$ and $D^a$) and the Keldysh equations

$$G^> = G^r \Sigma^> G^a \ ; \qquad D^> = D^r \Pi^> D^a \tag{24}$$

(and similar equations for $G^<$ and $D^<$). At steady state we focus on the Fourier transform to energy space of these functions. Approximate ways to calculate these functions are described in Sections 5 and 6 below. In particular, in the non-crossing approximation the self energy of any subsystem is made of additive contributions from different interactions that couple it to other components of the overall system. Once evaluated, the electronic GFs and SEs can be used to calculate important observables. The relaxation rates (in general matrices of rate coefficients) for electrons and phonons are given by

$$\Gamma(E) = i\left[\Sigma^r(E) - \Sigma^a(E)\right] \tag{25a}$$

$$\gamma(E) = i\left[\Pi^r(E) - \Pi^a(E)\right], \tag{25b}$$

the molecular spectral function (density of states projected on the molecular subspace) is

$$A(E) = i\left[G^r(E) - G^a(E)\right] \tag{26}$$

and the net steady state current, e.g., from the lead K into the molecule, is obtained

from[226, 227]

$$I_K = \frac{e}{\hbar} \int \frac{dE}{2\pi} \left[ \Sigma_K^<(E) G^>(E) - \Sigma_K^>(E) G^<(E) \right] \qquad (27)$$

where $\Sigma_K^{<,>}$ are lesser/greater projections of the self-energy due to coupling to the lead $K$ ($K = L,R$). The latter are given by

$$\Sigma_K^<(E) = i f_K(E) \Gamma_K(E) \qquad (28)$$

$$\Sigma_K^>(E) = -i[1 - f_K(E)] \Gamma_K(E) \qquad (29)$$

with $f_K(E)$ the Fermi distribution in the lead K, Eq. (20), and

$$[\Gamma_K]_{ij}(E) = 2\pi \sum_{k \in K} V_{ik} V_{kj} \delta(E - \varepsilon_k) \qquad (30)$$

The NEGF formalism provides a powerful, consistent and systematic framework for describing transport phenomena in interacting particle systems and has been extensively applied to electron tunneling in the presence of electron-phonon interaction.[228, 181, 229, 182] [192, 191] Its complexity, however, usually limits its usefulness to relatively simple molecular models and makes it necessary to explore approximate schemes for evaluating the needed GFs and SEs. The most common of these approximations is the Born approximation (BA) and its extension, the self consistent Born approximation (SCBA).[228, 230] [158, 231-239] In particular, Ueba and co-workers[231-233] and Galperin and co-workers[234, 235] applied the NEGF formalism to the resonant level model of phonon assisted tunneling. Similarly, Lorente and Persson[230] have generalized the Tersoff-Hamann approach to the tunneling in STM junctions, using many-body density functional theory in conjunction with the NEGF formulation of Caroli et al.[228] This formalism was later applied to formulate symmetry propensity rules for vibrationally inelastic tunneling.[158] A recently proposed simplified version of the BA approach[161, 166, 240] can handle relatively large systems within its range of validity.

The BA and the SCBA approximation schemes are very useful in weak electron-phonon coupling situations such as those encountered in analyzing inelastic tunneling spectra under off-resonance conditions. Important physical phenomena associated with strong electron-phonon coupling cannot be described within these approximations. Such strong coupling effects arise in resonance inelastic electron tunneling spectroscopy (RIETS) as well as in phenomena controlled by transient electronic population in the bridge. The latter situation is known as the Coulomb blockade regime in the nanodots literature[136-139, 241-246, 140] and its molecular analog is getting increasing attention in studies of molecular bridges with several accessible oxidation states.[28, 129, 247, 80, 86] On the theoretical side published works can be roughly divided into three groups:

(1) Works based on scattering theory considerations, either using multichannel scattering theory[204-211] or a Green function methodology.[203] These are the strong-coupling counterparts to the perturbation-theory based calculations discussed above, e.g. Refs. [155, 41, 44, 248]. As was pointed out above, including the Fermi statistics associated with the electronic population in the metal is done heuristically.

(2) Approaches based on many-body physics methodologies, in particular the non-equilibrium Green function technique. Some of these works[249-252] achieve simplification by disregarding the Fermi population in the leads, rendering them equivalent to the scattering theory approaches. Other workers, e.g. Král[253] (using a generalization of the linked cluster expansion to nonequilibrium situations[9]), Flensberg[254](using the equation of motion approach), Galperin et al[255] (using a small polaron (Lang-Firsov) Hamiltonian transformation within the NEGF framework) and Hyldgaard et al,[256] Mitra et all[257] and Ryndyk et al[258](based on the self consistent Born approximation) go beyond this simplification.[10] With the exception of Refs. [257] and [255], in the works mentioned above the phonon subsystem is assumed to remain in thermal equilibrium throughout the process.

Another important class of techniques is based on path integrals.[259, 260] This technique has been very useful in studies of equilibrium properties of electron-boson systems, e.g. in the context of dynamical image effects in scanning tunneling spectroscopy,[261] [262] [263] where marked differences of the dynamical image potential from its static analog were found when the tunneling time is of the order of or shorter than the inverse surface plasmon frequency. Also, tunneling suppression resulting from strong correlations associated with electron-electron and electron-phonon interactions in single electron traps in metal-oxide-semiconductor field-effect transistors was studied using this approach.[264] [265] Path integrals on the Keldysh contour were used to study effects of strong electron-phonon interactions in tunneling of electrons via magnetic impurities,[266] inelastic tunneling in quantum point contacts[267] and in

---

[9] This approach appears however to be unstable for diagrammatic expansion beyond the first order linked cluster expansion..

resonance tunneling,[268] in particular non-linear conduction phenomena associated with electron-phonon interactions.[269, 270, 92]

Finally, numerical renormalization group methodology was used to study inelastic effects in conductance in the linear response regime.[271, 272, 245, 273]

(3) Many workers[135, 274, 275, 137, 257, 241, 242, 276, 188, 277-280] treat strong electron-phonon coupling situations using kinetic equations that are based on the assumption that the time spent by the transporting electron on the molecule is long relative to decoherence processes (due to electron-electron interactions or the nuclear thermal environment) on the molecular subsystem. This assumption is expected to hold in the weak molecule-leads coupling, the so called Coulomb blockade limit of junction transport. It leads to a kinetic description of the electron hopping in and out of the bridge and coupled to the oscillator motion. For example, Gorelik et al[135] discuss a bridge-shuttle mechanism for electronic conduction in nanojunctions (see Sect. 10) using a classical damped harmonic oscillator model that couples to the electronic process through the bridge charging

$$m\ddot{x} = -kx - \gamma\dot{x} + \alpha\Phi q \qquad (31)$$

where the charge $q$ is obtained from the probability $P_n$ to have an excess number $n$ of electrons on the bridge, $q = e\sum_n nP_n$ which is assumed to satisfy the master equation

$$\frac{2}{\nu}\dot{P}_n = e^{-x/\lambda}\Gamma(n-1,n)P_{n-1} + e^{x/\lambda}\Gamma(n+1,n)P_{n+1} - \left[e^{-x/\lambda}\Gamma(n,n+1) + e^{x/\lambda}\Gamma(n,n-1)\right]P_n$$

(32)

In Eqs. (31) and (32) the oscillator mass $m$, force constant $k$, damping coefficient $\gamma$ and the parameters $\alpha$, $\lambda$ and $\nu$ are constants. The physical picture behind Eq. (32) is that of a junction in which electrons are injected onto the bridge from the source electrode and are absorbed by the drain electrode, at rates that depend on the oscillator coordinate $x$: the latter is assumed to alternately change the tunneling distances between the bridge and these electrodes, in a way that reflects center of mass motion in the tunneling direction.

Later works use a quantum mechanical oscillator model by invoking a master equation in both the electronic and nuclear state-spaces. Most relevant to our discussion is the work of Koch, von Oppen and coworkers,[276, 277, 188] who have used a master equation for the joint probability $P_q^n$ for the molecular bridge to be in a state with $n$ excess electrons and a vibrational level $q$

$$\frac{dP_q^n}{dt} = \sum_{n' \neq n, q'} \left[ W_{q' \to q}^{n' \to n} P_{q'}^{n'} - W_{q \to q'}^{n \to n'} P_q^n \right] - \frac{1}{\tau} \left[ P_q^n - P_{q,eq}^n \sum_{q'} P_{q'}^n \right] \quad (33)$$

---

[10] Note however that Refs. [256], [257] and [258] treat resonance situations in weak coupling situaions characterized by $M < \Gamma$.

The transition rates on the right are of two kinds: direct vibrational relaxation is taken to be characterized by a single relaxation time $\tau$. The rates $W$, calculated from the golden rule, correspond to processes that change the bridge electronic occupation and are proportional to the corresponding Franck Condon (FC) factors $|\langle q(n) | q'(n') \rangle|^2$ and the appropriate Fermi factor.

Such rate equation approaches are very useful in particular for Coulomb blockade situations with strong electron-phonon coupling, where the focus is on the state of the molecular oscillator (see Sections 9 and 10). While the more general NEGF methodology should in principle yield these equations in the appropriate limit, such bridging between the different approaches has not been achieved yet. A promising advance in this direction is offered by the recent work of Harbola et al[281] who have cast the desription of inelastic tunneling in molecular junctions in terms of the density matrix and its evolution in Liouville space.

A special manifestation of strong electron-phonon coupling in the operation of molecular junctions is the occurrence of molecular configuration changes caused by the induced current or by molecular charging.[69, 282, 70, 72, 75, 283-285] In favorable cases such configurational changes can lead to dramatic non-linear current/voltage behaviors such as switching, negative differential resistance and hysteresis in the I/Φ behavior. A full theoretical analysis of these phenomena is complicated by the need to account for the junction transport and the bridge configuration in a self consistent way. Several models and theoretical methods have been recently discussed,[286, 287, 252, 257, 270, 91, 92, 55, 56, 42, 43, 58, 190, 288] however a conclusive theoretical picture is still in formative stage. We return to these issues in Sections 8 and 10.

In addition to the I/Φ behavior, current noise characteristic is another observable that provides important information about the junction operation.[289] Vibrational effects on nanojunction noise were considered by several workers,[290, 275, 291-294] including related work on ac-driven junctions.[295, 296] Substantial work on this issue has been done within the scattering theory approach.[297-300] whose shortcomings were discussed above. NEGF treatments of this problem, discussing resonant shot noise spectra of molecular junctions were published by Zhu and Balatsky[250] and by Galperin et al.[301] Experimental noise studies in MTJs are also beginning to appear[302]. This subject is further discussed in Sect. 7.

## 3e. Numerical calculations

The importance of inelastic phenomena associated with electron transmission through molecular junctions, in particular the emergence of inelastic electron tunneling spectroscopy as a major diagnostic tool in need of theoretical support, has led to a considerable effort to develop relevant transport theories into practical numerical tools. Different numerical approaches to inelastic tunneling spectra[230, 158, 164, 156, 213, 162, 163, 167, 41, 155, 44, 165, 236-238, 303-305, 161, 166, 240] are reviewed in Sect. 5g. The same numerical methodologies have been used also to compute other consequences of inelastic electron transport in nanojunctions, such as mechanical effects including current induced forces[306-309, 160, 190, 219, 141, 310, 31, 157, 288] and junction heating.[212, 213, 165, 161, 166] The latter issues are discussed in Sections 9 and 10.

## 4. Incoherent vs. Coherent Transport

Two very important consequences of the electron interaction with its nuclear environment are the crossovers from tunneling to activated transport and from coherent to incoherent transmission under appropriate conditions. While these effects are not identical (e.g. thermal electron transfer from lead to molecule can be followed by coherent propagation along the molecule), energy and timescale considerations (Section 3c) indicate that they occur under similar conditions: when activated transport dominates it is likely that decoherence within the molecular bridge will be effective. The predicted experimental manifestations of these changes in the nature of the conduction process are, first, a transition from temperature independent to activated transport upon temperature increase, and second, an exponential drop with molecular chain length in the tunneling regime becoming an ohmic 1/length dependence for activated hopping conduction (or independence on length for activated band motion).

These phenomena where discussed by us elsewhere[7, 13] and here we give only a brief overview with emphasis on recent developments.

Early measurements of molecular transport junctions (the field is only a decade old) were made with relatively short molecules connecting metallic electrodes (usually gold).[311] Transport in these systems takes place in the coherent tunneling regime even at room temperature. Indeed, timescale estimates (Section 3c) suggest that the electron-molecule interaction time is in the sub-femtosecond range, implying weak effect of the electron-vibration interaction. The incoherent limit can be approached when gating is possible, such that the injection gap becomes small, giving sufficient time for decoherence resulting from electron-phonon interaction. Indeed, the onset of hopping conduction was recently seen[28] in a measurement of the heptamer of phenylenevinylene within a molecular transport junction. In this system, the long range of the transport, the presence of solvent and the relatively small injection gap partly caused by image effects in the electrodes, result in transient electron localization and phase loss. Another interesting demonstration is seen in DNA junctions, Figure 4,[89] where an exponential length dependence of tunneling through a DNA segment with a large injection gap is replaced by an inverse length dependence in the small gap, near resonance case.

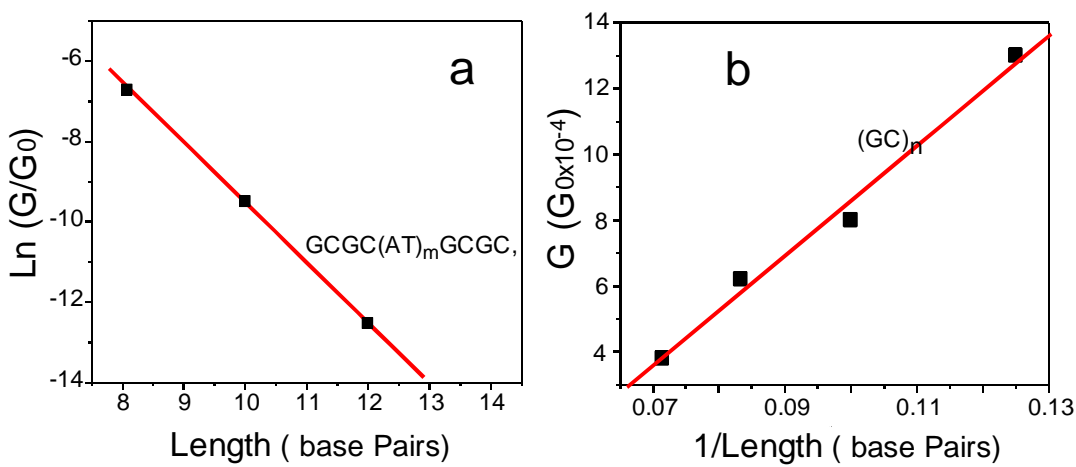

Figure 4. Mechanistic turnover from tunneling to hopping in short strand duplex DNA. The curve on the left shows tunneling through the AT segment, resulting in exponential decay of conductance with length. The segment on the right is for poly GC where transport occurs by hopping and the conductance scales like the inverse length (as it must for diffusion). From Ref. [89].

On the theory side, decoherence and thermal relaxation effects in junction transport have been described using the Buttiker probe technique,[186, 181-185] or by generalized master equations[312, 168, 170, 171, 173, 174] that were already mentioned in Section 3d. Such treatments predict the transition from exponential $\exp(-\beta x)$ to algebraic $(a+bx)^{-1}$ (with $\beta$, $a$ and $b$ constant parameters) bridge length ($x$) dependence of the junction conduction for a finite injection gap $|\Delta E|>0$ as well as the transition from tunneling to activated transport for increasing temperature. We note in passing that understanding decoherence and its proper description in the context of condensed phase transport is still an ongoing process.[313, 314]

With some variation in details, similar predictions are reached by invoking the finite temperature Fermi distribution of electrons in the leads without adhering to dynamical relaxation effects, see e.g. Ref. [315]. Indeed, the authors of Ref. [87] interpret their results on the temperature dependence of conduction in the large injection gap regime (Fig. 5) in the latter way.

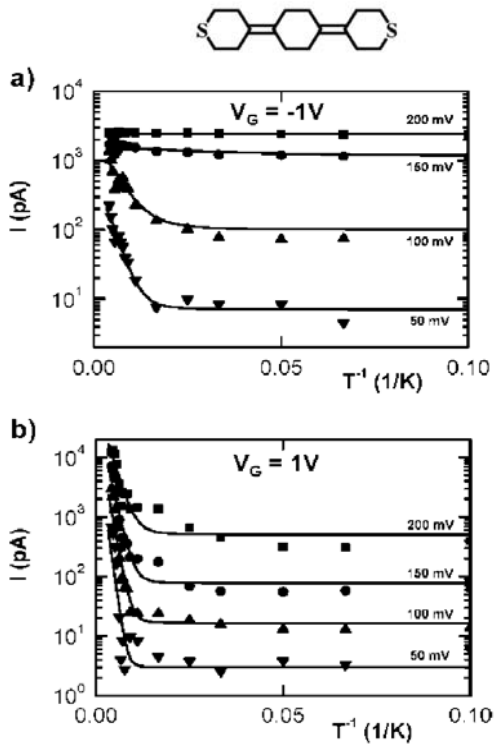

Figure 5. Onset of activated transport in a non-conjugated molecule. At low temperatures, quantum mechanical tunneling is seen, and the current is independent of temperature. As room temperature is approached, the transport becomes activated. The argument made by the authors is that the small activation energies describe the overlap of the Gaussian spectral density tail with the Fermi occupation tail. From Ref. [87].

The experimental studies described above are still quite rare in the molecular conduction literature. Many cases of such transitions have been seen in the closely related phenomena of intra-molecular electron transfer reactions.[316-320, 170, 18]

These are particularly common in biological or biomimetic systems, where clear transitions in the distance dependence of the rates and in the thermal behavior (from temperature independent in the tunneling limit to activated transport in the hopping limit) have been seen.[321]

## 5. Inelastic electron tunneling spectroscopy (IETS)

Inelastic electron tunneling spectroscopy was originally developed nearly one half century ago for studying metal-insulator-metal junctions.[322, 323] Its later and ongoing application in studies of MTJs has been of great importance.[29-45] Indeed, it is the most direct experimental manifestation of electron-vibration coupling in current-carrying molecular junctions and the most extensively studied consequence of this interaction. In addition, the combination of IETS measurements and parallel computational work has made this phenomenon a subject in which the experiments/theory interaction is the closest in all the MTJ literature. Indeed, computational results mirror experiments so well that theory has become a central tool for connecting IETS data to junction structure and dynamics. IETS is used both for demonstrating the presence of particular molecules within the junction and (in conjunction with propensity rules inferred from calculations) for obtaining structural information, e.g. molecular position and orientation in the junction. For example IETS has served to ascertain the presence of a hydrogen molecule in what appeared to be the smallest molecular junction[39], to distinguish between sigma and pi bonding at the molecular termini[44] and even allowing the monitoring of changes in the transport structure due to molecular reactions, such as the binding of water in thiol based gold junctions.[37]

### 5a. Experimental background

Experimental observations of inelastic electron tunneling may be classified according to its electronic resonance or non-resonance nature. In non-resonance inelastic tunneling[324, 322, 323, 32, 325-327, 100, 121, 328] [34, 329, 35] the energy of the incoming electron $E_{in}$ is far from any electronic energy difference $\Delta E$ between the original molecular state and the intermediate molecular ion. Consequently the interaction time between the molecule and the tunneling electron is short, of order $\hbar/|\Delta E - E_{in}|$, and inelastic effects are small. In the opposite resonance case[310, 83, 330] [331] $|\Delta E - E_{in}| < \Gamma$ where $\Gamma$ is the inverse lifetime of the intermediate molecular state. In this case the interaction time is of order $\hbar/\Gamma$ and strong inelastic effects are expected if $\Gamma$ is not too large. These inelastic processes are analogous to the corresponding optical phenomena, ordinary (non-resonance) and resonance Raman scattering (RS and RRS, respectively). In particular, the non resonance inelastic signal reflects the vibrational structure of the original molecular state – the ground electronic state in the RS case or the molecular state in the unbiased junction in the IETS process. The corresponding resonance signal reflects mostly the vibrational structure of the excited electronic state (in RRS) or the transient molecular ion (in RIETS). In spite of these similarities between the optical and tunneling processes some important differences exist as detailed next.

Consider first regular (non-resonance) IETS. In single electron language such processes involve tunneling of electrons whose energy is far from vacant molecular orbitals. (Some processes are more conveniently viewed as tunneling of holes far from resonance with occupied molecular orbitals). Inelastic signal associated with a molecular mode of frequency $\omega$ is observed in the current-voltage response of the junction at the inelastic threshold voltage $e\Phi_{sd} = \hbar\omega$, i.e. when the electron energy associated with the applied bias is just enough to excite the corresponding vibration. If the energies $E_{in}$ and $E_{out}$ of the incoming and outgoing electron could be resolved, we would have expected a peak in the electron flux plotted against the difference $|E_{in} - E_{out}|$ at the point where this difference equals $\hbar\omega$. This is similar to the analogous light scattering process or to inelastic electron scattering off molecular species in vacuum. In the language of the light scattering literature the peak at $E_{in} - E_{out} = \hbar\omega$ is a Stokes signal while that at $E_{in} - E_{out} = -\hbar\omega$ is an anti-Stokes signal whose intensity vanishes at $T \to 0$. These energies are however not resolved in the tunneling current, which is an integral over all incident and outgoing energies of the Fermi-weighted energy-resolved spectrum. Therefore the peak structure is expected, and often observed, in the second derivative, $d^2I/d\Phi_{sd}^2$.

In spite of the similarities described above, it is important to keep in mind that there is no full analogy between Raman scattering or vacuum electron scattering and IETS. An important difference stems from the fact that in the latter the incoming and outgoing state manifolds are partly occupied, given rise to important effects associated with the fermion nature of the scattered electrons. In particular, as discussed below, this results in contribution to the scattering intensity $d^2I/d\Phi_{sd}^2$ at $e\Phi_{sd}=\hbar\omega$ of quasi-elastically scattered electrons,[11] as well as interference between the elastic and quasielastic amplitudes. This can modify the observed feature which may appear as peaks (Fig. 6), dips (Figs. 7, 8), or derivative-like features (Fig. 9) in the $d^2I/d\Phi_{sd}^2$ spectrum.

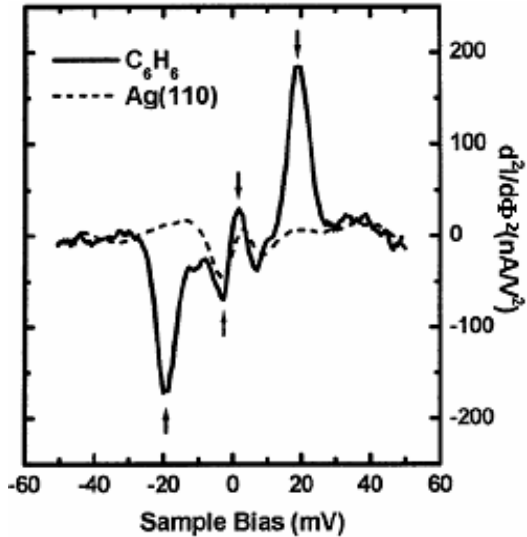

Fig. 6. Inelatic tunneling spectrum ($d^2I/d\Phi_{sd}^2$) acquired by STM on top of a single benzene molecule (continuous line) and on the bare silver surface (dashed line). The peaks at ±4 mV and ±19 mV represent a change of the junction conductance of about 1% and 8%, respectively. (From Ref. [327])

---

[11] The term quasielastic scattering is used to describe electrons that emerge at essentially the incoming energy following interaction with the phonon subsystem. For example, to second order in the electron-phonon interaction this is an electron that has (virtually) absorbed and emitted phonons of the same frequency. The implication of the Fermi function in the corresponding contribution to the scattering signal leads to a distinct spectral feature at $e\Phi_{sd}=\hbar\omega$.

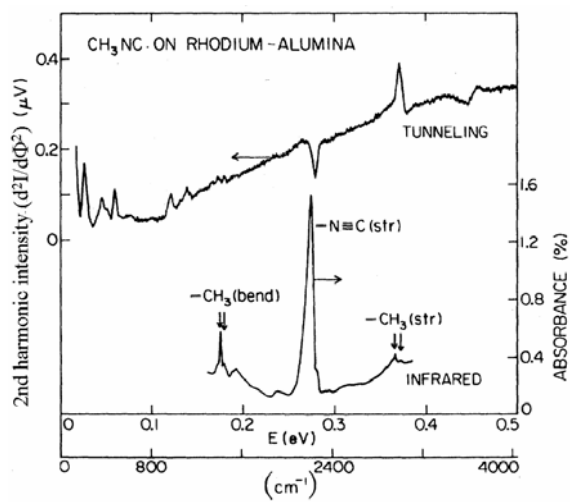

Figure 7. $d^2I/d\Phi_{sd}^2$ tunneling spectrum of CH$_3$NC (methyl isocyanide) molecules bonded to alumina supported rhodium. The infrared absorption spectrum (R.R. Cavanagh and J. T. Yates, Jr. Surf. Sci. 99, **L381** (1980)) is shown for comparison. The dip in the tunneling spectrum is seen at the same position where the IR spectrum has an intense peak due to the NC stretch vibration. (From Ref. [324]).

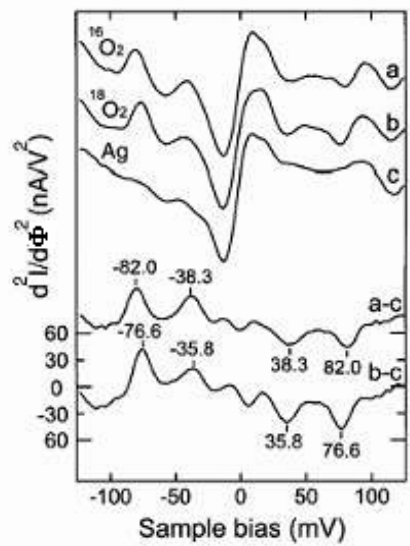

Figure 8. Single molecule inelastic tunneling spectra ($d^2I/d\Phi_{sd}^2$) obtained by STM-IETS for $^{16}$O$_2$ (curve *a*), $^{18}$O$_2$ (curve *b*), and the clean Ag(110) surface (curve *c*). The difference spectra (curve *a-c*, curve *b-c*) are also shown. (From Ref. [326]).

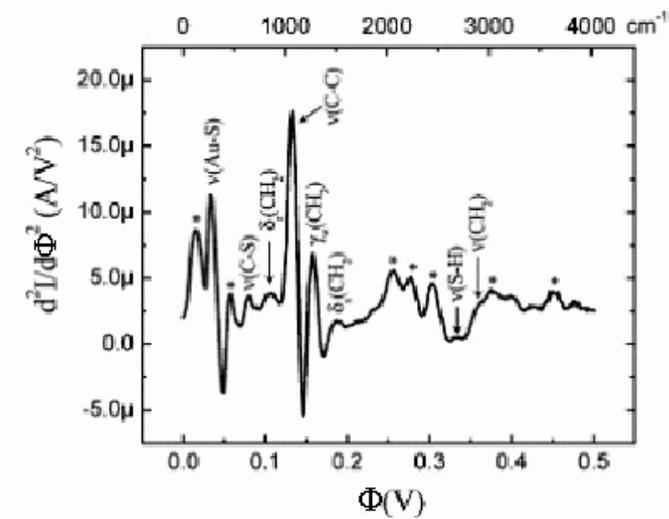

**Figure 9.** Inelastic electron tunneling spectra, $d^2I/d\Phi_{sd}^2$, of octane dithiol SAM obtained from lock-in second harmonic measurements with an AC modulation of 8.7 mV (RMS value) at a frequency of 503 Hz (T = 4.2 K). Peaks labeled * are assigned by the authors to background due to the encasing $Si_3N_4$. (From Ref. [34])

Next consider resonance inelastic electron tunneling. In this case the energy of the initial state, i.e. the original molecular state plus the incoming electron, is close to that of intermediate molecular state on the bridge, and the dominant vibrational structure is associated with the vibrational levels of the latter. In the language of single electron states, the energy of the incoming electron is close to that of the available electronic orbital (usually the lowest unoccupied or the highest occupied electronic orbital (LUMO or HOMO)[4]) in the molecular bridge. This happens when this orbital enters the window between the Fermi energies of the source and drain leads, a situation realized by imposing a higher potential bias or by shifting the molecular energy with a gate potential, and is marked by a step in the current when plotted against $\Phi_{sd}$, i.e. a peak in the conductance $dI/d\Phi_{sd}$. Vibrational states of the intermediate molecular ion serve as additional resonance levels,[12] however the corresponding conductance peaks are weighted by the corresponding Franck Condon (FC) factor.

Mathematically, provided that Γ is not too large, this process corresponds to the strong electron-vibration coupling limit, whose physical signature is multiple phonon peaks, sometimes referred to as phonon sidebands, in the $dI/d\Phi_{sd}$ spectrum when plotted against $\Phi_{sd}$ or $\Phi_g$.[331] This structure is often observed as satellite lines above the conduction thresholds in the diamond structures that represent the conductance plotted in the $\Phi_g - \Phi_{sd}$ plane[310, 83] [330] as seen in Fig. 10. It is of interest to note that vibrational effects in resonance tunneling were seen also in the observation of vibrational structure in the light emitted by molecules in biased STM junctions.[96]

---

[12] In analogy to resonance Raman spectroscopy one also expects overtone features associated with the vibrational structure of the neutral molecule. These features are expected to show mainly in the second derivative spectrum and their presence is not usually taken into account in analysis of RIETS spectra.

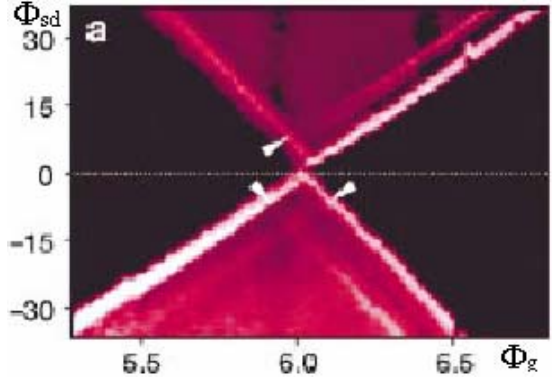

Fig. 10. Conduction of a $C_{60}$ molecule seated between gold leads, plotted against the source-drain and gate potentials. The bright purple areas indicate conducting regimes and the vibrational satellites (attributed to the center of mass oscillations of the $C_{60}$ species between the two leads) are the white lines indicated by white arrows. (from Ref. [310])

Such IETS and RIETS measurements provide effective fingerprints of the corresponding molecular junctions. However, their applicability, particularly for single molecule junctions, is limited by the variability between different measurements, which reflects the structural uncertainties and variance that characterize such junctions. This can be observed for molecules adsorbed at different sites[327] or in junctions prepared by different methods.[35, 34] Nevertheless, IETS and RIETS can provide important information on the presence of molecules in the junction (by its spectral fingerprint), on its orientation (by propensity rules on the electron-vibration coupling[41, 44, 45]) and on its dynamic characteristics (by the lineshapes and linewidths of inelastic features). Extracting this information relies strongly on theoretical interpretation, as discussed next.

**5b. Theoretical considerations – the weak vibronic coupling limit**

As discussed in Section 5b, the weak electron-phonon coupling limit is realized when the electron-phonon interaction (M in Eq. (7)) is small relative to the energy gap $\Delta E$ and/or the electronic lifetime broadening $\Gamma$ (Eq. (13)). In this limit we can apply a standard perturbation approach on the Keldysh contour, where the contour ordered exponent of the evolution operator

$$\hat{S} = T_c \exp\left[-\frac{i}{\hbar}\int_c d\tau \hat{\tilde{V}}_I(\tau)\right] \tag{34}$$

is expanded into Taylor series and truncated in some manner. Here $\hat{\tilde{V}}_I$ is the coupling, Eq. (10b), in the interaction picture defined with respect to the zero-order Hamiltonian (10a). Using the Hamiltonian separation scheme (10) is suggested by the fact that within the non-crossing approximation (or when M = 0) the electron self energy (SE) due to the molecule-leads coupling, and the primary vibration SE due to its coupling to the secondary phonon (thermal) bath can be obtained exactly, while the SE due to electron-vibration coupling can be obtained only using perturbation theory. The perturbation expansion itself can be done on different levels. The simplest approach is to expand the current, Eq. (27), to lowest order in the electron-phonon interaction.[13] A better strategy is to focus the expansion on the self energy. Truncating this expansion at second order in $\hat{\tilde{V}}_I$ leads to the Born approximation (BA) for the electron-vibration interaction. Pioneering considerations of inelastic effects on this level were presented by Caroli et al.[228] A higher level approximation is the Self Consistent Born Approximation (SCBA), where the electron self energy is expressed in a BA form in which the zero order GF is replaced by the full GF, then the GFs and SEs are calculated self-consistently by iterating between them.[14] These BA and SCBA methodologies were used in several theoretical studies.[332, 256, 165, 237, 57, 238, 257, 333, 240, 334, 258]

The SCBA scheme can be used also in a way that treats both electron and phonon Green functions self consistently. This makes it possible to account for the non-equilibrium distribution of the primary phonons in the biased junction. The SCBA expressions (on the Keldysh contour) for the electron and the primary phonon SEs, $\Sigma$ and $\Pi$, respectively, are given by

$$\Sigma_{ij}^{ph}(\tau_1,\tau_2) = i \sum_{l_1,l_2;\alpha_1,\alpha_2} M_{i,l_1}^{\alpha_1} D_{\alpha_1,\alpha_2}(\tau_1,\tau_2) G_{l_1,l_2}(\tau_1,\tau_2) M_{l_2,j}^{\alpha_2}$$
$$+ \delta(\tau_1,\tau_2) \sum_{l_1,l_2;\alpha_1,\alpha_2} M_{i,j}^{\alpha_1} M_{l_1,l_2}^{\alpha_2} \int_c d\tau' D_{\alpha_1,\alpha_2}(\tau_1,\tau')\left[G_{l_1,l_2}(\tau',\tau'+)\right]$$
(35)

and

$$\Pi_{\alpha_1,\alpha_2}^{el}(\tau_1,\tau_2) = -i \sum_{i_1,j_1,i_2,j_2} M_{i_1,j_1}^{\alpha_1} M_{i_2,j_2}^{\alpha_2} G_{j_1,i_2}(\tau_1,\tau_2) G_{j_2,i_1}(\tau_2,\tau_1) \tag{36}$$

---

[13] This implies an expansion of the Green functions in Eq. (27) to this lowest order. A similar expansion within the scattering theory approach[155], [41], [44] is discussed in Sect. 5g.

Π is sometimes referred to as the *polarization operator*, and *G* and *D* are the electron and phonon GFs, respectively. *G* and *D* are given in terms of these SEs using the Dyson equations

$$G(\tau,\tau') = G_0(\tau,\tau') + \int_c d\tau_1 \int_c d\tau_2\, G_0(\tau,\tau_1) \Sigma(\tau_1,\tau_2) G(\tau_2,\tau') \qquad (37)$$

$$D(\tau,\tau') = D_0(\tau,\tau') + \int_c d\tau_1 \int_c d\tau_2\, D_0(\tau,\tau_1) \Pi(\tau_1,\tau_2) D(\tau_2,\tau') \qquad (38)$$

Eqs. (35)-(38) provide a self-consistent scheme, where the GFs of the electrons and the primary vibrations are given in terms of the corresponding SEs while the latter depend on these GFs. These equations (or, at steady state, the Fourier transforms to energy space of their projections on the real time axis) are solved by iterations, starting from some reasonable choice for the GFs, e.g. their values in the absence of electron-vibration coupling, and proceeding until convergence is achieved.

The procedure was used in several theoretical studies of inelastic effects in molecular junctions.[257, 335, 336, 231, 258] We[235, 234] have recently applied this approach to describe generic features in the IETS lineshape, $d^2I/d\Phi_{sd}^2$ plotted against the applied voltage $\Phi_{sd}$. Fig. 11 compares results obtained using the three levels of approximation discussed above: the simplest perturbation theory, the BA and the SCBA. This comparison shows that simple perturbation theory can account for the positions of the fundamental inelastic peaks but not for their shapes, and may fail completely when interference phenomena, such as those giving rise to dips in the tunneling spectrum (see below), dominate the process. Quantitative difference between BA and SCBA is noted as well. It is seen that the inelastic features may appear as peaks (observed in the far off-resonant regime) and dips (observed at particular energetic situations when an electron tunnels from one lead to another through a wide tail of a broadened molecular orbital) as well as derivative-like features in $d^2I/d\Phi_{sd}^2$ (seen in intermediate situations). This interference behavior is reminiscent of Fano lineshapes,[337] known in atomic and molecular spectroscopy, which result from interference between transitions involving coupled discrete levels and continuous state manifolds. The observed signal depends on junction parameters, in particular on the energy and width of the bridge electronic orbital. This suggests that the shape of IETS features may depend on the applied gate voltage, the molecule-lead coupling and the way the bias potential falls on the bridge molecule, as shown in figures 12 and 13. In particular we see that such features may change from peak to dip through intermediate derivative-like shapes as junction parameters are changed.

---

[14] This corresponds to a partial resummation the Taylor series expansion, so that SCBA goes beyond the simple second-order approximation for the self energy.

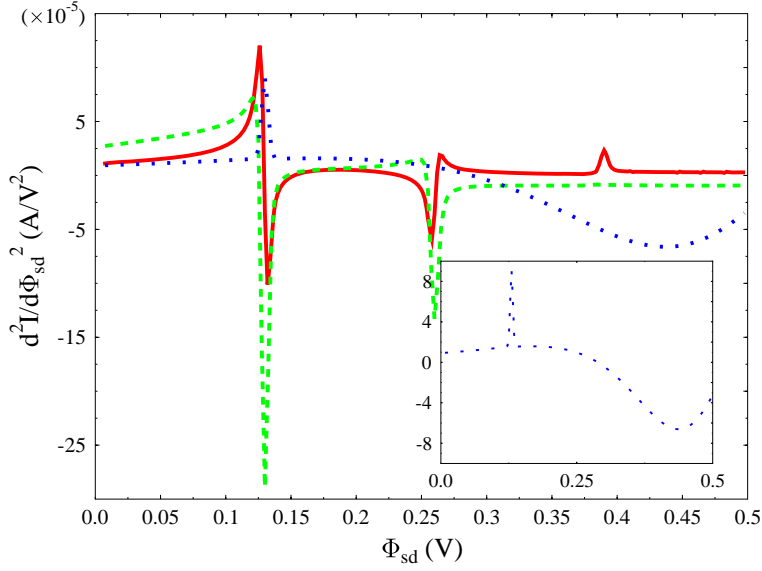

Figure 11. $d^2I/d\Phi_{sd}^2$ plotted against $\Phi_{sd}$ for the single resonant level model characterized by the parameters T=10K, $\varepsilon_0 = 0.6\,\text{eV}$, $\Gamma_L = 0.05\,\text{eV}$, $\Gamma_R = 0.5\,\text{eV}$, $E_F$=0, $\omega_0 = 0.13\,\text{eV}$, $\gamma_{ph} = 0.001$ eV and M=0.3 eV. The full line shows the result of the SCBA calculation; the dashed line – the BA result and the dotted line – the result of simple 2$^{nd}$ order perturbation theory. An expanded view of the latter is shown in the inset. (From Ref. [235])

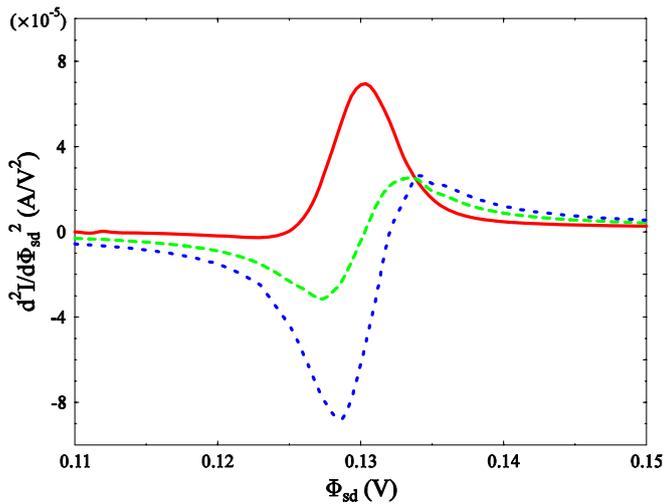

Figure 12. IETS threshold feature in $d^2I/d\Phi_{sd}^2$ for the one resonant level model with the parameters T=10K, $\Gamma_L = \Gamma_R = 0.5\,\text{eV}$, $E_F$=0, $\omega_0 = 0.13\,\text{eV}$, $\gamma_{ph} = 0.001\,\text{eV}$ and M=0.3eV. The different lines correspond to different positions of the resonance level relative to the Fermi

energy (as may be changed by a gate potential): $\varepsilon_0 = 0.7$ eV (solid line, red), 0.6 eV (dashed line, green), and 0.55eV (dotted line, blue). (From Ref. [235]).

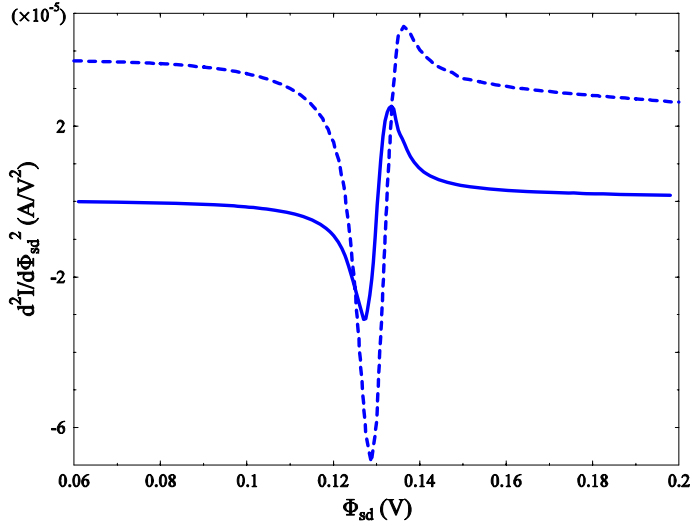

Figure 13. The IETS threshold feature in $d^2I/d\Phi_{sd}^2$ for the one resonant level model characterized by the parameters T=10K, $\varepsilon_0 = 0.6$ eV, $\Gamma_L = \Gamma_R = 0.5$ eV, $E_F$=0, $\omega_0 = 0.13$ eV, $\gamma_{ph} = 0.001$ eV, M=0.3eV. The full line corresponds to the case where the Fermi energies are shifted under the bias according to $\mu_L = E_F + (\Gamma_R/\Gamma)|e|\Phi_{sd}$, $\mu_R = E_F - (\Gamma_L/\Gamma)|e|\Phi_{sd}$, while the dashed line was produced for the model $\mu_L = E_F + |e|\Phi_{sd}$, $\mu_R = E_F$. (From Ref. [235]).

As in other spectroscopies, the widths of IETS features contain in principle information about the underlying dynamical processes. This information can be masked by thermal effects – the thermal width of the Fermi distribution in Eqs. (28) and (29) incorporates itself into threshold features obtained from the integral (27), as well as by inhomogeneous broadening in junctions containing many molecules. However, intrinsic linewidths may be uncovered by careful elimination of these factors.[121, 328, 34] Of particular interest is the nature of the relaxation process that dominates the intrinsic IETS width. The model (5)-(8) is characterized by two such processes: the relaxation of molecular (primary) vibrations to phonon baths in the leads and the rest of the surrounding environment (affected by the interaction parameters $U_{\alpha\beta}$ in Eq. (8)) and the relaxation of these vibrations to electron-hole pairs in the leads (via the interactions $M_{ij}^\alpha$ and $V_{ik}$ in Eqs. and (7) and (8)). Using order of magnitude estimates we have found[234] that the latter mechanism is dominant, contributing an order of ~ 1 meV to the width of IETS features in agreement with experimental data[121, 328, 34] and with analysis of infrared spectral linewidths of molecules adsorbed on metal surfaces.[338, 152] An alternative explanation of the widths observed in Refs. [121, 328, 34], in terms of congestion of unresolved IETS features, was offered by Seminario and Cordova.[159]

## 5c. Theoretical consideration – moderately strong vibronic coupling

The weak coupling methodology described above is used mostly in off resonance situations encountered in standard IETS experiments. SCBA has been used also in the resonant tunneling regime,[256, 257, 240, 334, 258] in cases where weak vibronic coupling results from strong electronic coupling to the leads (large electronic width $\Gamma$) that insures short electron lifetime on the bridge. Electron transport through the junction in the strong electron-vibration coupling case is different from the weak coupling limit discussed above both in the physical nature of the process and in the mathematical approach needed for its description. Physically, in the course of the transmission process the electron occupies the bridge long enough to affect polarization of the bridge and its environment. In the ultimate limit of this situation dephasing (decoherence) and thermal relaxation are sufficient to render the processes of bridge occupation and de-occupation, and often also transmission between different sites on the bridge, independent of each other. This makes it possible to treat the transmission process as a sequence of consecutive statistically decoupled kinetic events (see Section 4). Here we focus on intermediate situations of this strong coupling limit where effects of transient polaron formation on the bridge have to be accommodated, however dephasing is not fast enough to make simple kinetic description possible.[15] This is a difficult situation to describe theoretically, as standard perturbation theory of the kind described in Section 5c breaks down while simple kinetic schemes cannot be applied. In scattering situations (without the presence of the partially occupied electronic manifolds of the leads), in particular in models involving a bridge with a single electronic level, the solution is obtained by applying the small polaron (Lang-Firsov) transformation[150, 339] (see below) to the Hamiltonian[203, 249, 250], which replaces the additive electron-phonon coupling (third term on right-hand side in (7)) by a renormalization of the electronic coupling elements by phonon displacement operators. The renormalized electronic coupling now contains the effects of electron-phonon interaction to all orders, however the transformed Hamiltonian is not amenable to standard many-body perturbation theory techniques. Several attempts to approach this problem within the NEGF formalism[249, 250] introduce approximations which make them effectively equivalent to the scattering approach.

---

[15] Another reason for the treatment presented here to be possibly inadequate for very strong electron-phonon coupling is that vertex corrections are inadequately treated in the many body perturbation theory applied here.

We have recently advanced two computational schemes based on the equation-of-motion (EOM) approach to deal with these issues. In both we restrict our considerations of the model (5)-(8) to the case of a single molecular electronic orbital of energy $\varepsilon_0$ coupled to one primary vibrational mode of frequency $\omega_0$. The first approach,[91] discussed in Section 8, disregards vibrational dynamics and uses a mean-field approximation to describe the effect of electronic occupation on the nuclear configuration. Here we describe another approach[255] that takes into account both vibrational dynamics and (to some extent) electron-vibration correlations, that can account for the phenomenological aspects of resonance inelastic electron tunneling spectroscopy (RIETS).

This approach[255] is based on the EOM method applied on the Keldysh contour and treats both electron and vibrational degrees of freedom in a self-consistent manner. It is similar to the Non-equilibrium Linked Cluster Expansion (NLCE)[253] in using a cumulant expansion to express correlation functions involving the phonon shift operator in terms of phonon Green functions, but its present implementation appears to be more stable. This scheme is self-consistent (the influence of tunneling current on the phonon subsystem and vice versa is taken into account), and reduces to the scattering theory results in the limit where the molecular bridge energies are far from the Fermi energy of the leads. As in Ref. [254] the equations for the GFs are obtained using the EOM method, however we go beyond Ref. [254] in taking into account the non-equilibrium dynamics of the molecular phonon subsystem.

The starting point of this approach is again the small polaron (Lang-Firsov) transformation.[150, 339] Applied to the one level one primary phonon version of the Hamiltonian (5)-(8) it leads to

$$\hat{\bar{H}} = \bar{\varepsilon}_0 \hat{d}^\dagger \hat{d} + \sum_{k \in L,R} \varepsilon_k \hat{c}_k^\dagger \hat{c}_k + \sum_{k \in L,R} \left( V_k \hat{c}_k^\dagger \hat{d}\, \hat{X} + h.c. \right) \\ + \omega_0 \hat{a}^\dagger \hat{a} + \sum_\beta \omega_\beta \hat{b}_\beta^\dagger \hat{b}_\beta + \sum_\beta U_\beta \hat{Q} \hat{Q}_\beta \quad (39)$$

where

$$\bar{\varepsilon}_0 = \varepsilon_0 - \Delta \, ; \qquad \Delta \approx \frac{M^2}{\omega_0} \tag{40}$$

$\Delta$ is the electron level shift due to coupling to the primary phonon and

$$\hat{X} = \exp\left[i\lambda \hat{P}\right] \qquad \lambda = \frac{M}{\omega_0} \tag{41}$$

is the primary vibration shift generator. The operator $\hat{P}$ was defined by Eq. (12) Note that in this minimized model, indices associated with the bridge electronic state and primary phonons have been dropped.

As already mentioned, the Hamiltonian (39) is characterized by the absence of direct electron-phonon coupling present in (7). Instead, the bridge-contact coupling is renormalized by the operator $\hat{X}$. The electron GF on the Keldysh contour, Eq. (21), now becomes

$$G(\tau_1, \tau_2) = -i\left\langle T_c \hat{d}(\tau_1)\hat{X}(\tau_1)\hat{d}^\dagger(\tau_2)\hat{X}^\dagger(\tau_2)\right\rangle_{\bar{H}} \tag{42}$$

where the subscript $\bar{H}$ indicates that the system evolution is determined by the Hamiltonian (39). Next we make the approximation

$$G(\tau_1, \tau_2) \approx G_c(\tau_1, \tau_2)\mathcal{K}(\tau_1, \tau_2) \tag{43}$$

$$G_c(\tau_1, \tau_2) = -i\left\langle T_c \hat{d}(\tau_1)\hat{d}^\dagger(\tau_2)\right\rangle_{\bar{H}} \tag{44}$$

$$\mathcal{K}(\tau_1, \tau_2) = \left\langle T_c \hat{X}(\tau_1)\hat{X}^\dagger(\tau_2)\right\rangle_{\bar{H}} \tag{45}$$

which assumes that electron and phonon correlation functions can be decoupled.[16] Using second order cumulant expansion on the Keldysh contour to express the correlation function $\mathcal{K}$ in terms of the primary phonon GF leads to

$$\mathcal{K}(\tau_1, \tau_2) = \exp\left\{\lambda^2\left[iD_{PP}(\tau_1, \tau_2)\right] - \left\langle P^2 \right\rangle\right\} \tag{46}$$

$$D_{PP}(\tau_1, \tau_2) = -i\left\langle T_c \hat{P}(\tau_1)\hat{P}(\tau_2)\right\rangle \tag{47}$$

The EOM approach is next used to get Dyson-like equations for the electron and primary phonon GFs. It leads to[255]

---

[16] This approximation is the most sensitive step of this approach and is by no means obvious. In molecular physics it is usually justified by timescale separation between electronic and vibrational dynamics and is inherent in the Born-Oppenheimer approximation, however in problems involving electron transfer the timescale for the latter process is usually slower than that of molecular vibrations. The application of the Born-Oppenheimer approximation in such cases are done in the diabatic representation[22] and this is how the approximation (43) should be understood, however it has to be acknowledged that its use here has not been fully justified.

$$D_{PP}(\tau,\tau') = D_{PP}^0(\tau,\tau') + \int_c d\tau_1 \int_c d\tau_2 D_{PP}^0(\tau,\tau_1) \Pi_{PP}(\tau_1,\tau_2) D_{PP}^0(\tau_2,\tau') \tag{48}$$

$$G_c(\tau,\tau') = G_c^0(\tau,\tau') + \sum_{K \in L,R} \int_c d\tau_1 \int_c d\tau_2 G_c^0(\tau,\tau_1) \Sigma_{c,K}(\tau_1,\tau_2) G_c^0(\tau_2,\tau') \tag{49}$$

where the zero order Green functions are solutions of $-(2\omega_0)^{-1}(\partial^2/\partial\tau^2 + \omega_0^2) D_{PP}^0(\tau,\tau') = \delta(\tau,\tau')$ and $(i\partial/\partial\tau - \bar{\varepsilon}_0) G_c^0(\tau,\tau') = \delta(\tau,\tau')$ and where the functions $\Pi_{PP}$ and $\Sigma_{c,K}$ are given by

$$\Pi_{PP}(\tau_1,\tau_2) = \sum_\beta |U_\beta|^2 D_{P_\beta P_\beta}(\tau_1,\tau_2) \\ -i\lambda^2 \sum_{k \in L,R} |V_k|^2 \left[ g_k(\tau_1,\tau_2) G_c(\tau_1,\tau_2) \mathcal{K}(\tau_1,\tau_2) + (\tau_1 \leftrightarrow \tau_2) \right] \tag{50}$$

$$\Sigma_{c,K}(\tau_1,\tau_2) = \sum_{k \in K} |V_k|^2 g_k(\tau_1,\tau_2) \mathcal{K}(\tau_1,\tau_2); \quad K = L, R \tag{51}$$

Here $g_k$ is the free electron Green function for state $k$ in the contacts. The functions $\Pi_{PP}$ and $\Sigma_{c,K}$ play here the same role as self-energies in the Dyson equation.

Equations (46)-(51) constitute a closed set of equations for the non-equilibrium system under strong electron-primary vibration interaction. Their solution is obtained by an iterative procedure similar in principle to that applied for Eqs. (35)-(38). In addition one may consider the lowest order approximation obtained by stopping after the first iteration step, i.e. using the equations

$$G(\tau_1,\tau_2) \approx G_c^0(\tau_1,\tau_2) \mathcal{K}^0(\tau_1,\tau_2) \tag{52a}$$

$$\mathcal{K}^0(\tau_1,\tau_2) = \exp\left\{ \lambda^2 \left[ iD_{PP}^0(\tau_1,\tau_2) \right] - \langle P^2 \rangle_0 \right\} \tag{52b}$$

Some results obtained from this calculations are shown in figures 14 and 15. Figure 14 depicts the projected density of states $A(E)$ (Eq. (26)) of the equilibrium junctions for different positions (controllable by a gate potential) of the molecular electronic level relative to the leads Fermi energy while Fig. 15 shows the conductance-voltage spectrum of this junction. The following points are noteworthy:

(a) Strong coupling situations are characterized by pronounced progressions of vibrational peaks observed in the spectral function and in the conductance spectrum ($dI/d\Phi_{sd}$ plotted against the bias voltage $\Phi_{sd}$) As discussed above, this structure is associated with the vibrational levels of the intermediate bridge electronic state.

(b) The combination of electron-phonon interaction plus the interaction with the partially occupied electronic manifolds in the leads has a profound effect on the spectral lineshape. In particular, the relative spectral shifts displayed in Figs 14a,b,c result from the phonon-induced renormalization of the bridge electronic energy that depends on its electronic occupation. Obviously, this shift cannot be obtained in a scattering theory-based approach (See Section 5d).

(c) As seen in Fig. 14, the lowest order approximation can account for qualitative aspects of the spectrum, however in the strong coupling case it fails quantitatively, in particular for the partially-occupied bridge situation.

(d) The non-equilibrium electronic process affects the primary phonon distribution. Ultimately this results in heating the phonon subspace (see Section 9). In the present calculation this is seen by the appearance of a phonon peak on the negative energy side of the elastic signal (corresponding to phonon absorption) in Fig. 15. Obviously, this feature cannot be obtained in the lowest order calculation.

(e) The relative intensities of the observed vibronic lines reflect the overlap (Franck-Condon factors) between nuclear wavefunctions in the initial and intermediate bridge electronic states. As in molecular optical spectroscopy such "renormalization" of spectral intensities results in a shift of the signal peak from its electronic origin. This phenomenon has been nicknamed "Franck-Condon blockade" in the nanojunction conduction literature.[276, 340] For very strong electron-phonon coupling, the width of the vibronic components in Figures 14 and 15 may exceed their spacing. In this limit, as in molecular spectroscopy, the vibronic lineshape appears as a broad Franck-Condon envelope as seen, e.g. in Ref. [341].

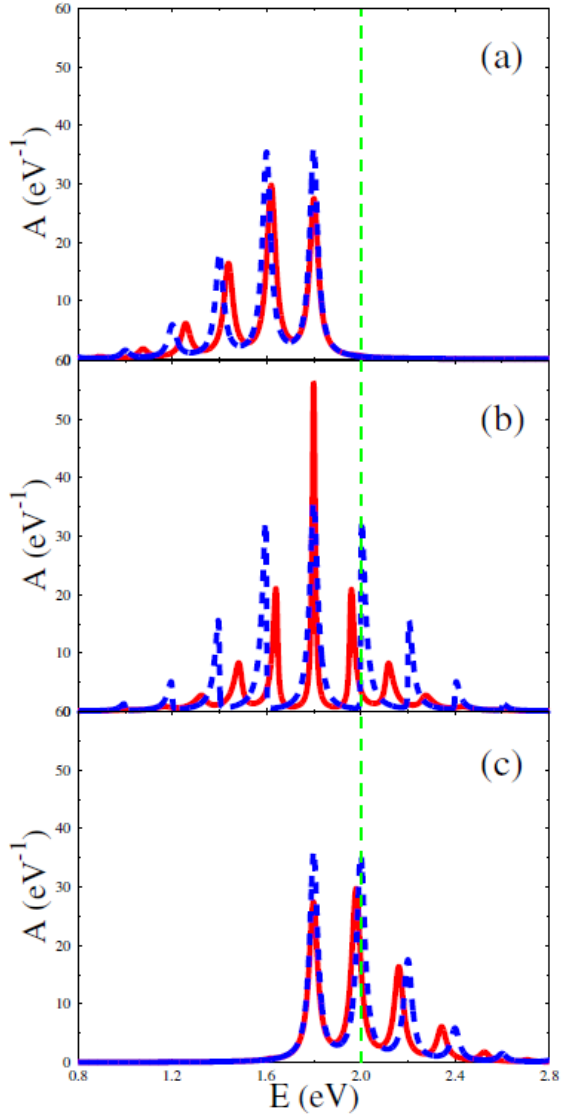

Figure 14. The equilibrium DOS for a system characterized by the parameters $T$=10K, $\varepsilon_0 = 2$ eV, $\Gamma_K = 0.02\,\text{eV}$; $K = L, R$, $\omega_0 = 0.2$ eV, M=0.2eV and $\gamma_{ph} = 0.01\,\text{eV}$. Solid line: self-consistent result. Dashed line: lowest order result. The dashed vertical line indicates the position of the DOS peak in the absence of coupling to vibrations. Shown are cases of filled (a), partially filled (b), and empty (c) electron levels obtained for different positions of the lead Fermi energy $E_F$ which is placed at 3.8 eV in (a), 1.8 eV in (b) and -0.2 eV in (c). (This figure follows Fig. 2 of Ref. [255] but with the parameters used in Fig. 4 in that paper).

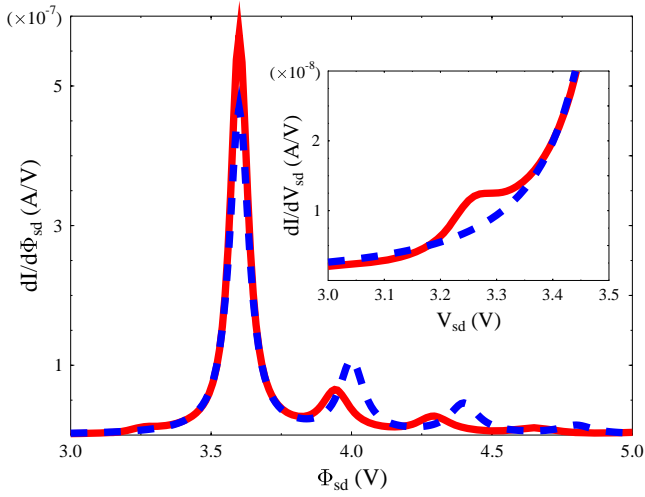

Figure 15. (Fig. 4 of Ref. [255]) Differential conductance vs. source-drain voltage for the junction model of Fig. 14 with $E_F = 0$. Shown are the self-consistent (solid line) and zero-order (dashed line) results. An expanded view of the phonon absorption peak on the negative side of the elastic signal is shown in the inset.

The qualitative similarity of these results to optical spectra, in particular resonance Raman scattering (RRS), is evident, however important differences exist as is seen by the relative spectral shifts seen in Fig. 14. Another interesting difference is related to the trivial observation that in RRS changing the incident frequency (the analog of changing the bias voltage $\Phi_{sd}$ in the junction case) and changing the energy of the excited electronic state (the analog of imposing a gate potential $\Phi_g$) are completely equivalent. In the junction case they are not. Indeed the vibrational structure seen when the conductance is plotted against $\Phi_{sd}$ is absent in the conductance vs. $\Phi_g$ spectrum at small bias potentials. This observation, first made in Ref. [257] in contradiction to earlier assertions,[203, 249, 250] is confirmed by the calculations described above. This statement, while seemingly profound by optical spectroscopy standards is a common experimental observation in junction inelastic spectroscopy and would probably be regarded as trivial by experimentalists used to display conduction data in the $\Phi_{sd} / \Phi_g$ plane.

## 5d. Comparison of approximation schemes

Above (see also Section 5g below) we have described different approximations used in the literature to evaluate, analyze and predict IETS spectra. It should be emphasized at the outset that the quantitative success of such calculations depends to a large extent on the quality of the underlying electronic structure calculation that determines the junction configuration and the relevant molecular electronic energies and vibrational frequencies. Here we briefly address another factor – the approximation used in the transport calculation. We will consider several approximation schemes, already discussed above:

(a) The scattering approach, using standard scattering theory methods or the Tersoff-Hamann approach[93, 94], is frequently used in IETS calculations. The basis for this approximation is the Landauer conduction formula that expresses elastic conduction in terms of a scattering property – the transmission coefficient $T(E)$ that is calculated for the isolated molecular target. The actual current is calculated by weighting this transmission with the proper Fermi occupation numbers in the leads,

$$I = \frac{2e}{\hbar} \int \frac{dE}{2\pi} T(E) \left[ f_L(E) - f_R(E) \right] \tag{53a}$$

It is remarkable that this result is exact for elastic transmission even though a scattering calculation disregards the fact that at steady state of the conduction process the electronic molecular levels involved can be partly occupied. Eq. (53a) can be written in the form

$$I = \frac{2e}{\hbar} \int \frac{dE}{2\pi} T(E) \left[ f_L(E)(1 - f_R(E)) - f_R(E)(1 - f_L(E)) \right] \tag{53b}$$

where the integrand has the appealing form of a difference between two fluxes, each written as a product of the transmission coefficient and the probability that the initial level is occupied while the final is not.

In the scattering theory approach to inelastic conduction Eq. (53b) is generalized to take the electronic energy change into account

$$I = \frac{2e}{\hbar} \int \frac{dE_i dE_f}{2\pi} T(E_i, E_f) \left[ f_L(E_i)(1 - f_R(E_f)) - f_R(E_i)(1 - f_L(E_f)) \right] \tag{54}$$

where $T(E_i, E_f) dE_f$ is the probability that an electron incident on the molecular target with energy $E_i$ will be transmitted with energy $E_f$. For elastic transmission $T(E_i, E_f) = T(E_i) \delta(E_i - E_f)$.

Eq. (54) is often used in the IETS literature, where $T(E_i, E_f)$ is usually computed by low order perturbation theory (see, e.g. Refs. [155, 41, 248]) but can be, for simple models, also calculated exactly[202, 203, 200] or to high order in the electron-phonon coupling.[204, 205, 211, 342] It is well known however[199] that the cancellation of terms that makes Eq. (53) exact for elastic process does not happen for inelastic conduction. Also, for large bias potential the transmission becomes bias dependent, which is of course missed in the scattering calculation. The resulting error is hard to assess because (54) is not a first term in a systematic expansion. In what follows we refer to (54) as the '$f(1-f)$ approximation'.

(b) The lowest order perturbation theory (LOPT) approximation to the NEGF-based conduction. In this approximation Eq. (27) is evaluated keeping only terms up to the lowest (second) order in the electron-phonon coupling. The phonon distribution is assumed to remain in thermal equilibrium.

(c) The Born approximation (BA). Here the GFs needed in Eq. (27) are calculated by applying the lowest (second) order approximation to the self energies used in (23).

(d) The self consistent Born approximation (SCBA). In the second order expressions for the SEs used in the BA, the zero order electron and phonon GFs are replaced by their exact counterparts. The resulting set of equations is solved self consistently by iterations. In one variant of this approximation only the electronic GF is evaluated self consistently and the phonons are assumed to remain at thermal equilibrium. In the other the calculation is performed self consistently for both electron and phonon subsystems.

Approximations (b)-(d) are expected to be appropriate in weak coupling situations, with the SCBA doing better for stronger electron-phonon coupling. Intuitively we expect also that Eq. (54) may provide a reasonable approximation for small coupling. When the electron-phonon coupling becomes stronger, e.g. under Coulomb blockade conditions (near resonance tunneling in systems where the bridge-leads coupling is small) we need to use a formalism that accounts for changes in the nuclear configuration associated with the electron-phonon interaction:

(e) The self consistent strong coupling (SCSC) scheme of Section 5c.[255] In this approximation the polaron transformation is applied to the Hamiltonian (5)-(8) and the factorization approximation (43)-(45) is made for the resulting GF (42). This leads to the set of Eqs. (46)-(51) that are solved self consistently by iterations.

(f) Strong coupling lowest order (SCLO) approximation to Eqs. (46)-(51). This is obtained by using the zero-order expressions for $G_c$ and $K$ (i.e. their forms in the absence of electron-phonon coupling) in Eqs (43)-(45).[17]

It is instructive to obtain the $f(1-f)$ approximation from the full NEGF formalism. This can be done by noting that in scattering theory we do not evaluate the net flux but (separately) the scattering flux for the left-to-right and the right-to-left processes. Indeed, the term involving $\Sigma_L^>(E)G^<(E)$ in (27) reflects electronic population in the molecular target ($G^<(E) = in(E)A(E)$ where $n(E)$ and $A(E)$ are, respectively, the population and the electronic density of states associated with the molecular resonance level), which should be taken 0 (or 1 if we consider hole transport) in a scattering calculation. For example, it can be shown[343] that the $f(1-f)$ approximation to the left-to-right flux is obtained from (cf. Eq. (27)) $I_L = (e/\hbar)\int(dE/2\pi)\Sigma_L^<(E)G^>(E)$, provided that we disregard back-reflection to the L lead, i.e. terms containing $f_L(1-f_L)$ (this implies that the contribution of $\Sigma_L^>$ to $G^>$ is disregarded), and take the phonon bath to be at thermal equilibrium.

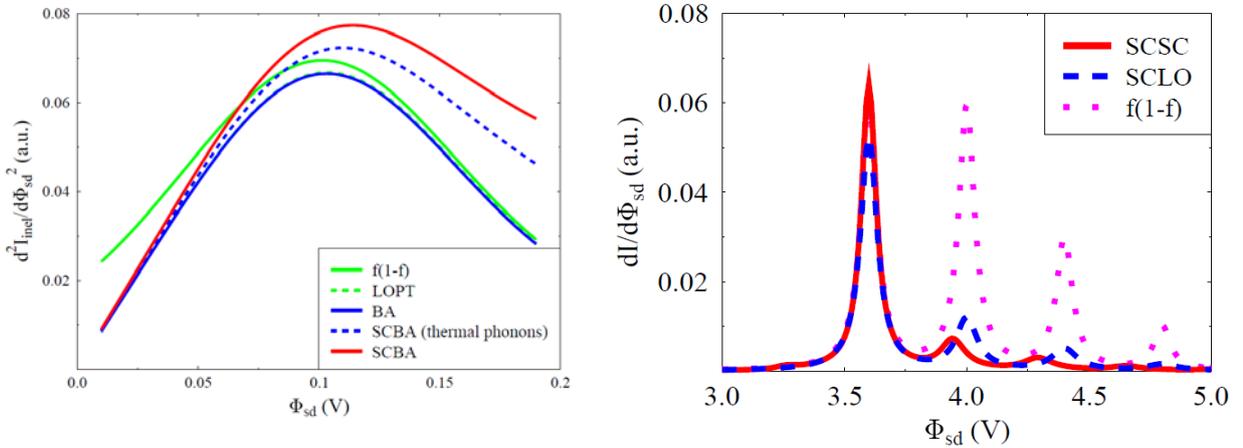

Fig. 16. Comparison between different approximations used in IETS calculations. Left – a weak coupling (off resonance) case: $T$=300K, $\varepsilon_0 = 1$ eV, $E_F$=0, $\Gamma_L = \Gamma_R = 0.1$ eV, $\omega_0 = 0.1$ eV, $M$=0.5 eV, $\gamma_{ph} = 0.005$ eV. Lines are indicated according to their order from top to bottom on the right side of the figure: Full NEGF-SCBA calculation, SCBA with phonon kept at

---

[17] The difference between this approximation and the version of scattering theory based on exact evaluation of the inelastic vacuum transmission coefficient is that here information about the steady state population of the bridge electronic level is taken into account. Therefore standard scattering theory deals with either electron or hole transport, while SCLO accounts for both additively.

thermal equilibrium, $f(1-f)$ (scattering theory) calculation, NEGF LOPT calculation and NEGF-BA calculation. The LOPT curve is almost not seen because it overlaps the BA calculation. Right – Strong coupling (resonance) situation: $T$=10K, $\varepsilon_0 = 2$ eV, $E_F$=0, $\Gamma_L = \Gamma_R = 0.02$ eV, $\omega_0 = 0.2$ eV, $M$=0.2 eV, $\gamma_{ph} = 0.01$ eV. Full line – SCSC calculation, Dashed line – SCLO calculation, dotted line – $f(1-f)$ (with exact scatteing theory evaluation of transmission) calculation.

Figure 16 shows how these approximations perform. Both figures were obtained using a single electronic level/ single ocsillator model. The left panel depicts a off resonance weak coupling situation that characterizes regular IET spectra. It is seen that despite its fundamental limitation, the scattering theory approach performs well in this limit where the intermediate level is essentially unoccupied. Indeed the different approximation schemes yield practically identical results for the I/Φ characteristic and small differences are seen only in the derivatives. The right panel corresponds to a strong coupling (resonance) situations. Here the difference between different calculations is considerable, and in particular a scattering theory approach fails to account for the line intensities.

We have pointed out that the quality of computed IET spectra depends on the input from electronic structure calculation. We end by emphasizing that in principle the electronic structure calculation should be carried out at the non-equilibrium steady state relevant to the experimental conditions. Standard (non-resonant) IETS experiments are done in low bias situations where equilibrium structures are relevant, however calculations pertaining to high bias far from equilibrium systems have to take this into account, as done (with uncertainties related to the use of the Hellman-Feynman theorem in far from equilibrium situations) in the more advanced NEGF-based molecular transport codes available.

## 5e. Asymmetry in IETS

The prediction and observation of rectification, i.e. asymmetry in current-voltage response of molecular junctions beyond the linear response regime have been driving forces in the development of molecular electronics.[344-347] Normal IETS measurements are usually carried for the low-voltage linear response regime, where the elastic contribution to junction conduction is symmetric with respect to voltage reversal. Inelastic features, normally observed in the second derivative $d^2I/d\Phi_{sd}^2$ often preserve this symmetry, i.e, the relationship $\left(d^2I/d\Phi_{sd}^2\right)_{\Phi_{sd}} = -\left(d^2I/d\Phi_{sd}^2\right)_{-\Phi_{sd}}$ is satisfied within experimental deviations, however asymmetry may be observed and, as discussed below, may convey interesting implications.

To elucidate the possible source of such asymmetry we follow Ref. [348] where the model (5)-(8) is again restricted to the case of a single bridge electronic orbital and a single primary vibrational mode. In the Born approximation (which is applicable in the weak electron-phonon coupling limit relevant for normal IETS experiments) and in the non-crossing approximation, the overall current, Eq. (27), is made of additive contributions of elastic and inelastic components.[235] In particular, the inelastic contribution takes the form[348]

$$I_{inel} = \frac{2e}{\hbar}M^2 \int_{-\infty}^{\infty}\frac{dE}{2\pi}\int_0^{\infty}\frac{d\omega}{2\pi}\rho_{ph}(\omega)\left|G^r(E)\right|^2\left|G^r(E-\omega)\right|^2$$
$$\times\left\{f_L(E)\left[1-f_R(E-\omega)\right]\Gamma_L(E)\Gamma_R(E-\omega) - f_R(E)\left[1-f_L(E-\omega)\right]\Gamma_L(E-\omega)\Gamma_R(E)\right\}$$
(55)

The first term in the curly brackets in (55) is responsible for the current at $\Phi = \mu_L - \mu_R > 0$, while the second term contributes at $\Phi < 0$. It is evident from (55) that if $\Gamma_L$ and $\Gamma_R$ are constants independent of energy, or even if $\Gamma_L(E) = \lambda\Gamma_R(E)$ with an energy independent constant $\lambda$, these two terms are mirror images of each other, implying a symmetric inelastic current, $I_{inel}(\Phi) = -I_{inel}(-\Phi)$. On the other hand asymmetry may reflect a difference between the energy dependence of $\Gamma_L$ and $\Gamma_R$ in the energy range between the two chemical potentials. Note that the elastic part of the current is symmetric, because the elastic transmission coefficient $T(E)$ in Eq. (53) depends only on the product $\Gamma_L(E)\Gamma_R(E)$ and sum $\Gamma_L(E) + \Gamma_R(E)$.[18]

The asymmetry in the inelastic signal may be cast explicitly by considering the conduction derivative, $d^2I/d\Phi_{sd}^2$. Assuming symmetric distribution of the bias potential along the junction, Eq. (55) yields

$$\frac{d^2I}{d\Phi_{sd}^2} \sim \text{sgn}(\Phi_{sd})M^2\frac{2e}{\hbar}\rho_{ph}(\Phi_{sd})\left|G^r(\mu_L)\right|^2\left|G^r(\mu_R)\right|^2\Gamma_L(\mu_L)\Gamma_R(\mu_R) \qquad (56)$$

in addition to smooth background terms, with $\mu_L = E_F + \Phi_{sd}/2$ and $\mu_R = E_F - \Phi_{sd}/2$. Because the phonon DOS $\rho_{ph}(\omega)$ is sharply peaked near $\omega = \omega_0$ this leads to corresponding features at $\Phi_{sd} = \pm \omega_0$ with intensities proportional to $\Gamma_L(E_F \pm \omega_0/2)\Gamma_R(E_F \mp \omega_0/2)$. This implies that asymmetry to voltage reversal in the inelastic structure follows from, and indicates the presence of, different energy dependencies of $\Gamma_L(E)$ and $\Gamma_R(E)$.

---
[18] This can be shown to hold also for higher order contributions to the elastic flux that enter in Eq. (60)

As a specific example assume that the junction can be modeled as a rectangular double barrier, where the molecular site is separated from the leads by barriers of height $U_K$ and width $D_K$ ($K = L, R$). In this case the energy dependence of $\Gamma_K$ reflects the tunneling probability

$$\Gamma_K(E) = A_K \exp\left(-2\sqrt{\frac{2m}{\hbar^2}(U_K - E)}\, D_K\right); \qquad K = L, R \qquad (57)$$

For $U_L = U_R$ a different energy dependence of $\Gamma_L$ and $\Gamma_R$ arises from different barrier widths $D_L$ and $D_R$, i.e. from asymmetric positioning of the molecule (or the molecular level) between the two leads. Figure 17 shows a specific example of this phenomenon.

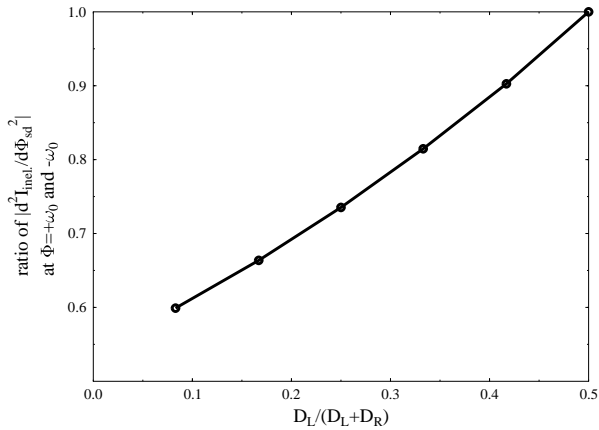

Figure 17. Asymmetry in a model calculation of IETS intensities: The ratio between $\left|d^2I/d\Phi_{sd}^2\right|$ evaluated at $\Phi_{sd} = +\omega_0$ and $\Phi_{sd} = -\omega_0$ is plotted against the positional asymmetry $D_L/(D_L + D_R)$ using the model (57) with the parameters $\omega_0 = 0.1\,\text{eV}$, $\varepsilon_0 = 1\,\text{eV}$, $E_F = 0$, $M = 0.3\,\text{eV}$, $\gamma_{ph} = 0.001\,\text{eV}$. (From Ref. [348]).

### 5f. The origin of dips in IETS signals

At the threshold for inelastic tunneling, $e\Phi_{sd} = \hbar\omega_0$, an inelastic transmission channel opens up showing as a step in conductance. It is natural to expect that this step will be positive, showing as a peak in the conduction derivative $d^2I/d\Phi_{sd}^2$. While this is quite common, dips in the IETS signal have been observed,[324, 326, 39, 349, 350] as well as more complex features,[34] as shown in Figures 6-9. A common explanation, first advanced by Davis[351] and later elaborated on by many workers[352, 338, 353, 231-233, 235, 165, 354, 161, 240] attributes this observation to renormalization of elastic channel upon opening of the additional inelastic one. As a simple demonstration consider the single level-single vibration bridge model connecting two free electron reservoirs $L$ and $R$. The non-equilibrium GF formalism in the Born and the non-crossing approximations leads to the following result for the current up to second order in the electron-phonon interaction[19]

$$I^{(2)} = I_{el}^{(0)} + \delta I^{(2)} \tag{58}$$

$$I_{el}^{(0)} = \frac{2e}{\hbar} \int \frac{dE}{2\pi} T_0(E)\left[f_L(E) - f_R(E)\right] \tag{59}$$

$$\delta I^{(2)} = \delta I_{el}^{(2)} + \delta I_{inel}^{(2)}$$
$$= \frac{2e}{\hbar} \int \frac{dE}{2\pi} T_0(E) \frac{\Gamma_{ph}(E)}{\Gamma(E)} \left\{ 1 - \frac{\Gamma^2(E)}{2\Gamma_L(E)\Gamma_R(E)} T_0(E) \right\} \left[f_L(E) - f_R(E)\right] \tag{60}$$

In Eqs. (58)-(60) $T_0(E) = \Gamma_L(E)G_0^r(E)\Gamma_R(E)G_0^a(E)$ is the elastic transmission coefficient, $G_0^r(E) = G_0^a(E)^* = (E - \tilde{\varepsilon}_0 + i\Gamma(E)/2)^{-1}$, $\tilde{\varepsilon}_0$ is the resonance level energy shifted by the real part of the electron self energy and $\Gamma(E) = \Gamma_L(E) + \Gamma_R(E)$ is obtained from the imaginary part of the same self energy. $\Gamma_{ph}(E)$ is twice the imaginary part of the phonon contribution to the electron self energy, which on the SCBA level takes the form

$$\Gamma_{ph}(E) = M^2 \int_0^\infty \frac{d\omega}{2\pi} \rho_{ph}(\omega)$$
$$\times \left\{ \left[1 + N_{eq}(\omega)\right]\left(\left[1 - n(E - \omega)\right]\rho_{el}(E - \omega) + n(E + \omega)\rho_{el}(E + \omega)\right) \right. \tag{61}$$
$$\left. + N_{eq}(\omega)\left(\left[1 - n(E + \omega)\right]\rho_{el}(E + \omega) + n(E - \omega)\rho_{el}(E - \omega)\right) \right\}$$

where $\rho_{el}(E)$ and $n(E)$ are respectively the electronic density of states associated with the resonance level and the steady state occupation of this level. In lowest order they are given by

$$\rho_{el}(E) = \Gamma(E)\left|G_0^r(E)\right|^2 = \frac{\Gamma(E)}{\Gamma_L(E)\Gamma_R(E)} T_0(E) \tag{62}$$

$$n(E) = \frac{\Gamma_L(E)}{\Gamma(E)} f_L(E) + \frac{\Gamma_R(E)}{\Gamma(E)} f_R(E) \tag{63}$$

and $N_{eq}(\omega) = \left[\exp(\omega/k_B T_K) - 1\right]^{-1}$ is the phonon thermal population. Note that the second order correction $\delta I^{(2)}$ contains the inelastic current as well as the additional elastic contribution that can be identified as arising from interference between the zero order elastic amplitude and the second order amplitude of the process in which a phonon quantum was created and destroyed. For small bias and low temperature, and when the resonance energy $\tilde{\varepsilon}_0$ is far from the Fermi window, the integrand in (60) apart from the Fermi functions depends weakly on energy. Approximating in (61) $\rho_{ph}(\omega) \approx 2\pi\delta(\omega - \omega_0)$, $\rho_{el}(E \pm \omega_0) \approx \rho_{el}(E_F)$, $\Gamma_K(E \pm \omega_0) \approx \Gamma_K(E_F)$, $N_{eq}(\omega_0) = 0$ and $f_K(E) \approx \theta(\mu_K - E)$, leads to $\Gamma_{ph}(E) = M^2 \times \left\{\xi_L\left[\theta(E - \omega_0 - \mu_L) + \theta(\mu_L - E - \omega_0)\right] + \xi_R\left[\theta(E - \omega_0 - \mu_R) + \theta(\mu_R - E - \omega_0)\right]\right\}$, where $\xi_K = \Gamma_K(E_F)/\Gamma(E_F)$, $K = L, R$. Substituting this this into (60) then leads to

$$\delta I^{(2)} = \frac{e}{\pi\hbar} T_0^2(E_F) \left\{ 1 - \frac{\Gamma^2(E_F)}{2\Gamma_L(E_F)\Gamma_R(E_F)} T_0(E_F) \right\} \frac{M^2}{\Gamma_L(E_F)\Gamma_R(E_F)} \tag{64}$$
$$\times \left\{ \theta(\Phi)(\Phi - \omega_0) - \theta(-\Phi)(|\Phi| - \omega_0) \right\}$$

---

[19] These relatively simple fiorms are obtained under the simplifying assumption $\Gamma_R(E) = c\Gamma_L(E)$,

Since $T_0(E_F) \ll 1$ in this off resonance situation this current correction is positive. On the other hand, when the resonance level is inside the Fermi window the dominant contribution to (60) comes from the neighborhood $\tilde{\varepsilon}_0$. If $\Gamma$ is not too large the sign of $\delta I^{(2)}$ is then determined by the sign of $1 - \frac{\Gamma^2(\tilde{\varepsilon}_0)}{2\Gamma_L(\tilde{\varepsilon}_0)\Gamma_R(\tilde{\varepsilon}_0)} T_0(\tilde{\varepsilon}_0)$ and because $T_0(\tilde{\varepsilon}_0)$ is of order 1, it can be negative. It is seen that inelastic threshold dips are expected near resonance tunneling.

An alternative explanation of inelastic threshold dips, pertaining to situations encountered in point contact spectroscopy, has been given by Agraït et al[355] and Smit et al.[39] following Refs. [356] and [357]. Point contact spectroscopy is inelastic electron tunneling spectroscopy carried in open channel situations where the elastic

---

with c a constant independent of $E$.

transmission probability (per channel) is close to 1, independent of energy (i.e. voltage). Obviously, when the electron is transmitted with probability 1 the effect of inelastic interactions can only be reduction of current beyond the inelastic threshold, as indeed observed. The elegant argument provided by the above papers relies on the assumption that when the transmission probability is unity the electronic distribution in the leads at energies within the Fermi window is such that only forward moving electronic states are populated. In other word, it is assumed that thermal relaxation within the leads is not fast enough to relax the biased distribution formed at the contact region. (In contrast, development that leads to Eq. (58) relies on the assumption that the leads are in thermal equilibrium). Beyond the inelastic threshold, an electron that loses energy $\hbar\omega$ to a phonon must end up in a backward going state (since all the forward states are occupied), leading to current reduction.

It is interesting to note that while the two proposed mechanisms for current reduction appear to be based on different physical models, they lead to similar qualitative predictions under similar conditions (large transmission) and can not therefore be unambiguously confirmed by available experimental results. It should also be kept in mind that observed vibrational features in MTJ conduction can result from more complex situations. A very recent observation of spikes in the conductance ($dI/d\Phi$) spectrum at the threshold for vibrational excitation in Pt-CO-Pt and Pt-$H_2$-Pt junctions[358] was interpreted by the authors in this spirit as due to vibrationally induced transitions between two molecular configurations in the junction.

## 5g. Computational approaches

The wealth of data obtained from IETS experiments has been a strong motivating force for numerical calculations aimed at interpreting and predicting the observed spectra. In general, such spectra reflect the effects of several factors, including the vibration of the bare molecule, the bonding between the molecule and the metallic surfaces, the overall junction structure and symmetry, the electron-phonon interaction and the non-equilibrium nature of the transport process. Anticipating that non-equilibrium effects are small in the low bias conditions used in IETS, many workers[359] [159, 359, 360] have focused on evaluating equilibrium vibrational spectra of molecules adsorbed on metal surfaces or on metallic clusters. Such studies can be compared with IETS data or with other methods of surface vibrational spectroscopy such as surface Raman spectroscopy[361] or high resolution electron energy loss spectroscopy (HREELS).[362-364]

In these and most other studies described below, the main tool for electronic structure calculations is density functional theory (DFT). The applicability of DFT for transport problems is questionable,[365-367] however it is the only first-principle technique available today to general practitioners that is capable of dealing with large organic molecules often used as molecular wires.

Going beyond equilibrium cluster and surface calculations, the simplest way to account for the electron-phonon interaction and its dynamical implications is to cast the calculation within the framework of scattering theory. This has been done either using the Tersoff-Hamann approach,[93, 94] which has been developed for tunneling in STM configurations, or using extensions of the Landauer methodology.[196, 197] Such approaches to transport are single-particle in nature, and disregard many-body effects (important even for non-interacting electrons because of their fermion nature) in the transport process.

In the Tersoff-Hamann approach,[93, 94] which is applicable to STM configurations, the signal is dominated by the phonon-modified local density of states (DOS) at the tip position and by the occupation of the tip and substrate electronic states. Early applications[338] of this approach to inelastic tunneling have used phenomenological models with parameters fitted to experimental data.[324] First principles calculations based on this approach (with the required DOS obtained within a NEGF-DFT framework) were carried by Lorente, Persson and coworkers[230, 158, 164] and used to interpret experimental data of Ho and coworkers.[368, 369, 325, 361] In particular, Ref. [158] represents a first attempt to formulate propensity rules for inelastic tunneling spectra.

In a generalized Landauer theory one considers the generalization to inelastic scattering of 1-particle transmission probability through the junction. The current is related to this transmission probability weighted by the electronic state population (Fermi function $f$ and its complement $1-f$) of the source and drain leads, Eq. (54). Electron-phonon interactions can be taken into account exactly in such approach, as is done in the multi-channel mapping method.[204, 342] Interesting attempts to generalize this approach so as to take into account Pauli exclusion of transmission/reflection events into the same state were made, and a corresponding self-consistent procedure was proposed,[370] still the approach is essentially a one-particle picture in which scattering channels are considered independent of each other.

Other scattering-type calculations[156, 213] [162, 163, 167, 43, 41, 155, 44, 45] use perturbation expansions in the electron-phonon interaction about the Landauer theory for elastic transmission. In contrast to the multi-channel mapping method this is a low order calculation, however good agreement with experiment is reported for off resonance inelastic tunneling properties of various molecular bridges. Two general observations of these calculations are the high sensitivity of the computed spectra to the structure of the molecular bridge[156, 163, 371] and the significance of modes with large longitudinal component, i.e. motion along the tunneling direction.[156, 213, 371] In particular, Chen et al,[213] studying inelastic tunneling in alkanethiols, have found an interesting alternating behavior with chain length, in agreement with HREELS data[364] that is due to the alternating direction of the $CH_3$ group motion with respect to the tunneling direction. Jiang et al[163] have explained the difference between experimental inelastic tunneling spectra of alkanethiols in Refs.[34, 35] by different molecular conformations, postulating linear[34] and twisted[35] molecular backbone structures. Propensity rules for the importance of vibrational modes in the inelastic signal were recently formulated by Troisi and Ratner.[44, 45] Again the importance of modes with large component in the tunneling direction is emphasized, so for a linear chain with one orbital per atom only totally symmetric modes contribute to IETS signal. For molecules with side chains any normal mode dominated by side-chain motion will contribute only weakly to IETS. The authors employ group theory to identify the main normal modes for planar conjugated molecules with $C_{2h}$ symmetry. This concept appears to be useful in spite of its application to the isolated molecules, which disregards the contact effects on the electronic structure and the molecular symmetry. On the other hand the mechanism observed by Grobis et al[162] in STM studies of Cd@$C_{82}$ on Ag(001) associates the observed feature at 52mev with a cage vibration that affects the localization length of the electronic wavefunction, stressing the point that modes that affect electron localization will be important in IETS irrespective of the spatial extent of the mode itself.

A significant point regarding this approach to computational IETS is the fact that perturbative scattering theory based approximation provides a useful and practical tool for non-resonant inelastic tunneling in spite of the weakness of its theoretical foundation (see Section 5d). As a case in point we focus on the Troisi-Ratner approach,[41, 155,

44, 45] (see also Ref. [248]). The starting point is the Landauer expression for the elastic conductance

$$g^{el}(E_F) = g_0 \text{Tr}\left[\Gamma^L(E_F)G^r(E_F)\Gamma^R(E_F)G^r(E_F)^\dagger\right] \quad (65)$$

with $g_0 = e^2/\pi\hbar$. Connection to IETS spectra is made by evaluating the dependence of the GFs in (65) on the nuclear configuration and using the lowest order expansion of these in the normal mode coordinates,

$$G_{ij}^\alpha = \frac{\sqrt{2}}{2}\left(\frac{\partial G_{ij}^r(E,\{Q_\alpha\})}{\partial Q_\alpha}\right) \quad (66)$$

The electronic and vibrational structure needed as input are obtained from Hartree-Fock[155] or DFT[41, 44] calculations. The resulting IETS spectrum ($d^2I/d\Phi^2$ plotted against $\Phi$) consists of a series of peaks, whose position is determined from $e\Phi = \hbar\omega_\alpha$ and their intensity is given by

$$W_\alpha = g_0 \text{Tr}\left[\Gamma_L(E_F)G^\alpha(E_F)\Gamma_L(E_F)G^\alpha(E_F)^\dagger\right] \quad (67)$$

The individual peak intensities and positions are obtained directly. The lineshape, however, cannot be obtained from this low order calculation and needs to be fitted. It should be emphasized that these results are valid only in the Landauer-Imry regime, far from the electronic resonance. In particular, when electronic resonances are approached, Eq. (67) is no longer valid and other effects such as strong vibronic coupling can dominate the I/Φ characteristic.

Actual calculations can be done for particular bridge models. The simplest model assigns only molecular modes to the primary set, but more extended analysis can be done.[45] One first optimizes the structure and does vibrational analysis on the isolated molecule; this includes the evaluation of the coupling elements (66). This optimized structure and normal modes can be translated directly into the geometry of the junction after choosing the molecular orientational placement. This approach provides a simple computational tool for non-resonant inelastic tunneling that was successfully applied to several organic molecules. Because the expansion in Eq. (66) is in normal modes, contributions to the IETS spectrum can be classified according to the point groups of the molecular entity within the junction. Because the expression for IETS can be written as a pathway sum, it is clear that particular pathways will make larger contribution to the overall sum than will others. The combination of the pathway aspect and the symmetry aspect leads to a set of propensity rules – these are analogous to the symmetry selection rules seen in vibrational spectroscopy (Raman, infrared), but are based not on the symmetry of the radiation/matter interaction, but rather on global and local symmetries and on dominance of certain pathways. These rules are developed and discussed elsewhere.[44] They are important because they can help lead to understanding both the pathways along which electrons travel within the molecule, and the possible geometries of the molecule within the junction.

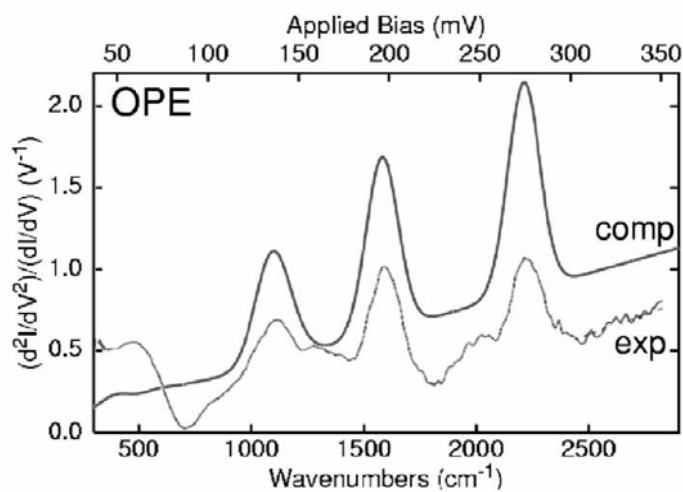

Figure 18. IETS spectrum of phenyleneethynylene trimer. The computations and experiment[35] agree nicely with respect to position and strength of the various IETS peaks, that can be assigned to specific normal coordinates. Line shapes are more complicated, and require more sophisticated theory. From reference [41].

Figure 18 demonstrates the accuracy of these calculations – the trimer of phenyleneethynylene has been measured using IETS. The figure compares experimental and computational results. The linewidths in the computation are arbitrary, but the intensities (the area under each curve) and the positions come directly out of the DFT calculation. Note that the higher frequency regime (above 400 cm$^{-1}$) is dominated by totally symmetric ($a_{1g}$) modes. Only in the low frequency regime do modes others than $a_{1g}$ appear with real intensities, as indeed follows from the propensity rules.

The calculation appears to be quite accurate for describing IET spectra. This accuracy makes this spectroscopy, in combination with theory, a useful diagnostic tool. Observing the molecular signal in the IETS spectrum indicates the molecular involvement in the tunneling process. Of greater utility, probably, is the geometrical information. Figure 19 shows calculations on pentane dithiol, once again using DFT methods (B3LYP, dzp basis). The comparison between calculations and experiment demonstrates that the alkane dithiol is not aligned perpendicular to the electrodes, but rather at an angle of roughly 50 degrees. This information is consistent with what is generally believed about alkane thiols in many situations, but constitutes an experimental observation whose computational explanation provides information on the in-junction geometry that is very difficult to obtain in any other way.

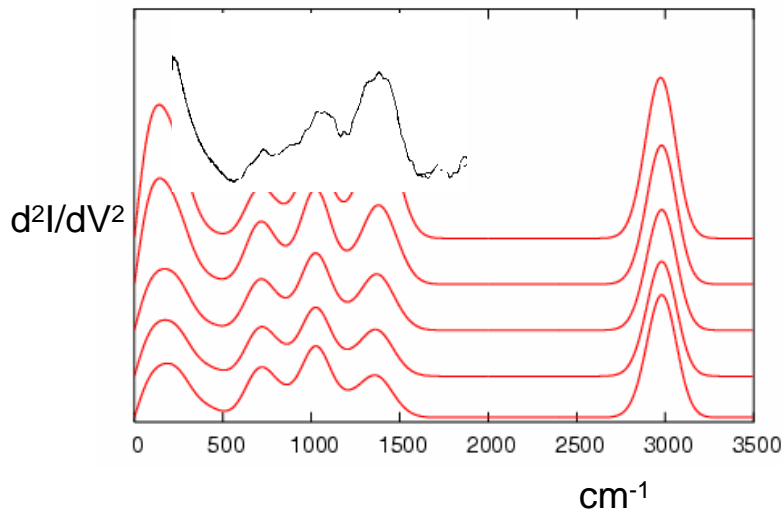

Figure 19. Use of IETS to describe geometry. The red curves are calculated at different tilting angles in the plane. It is clear that the experimental spectrum[35] (black trace) best fits the top computed spectrum, corresponding to a tilt angle of 40° (the lower ones are for 30°, 20°,10° and 0° respectively). From Troisi and Ratner Phys Chem. Chem. Phys, submitted for publication.

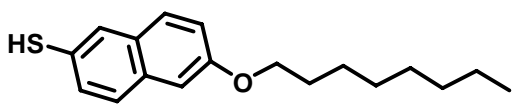

Figure 20. Alkyl naphthylthiol ether molecule. Its experimental IETS study combined with theoretical analysis, demonstrate clearly the nature of the tunneling pathways. J. Kushmerick et.al to be submitted.

From the point of view of chemical transport mechanism, new work on actual pathways is perhaps the most interesting. The naphthalene chromophore in Fig. 20 is linked to a thiol on one end, and an alkyl ether on the other. By comparison of the measured IETS spectrum of this molecule with the computed one, it is possible to examine the relative intensities of the different vibrational normal modes, thereby to deduce the pathway for transport. We find that the electrons are injected through the terminal methyl group, tunnel through the sigma bridge to the etheric linkage, mix with the pi electrons, pi tunnel through the aromatic, and switch back to the sigma tunneling, through the thiol and out onto the counter electrode. This pathways information is more detailed than has been obtained for other important applications of pathways concepts, such as motion through proteins and peptides, and constitutes (we believe) a sweet application of vibronic coupling theory associated with molecular transport.

Putting computational simplicity aside, the most sophisticated ab-initio approach used today for inelastic tunneling spectroscopy is based on the combination of electronic structure calculations using DFT (or its tight-binding based variation, TFTB) and evaluation of transport properties using the NEGF framework.[165, 236-238, 161, 166, 240, 303-305] While the theoretical framework is general, current applications to IETS do not allow for charge transfer to/from the molecule, and the non-equilibrium character is introduced as potential boundary conditions. The effects of electron-phonon coupling are usually treated at the BA or the SCBA level. Again, correlation between inelastic signal intensity and mode motion in the tunneling direction is reported[165, 305] although Solomon et al[238] also point out an interesting correlation with modes that reflect motion in regions where the electron density is high in the low bias limit. Generally good agreement with experiment is found. Thus, Frederiksen et al [165] report quantitative agreement between their calculation of the IETS signal for atomic gold wires and experimental results.[355] Another interesting result of this calculation is the decrease in conductance with increase in the inelastic signal and softening of the inelastic mode resulting from straining the wire. Similarly, using DFT based tight-binding approach to electronic structure, Pecchia et al [237] report reasonable agreement between their calculations on Au-octanethiolate-Au junction IETS results of Ref. [34], while calculations of Mehl and Papaconstantopoulos[304, 305] on atomic gold chains compare well with the experimental results of Agrait and coworkers.[350] Finally, Paulsson et al[161, 166] have proposed a computationally inexpensive scheme based on the lowest order expansion in the electron-vibration coupling within a SCBA calculation. This simplification, valid for weak electron-vibration coupling and slow variation of the leads DOS in a range of a few vibrational frequencies around the Fermi energy, was used to calculate IETS spectra of gold wires as well as $H_2$ molecular junction. With one fitting parameter good agreement is obtained with the experimental results of Refs. [350] for gold wires, [39] [349] for $H_2$ junctions and [35] for several organic bridges. A similar approach (expanding the Born approximation rather than SCBA expressions) was used in Ref. [240].

# 6. Effects of electron-electron (e-e) interactions

Our discussion of electron-phonon interactions in molecular junctions has disregarded so far e-e interactions. Such interactions are the source of Coulomb blockade phenomena, where transport through a bridge that is coupled weakly to the contacts can be blocked due to the needed charging energy, or in the Kondo effect where strong correlation between bridge and contact electrons leads to formation of a tunneling channel in the zero-bias region. The rational for disregarding such interactions in much of the molecular electronics literature is that the small molecular size makes charging by more than one electron energetically too costly, and that the Pauli principle is enough to take care of restrictions relevant to single electron transport. However, recent work from several groups[311, 83, 372, 84, 373, 28, 330, 244, 87, 86, 374, 88] has observed Coulomb blockade (CB), Kondo effect or both in molecular junctions. These are often accompanied by vibrational features that indicate effects of electron-vibration coupling. Such features may correspond to the center-of-mass motion of the bridge[310] or to intramolecular vibrations.[372, 84, 373, 330, 244] Furthermore, electron-phonon interaction may cause the effective e-e interaction to become attractive, with interesting consequences. [251, 252, 375, 376, 278]

Early theoretical approaches to transport in the CB regime were based either on linear response theory carried near equilibrium[377-379, 271, 272] or on treating transport at the level of quasiclassical rate equations.[380-389]. These approaches are valid close to equilibrium and/or for weak molecule-lead coupling at relatively high temperatures. Stronger molecule-leads coupling relevant (for example) to the observation of nonequilibrium Kondo resonance should be treated at a more sophisticated level. Recent approaches to this problem are based on the slave-boson technique,[390-395] on the equation-of-motion method,[394, 396-401] or on perturbation theory on the Keldysh contour.[402-412]

Such studies are usually carried in the framework of a Hubbard-type Hamiltonian. A model of this type, generalized to include phonons and electron-phonon interaction can be described by the Hamiltonian

$$\hat{H} = \sum_{K=L,R} \sum_{k \in K, \sigma} \varepsilon_{k\sigma} \hat{c}_{k\sigma}^\dagger \hat{c}_{k\sigma} + \sum_\sigma \varepsilon_{0\sigma} \hat{d}_\sigma^\dagger \hat{d}_\sigma + \omega_0 \hat{a}^\dagger \hat{a} + \sum_\beta \omega_\beta \hat{b}_\beta^\dagger \hat{b}_\beta$$
$$\sum_{K=L,R} \sum_{k \in K, \sigma} \left( V_{k\sigma} \hat{c}_{k\sigma}^\dagger \hat{d}_\sigma + h.c. \right) + U \hat{n}_\uparrow \hat{n}_\downarrow + M \hat{Q} \sum_\sigma \hat{n}_\sigma + \sum_\beta U_\beta \hat{Q} \hat{Q}_\beta \quad (68)$$

Eq. (68) is written for a bridge characterized by one electronic level coupled to a single primary phonon. Here $\sigma=\uparrow,\downarrow$ is the electron spin index, $\hat{c}_{k\sigma}^{\dagger},(\hat{c}_{k\sigma})$ and $\hat{d}_{\sigma}^{\dagger},(\hat{d}_{\sigma})$ create (destroy) electrons in the leads and the molecular level, respectively, $\hat{a}^{\dagger},\hat{a}$ and $\hat{Q}=\hat{a}^{\dagger}+\hat{a}$ are operators for the molecular vibration while $\hat{b}_{\beta}^{\dagger},\hat{b}_{\beta}$ and $\hat{Q}_{\beta}=\hat{b}_{\beta}^{\dagger}+\hat{b}_{\beta}$ are similar operators for the bosonic thermal bath. $\hat{n}_{\sigma}=\hat{d}_{\sigma}^{\dagger}\hat{d}_{\sigma}$ is the electron number operator on the molecular bridge for spin $\sigma$. The parameters $M$ and $U$ characterize respectively the electron-phonon and electron-electron interactions on the bridge while $U_{\beta}$ represents the interaction between bridge and environmental phonons. A small polaron (Lang-Firsov) transformation identical to that used to obtain (39) now leads to

$$\hat{\bar{H}} = \sum_{K=L,R}\sum_{k\in K,\sigma}\varepsilon_{k\sigma}\hat{c}_{k\sigma}^{\dagger}\hat{c}_{k\sigma} + \sum_{\sigma}\bar{\varepsilon}_{0\sigma}\hat{d}_{\sigma}^{\dagger}\hat{d}_{\sigma} + \omega_0\hat{a}^{\dagger}\hat{a} + \sum_{\beta}\omega_{\beta}\hat{b}_{\beta}^{\dagger}\hat{b}_{\beta}$$
$$\sum_{K=L,R}\sum_{k\in K,\sigma}\left(\bar{V}_{k\sigma}\hat{c}_{k\sigma}^{\dagger}\hat{d}_{\sigma}+\text{h.c.}\right)+\bar{U}\hat{n}_{\uparrow}\hat{n}_{\downarrow}+\sum_{\beta}U_{\beta}\hat{Q}\hat{Q}_{\beta} \quad (69)$$

where

$$\bar{\varepsilon}_{0\sigma} = \varepsilon_{0\sigma} - M^2/\omega_0 \quad (70)$$

$$\bar{U} = U - 2M^2/\omega_0 \quad (71)$$

$$\bar{V}_{k\sigma} = V_{k\sigma}\hat{X} \quad (72)$$

and where $\hat{X}$ is the phonon shift generator (41) that now takes the form

$$\hat{X} = \exp\left[\frac{M}{\omega_0}\left(\hat{a}-\hat{a}^{\dagger}\right)\right] \quad (73)$$

In addition to renormalization of parameters discussed in Sect. 5c, the electron-phonon interaction is seen, Eq. (71), to induce an effective attractive interaction between electrons. Although no conclusive observations of this effect in molecular transport junctions have been so far reported, this bi-polaronic attraction can potentially change the physics of the transmission process[251, 252, 375, 376, 278].

Vibrational features of single electron transistors were considered within this or similar models, using approximations based on either near-equilibrium considerations[252, 271, 272, 413] or master equation approaches. [274, 414, 276, 415, 136] Alexandrov and Bratkovsky [252] use exact results for the isolated molecule in an expression for the current obtained by coupling it to the leads. Cornaglia and coworkers[271, 272] use numerical renormalization group to describe the linear response regime of junction conductance, while Al-Hassanieh et al[413] use exact diagonalization supplemented by a Dyson equation embedding procedure to study the influence of center-of-mass motion on linear conductance of the junction. Braig and Flensberg[274, 414] use a quasiclassical master equation approach to study the $U \to \infty$ limit of Coulomb blockade in the presence of equilibrium vibrations. Similar approaches were used by Koch and von Oppen[276] to predict a significant current suppression (Franck-Condon blockade) at low bias and large noise enhancement at higher bias due to strong electron-phonon coupling, and to study vibrational heating[188] and anharmonic effects,[277] in model molecular junctions, by Siddiqui et al[415] to discuss similar effects in nanotube quantum dots and by Armour and MacKinnon[136] to study the effect of quantized vibrational mode (center-of-mass motion) on electron tunneling within Coulomb blockade regime. Finally, a more advanced approach was recently proposed in Ref. [273] to study vibrational sidebands of the Kondo resonance. The authors use a perturbative renormalization group (in the limit of weak electron-vibration coupling) to study an STM-like situation, where the molecular level is in equilibrium with the substrate side of the junction.

We[195] have recently extended the equilibrium equation-of-motion approach used in Refs. [378] and [416] to the case of nonequilibrium transport, and have used an approximate scheme akin to the Born-Oppenheimer approximation to further generalize it to the presence of electron-phonon interactions. This leads to a generalization of the computational scheme discussed in Section 5c[255] to the model (68). This approach is capable of grasping the main vibrational features observed in Coulomb blockade transport situations in molecular junctions. Inelastic resonant cotunneling peaks appear in the conductance plot (left panel of Figure 21) as satellites parallel to the diamond boundaries, and inelastic non-resonant cotunneling peaks are better seen in the $d^2I/d\Phi^2$ plot (right panel of Figure 21) as horizontal features in the blockaded regions of the plot. It can also reproduce inelastic satellites of the Kondo peak in the limit of small population fluctuations when the Kondo effect is due mainly to spin fluctuations.[195]

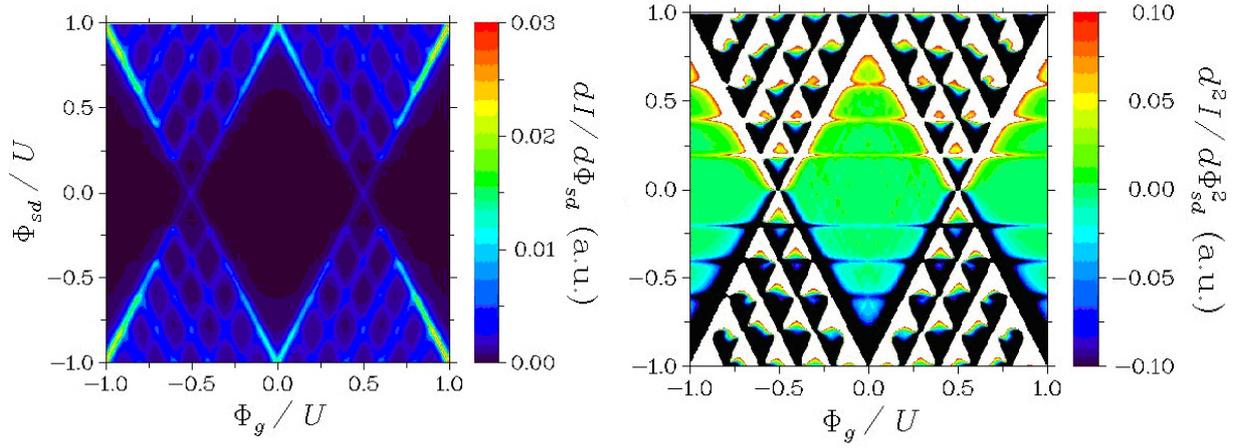

Figure 21. A countour map of the differential conductance, $dI/d\Phi_{sd}$ (left) and the conductance derivative, $d^2I/d\Phi^2$, plotted in the $\Phi_{sd}-\Phi_g$ plane for the model (68) characterized by the parameters T=10K, $\varepsilon_{0\sigma}=0.3\,\text{eV}$, $\Gamma_K^{(0)}=0.01\,\text{eV}$, $U$ = 2.6eV, $E_F$ = 0, $\omega_0=0.2\,\text{eV}$, M=0.4eV and $\gamma_{ph}=0.01\,\text{eV}$. Non-resonance (IETS) features are seen in the conductance derivative map (right) in the non conducting regions of this diamond diagram.

## 7. Noise

In addition to the current-voltage characteristics, noise in the current signal provides an important source of additional information on junction transport properties.[289] Inelastic effects in the noise spectrum were studied recently, first in connection with nanoelectromechanical systems (NEMS).[290, 275, 291-294] Several works use the scattering theory approach[297-300] and a recent work utilizes the NEGF methodology.[250] Here we discuss several important aspects of this issue following on our recent NEGF-based analysis[301] and focusing on inelastic effects on the zero frequency noise in the tunneling current.

Standard analysis of current noise usually considers its spectrum, defined as the Fourier transform of the current correlation function

$$S(\omega) = 2\int_{-\infty}^{\infty} dt\, S(t) e^{i\omega t} \qquad (74)$$

$$S(t) = \frac{1}{2}\langle \Delta \hat{I}(t)\Delta \hat{I}(0) + \Delta \hat{I}(0)\Delta \hat{I}(t)\rangle \qquad (75)$$

where

$$\Delta \hat{I}(t) = \hat{I}(t) - \langle \hat{I} \rangle \qquad (76)$$

While at steady state the average current $\langle \hat{I} \rangle$ does not depend on position along the wire, more care is needed when time dependent fluctuations are considered. Following Ref. [417] we write

$$\hat{I}(t) = \eta_L \hat{I}_L(t) + \eta_R \hat{I}_R(t) \qquad (77)$$

where

$$\hat{I}_K(t) = \frac{2ie}{\hbar} \sum_{k=K;i} \left( V_{ki} \hat{c}_k^\dagger(t) \hat{d}_i(t) - V_{ik} \hat{d}_i^\dagger(t) \hat{c}_k(t) \right) \qquad (78)$$

and

$$\eta_L = \frac{C_R}{C}; \quad \eta_R = -\frac{C_L}{C} \qquad C = C_L + C_R \qquad (79)$$

where $C_L$ and $C_R$ are junction capacitance parameters that describe the response to charge accumulation at the corresponding bridge-lead interfaces. Other important parameters are the voltage division factor $\delta$ that describes the way in which the bias voltage $\Phi$ is distributed between the two molecule-lead interfaces

$$\mu_L = E_F + \delta e\Phi \equiv E_F + e\Phi_L \qquad (80a)$$

$$\mu_R = E_F - (1-\delta)e\Phi \equiv E_F + e\Phi_R \qquad (80b)$$

and the asymmetry in the leads-molecule couplings defined by

$$\alpha = \frac{\Gamma_L}{\Gamma} \qquad 1 - \alpha = \frac{\Gamma_R}{\Gamma} \qquad (81)$$

Here the wide band approximation is invoked by disregarding the energy dependence of the latter parameters. Equivalent circuit arguments[301] suggest the following relationship between these parameters

$$\frac{\alpha}{\alpha - 1} = \frac{1-\delta}{\delta} \frac{1-\eta}{\eta} \qquad (82)$$

This leaves two undetermined parameters in the theory. In what follows these will be represented by $\alpha = \Gamma_L / \Gamma$ and $\eta = \eta_L = C_L / C$.

Within the NEGF formalism and the non-crossing approximation with respect to coupling to the leads, the noise spectrum is obtained in the form[301]

$$S(\omega) = \frac{2e^2}{\hbar} \sum_{K_1,K_2=L,R} \eta_{K_1}\eta_{K_2} \int_{-\infty}^{\infty} \frac{dE}{2\pi} \text{Tr}\Big\{ \delta_{K_1,K_2}\Big[ G^<(E-\omega)\Sigma^>_{K_1}(E) + \Sigma^<_{K_1}(E)G^>(E-\omega) \Big]$$
$$+ G^<(E-\omega)\Big[\Sigma_{K_1}G\Sigma_{K_2}\Big]^>(E) + G^>(E-\omega)\Big[\Sigma_{K_1}G\Sigma_{K_2}\Big]^<(E)$$
$$- \Big[\Sigma_{K_1}G\Big]^<(E-\omega)\Big[\Sigma_{K_2}G\Big]^>(E) - \Big[G\Sigma_{K_1}\Big]^<(E-\omega)\Big[G\Sigma_{K_2}\Big]^>(E) + (\omega \leftrightarrow -\omega) \Big\}$$

(83)

where we use the notation

$$[AB]^{>,<}(E) = A^{>,<}(E)B^a(E) + A^r(E)B^{>,<}(E)$$
$$[ABC]^{>,<}(E) = A^{>,<}(E)B^a(E)C^a(E) + A^r(E)B^{>,<}(E)C^a(E) \quad (84)$$
$$+ A^r(E)B^r(E)C^{>,<}(E)$$

An essentially equivalent result was obtained by Bo and Galperin[418]. We restrict our discussion to the zero frequency noise which is the relevant observable when the measurement time is long relative to the electron transfer time, a common situation in usual experimental setups. The results presented below correspond to a bridge characterized by one electronic level coupled to a single (primary) vibrational mode. In the absence of electron phonon interaction it can be shown that (83) simplifies to a sum of a thermal contribution due to thermal excitations in the contacts and a shot noise term associated with the discrete nature of the electron transport.[289] Also, in this ballistic transport process the zero-frequency noise does not depend on the junction capacitance factors $\eta_K$. Fig. 22 shows the results obtained from (83) with $\omega = 0$. The shape of the noise characteristic as a function of applied source-drain voltage depends on asymmetry parameter $\alpha$, and changes from a double-peak structure for symmetric coupling to single-peak shape for a highly asymmetric junction. Specifically, the condition for double-peak structure is found[301] to be

$$\alpha^2 - \alpha + \frac{1}{8} < 0 \tag{85}$$

and the asymmetry in this structure results from the thermal noise contribution. The difference between peak heights can be shown to be of order $\sim T/\Gamma$.

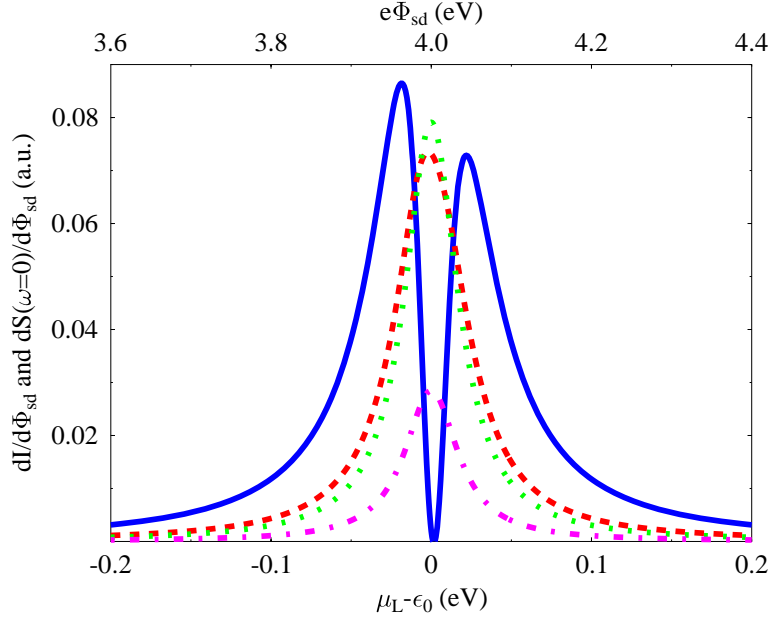

Figure 22. (Fig. 1 of Ref. [301]) Conductance and differential noise vs. applied source-drain voltage in the elastic transmission case. Shown are the conductance $dI/d\Phi_{sd}$ (dotted and dash-dotted lines) and the differential noise $dS(\omega=0)/d\Phi_{sd}$ (solid and dashed lines) for $\alpha=0.5$ and $\alpha=0.1$ respectively. Other parameters of the calculation are $\Gamma^{(0)}=0.04$ eV, $\varepsilon_0=2$ eV, T=10K.

To account for phonon effects on the noise spectrum we consider Eq. (83) in the $\omega=0$ limit. The GFs and SEs that appear in this expression are evaluated for the one bridge state/ one bridge oscillator version of the model (5)-(8), using the weak[235] or strong[255] coupling procedures described in Sections 5b,c. It is found that the noise spectrum can no longer be cast in terms of additive thermal and shot noise contributions, and that its character depends strongly on the electron-phonon interaction in addition to the junction parameters α and η. Some examples are shown in Figures 23-25. Figs. 23 and 24 show the current $I$, the zero frequency noise $S$ and the Fano factor $S/I$, normalized by their counterparts in the absence of electron-phonon coupling ($I_0$, $S_0$ and $F_0$) and plotted against the applied voltage. These results belong to the weak coupling case ($M<\Gamma$), however Fig. 23 corresponds to the off resonance situation ($E-\varepsilon_0>\Gamma$, where $E$ is the energy of the tunneling electron) while Fig. 24 corresponds to the resonant case where the opposite inequality is satisfied. In the off-resonant limit the $F/F_0$ ratio is smaller than 1 for any choice of parameters α and $\eta_K$. In contrast, in the resonant case, this ratio is greater than 1 and increases with $\Phi_{sd}$ in symmetric junctions with $\alpha=0.5$, but is smaller than 1 and decreases with $\Phi_{sd}$ in the highly asymmetric junctions.

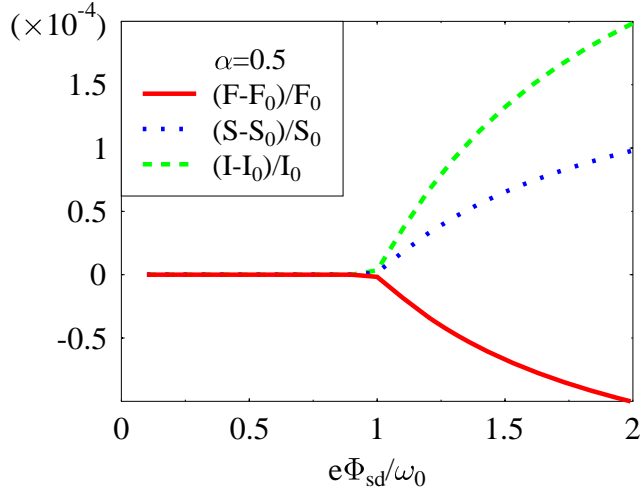

Figure 23. Ratios of Fano factors (solid line), zero frequency noises (dotted line), and currents (dashed line) with and without electron-vibration coupling, plotted against the applied voltage in the off-resonant tunneling regime. Parameters of the calculation are T=10K, $\varepsilon_0 = 5\,\text{eV}$, $\Gamma = 0.5\,\text{eV}$, $\alpha = 0.5$, $E_F$=0, $\omega_0 = 0.1\,\text{eV}$, M=0.1eV, $\gamma_{ph} = 0.01\,\text{eV}$. (From Ref. [301]).

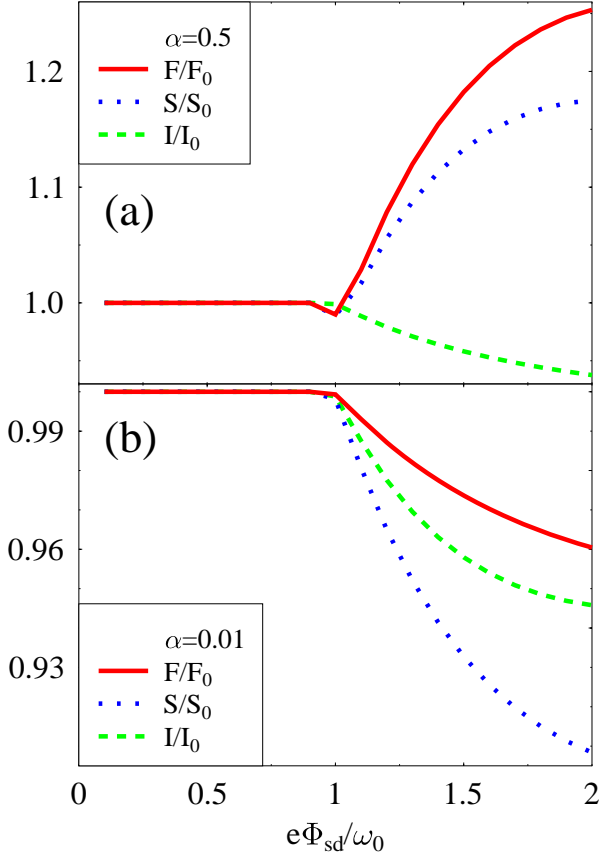

Figure 24. Ratios of Fano factors (solid line), zero frequency noises (dotted line), and currents (dashed line) with and without electron-vibration coupling, plotted against the applied voltage in the resonant tunneling regime. Shown are (a) symmetric $\alpha = 0.5$ and (b) asymmetric $\alpha = 0.01$ coupling cases. Parameters of the calculation are the same as in Figure 23 except $\varepsilon_0 = 0.05$ eV. (From Ref. [301])

Figure 25 shows an example for the noise characteristic in the strong electron-vibration coupling case, $M > \sqrt{(E_F - \varepsilon_0)^2 + (\Gamma/2)^2}$ which characterizes many resonance tunneling situations. The general shape of the $dS(\omega = 0)/d\Phi_{sd}$ vs. $\Phi_{sd}$ spectrum is similar to the conductance-voltage spectrum, with a central elastic feature at the energy of the bridging orbital accompanied by phonon sidebands. An important difference between the noise and conductance lineshapes is the form of the elastic feature, which appears as a single peak in the conductance spectrum. In the differential noise spectrum this feature crosses from a double-peak structure to a single-peak shape as the coupling $M$ or the coupling asymmetry parameter $\alpha$ increase. This lineshape dependence on junction properties can be used to estimate junction coupling parameters from noise spectra.

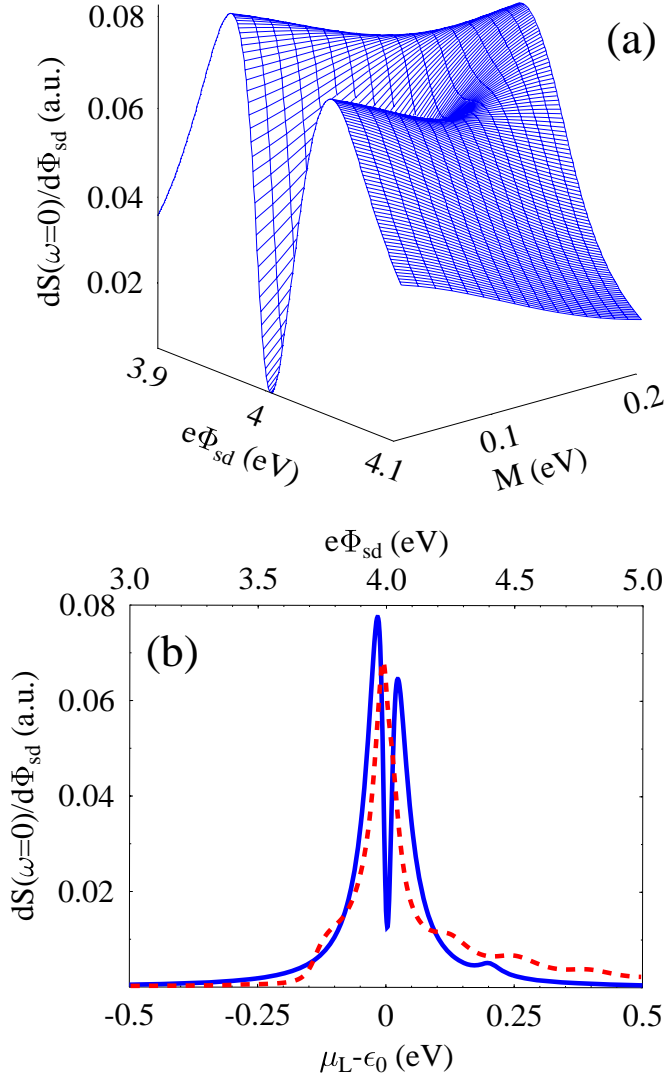

Figure 25. (Fig. 11 of Ref. [301]) Differential noise vs. applied source-drain voltage for a junction characterized by the symmetry factor $\alpha = 0.5$. (a) Surface plot of the differential noise, $dS(\omega=0)/d\Phi_{sd}$, as function of $\Phi_{sd}$ and $M$ as obtained from a lowest order calculation. (b) Differential noise plotted against gate potential for weak ($M = 0.04$ eV, solid line) and strong ($M = 0.3$ eV, dashed line) electron-vibration coupling, obtained from a self-consistent calculation using the strong coupling procedure described in Sect. 5c. Other parameters of the calculation are $T = 10$K, $E_F = 0$, $\varepsilon_0 = 2$ eV, $\Gamma^{(0)} = 0.04$ eV, $\omega_0 = 0.2$ eV, $\gamma_{ph} = 0.01$ eV.

Strong electron-phonon coupling can give rise to noise phenomena that are not described by the model presented above. For example, such strong coupling may lead to multi-stable behavior that can appear as intermittent noise in the junction current (see Section 8). More generally, large amplitude conformational fluctuations, e.g. dynamical structural variations that are not described by our harmonic model, may contribute to the observed noise. Finally, it has been pointed out[276, 340] that strong electron-phonon coupling associated with pronounced Franck-Condon blockade behavior leads to electron transport by avalanches. Such avalanche behavior results from repeated sequences of junction heating followed (once the blockade threshold is overcome) by transfer of a large number of electrons across the junction, and is manifested by noise behavior characterized by very large ($10^2$-$10^3$) Fano factors.

## 8. Non-linear conductance phenomena

In section 5c we discussed spectroscopic manifestations of strong electron-phonon coupling. Here we address other possible consequences of such strong coupling, where charging of the molecular bridge (stabilized by this interaction) can lead to non linear transport behavior. Indeed, 'stabilization of molecular charging' may often appear as modification of molecular geometry and can therefore give rise to substantial and sometimes striking effects of negative differential resistance, multistability and hysteresis phenomena. Such structural changes are characteristic of molecular entities, and will therefore be of major importance in the ongoing investigation of molecular, as opposed to solid state or mesoscopic, transport structures.

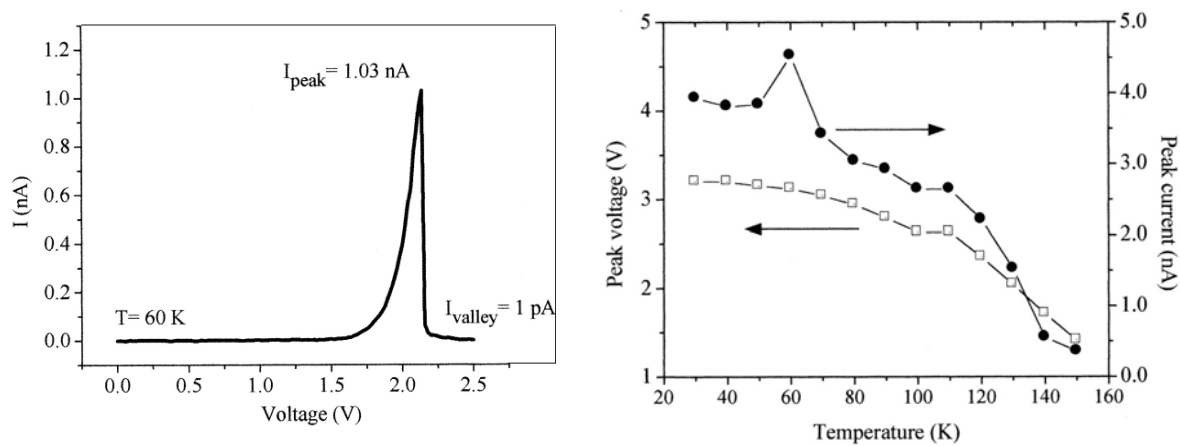

Figure 26 NDR of a junction based on monolayer of 2'-amino-4-ethynylphenyl-4'-ethynylphenyl-5'-nitro-1-benzenethiolate embedded between gold wires at 60K. Shown on the left is the reported $I$-$\Phi$ characteristic with NDR peak to valley ratio 1030:1. The temperature dependence of the current and voltage values at the peak is shown on the right. (From Ref [419]).

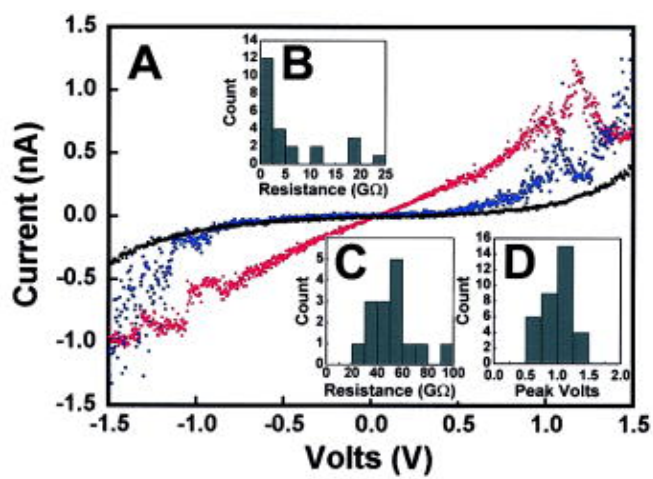

Figure 27. Conducting atomic force spectroscopy measurements of current-voltage characteristics (A) for 1-nitro-2,5-di(phenylethynyl-4'-thioacetyl)benzene (red and blue curves) and 2,5-di(phenylethynyl-4'-thioacetyl)benzene (black curve) molecules. The first molecule exhibits both NDR and wide range of background ohmic currents, distribution of resistances is shown by the histogram inset (B). The second molecule shows no NDR-like features and resistance in the ohmic region is much more tightly clustered, see inset (C). Distribution of NDR peak voltages for the first molecule is shown in inset (D). (From Ref. [420]).

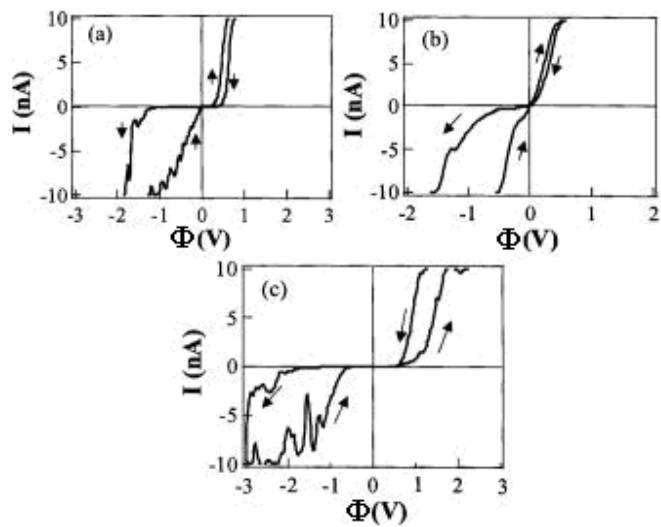

Figure 28. Typical current-voltage curves of the Pd/molecular wire/Au SAM junctions on Si/SiO$_2$ substrate for molecules containing electron-withdrawing nitro or pyridine groups. (From Ref. [421])

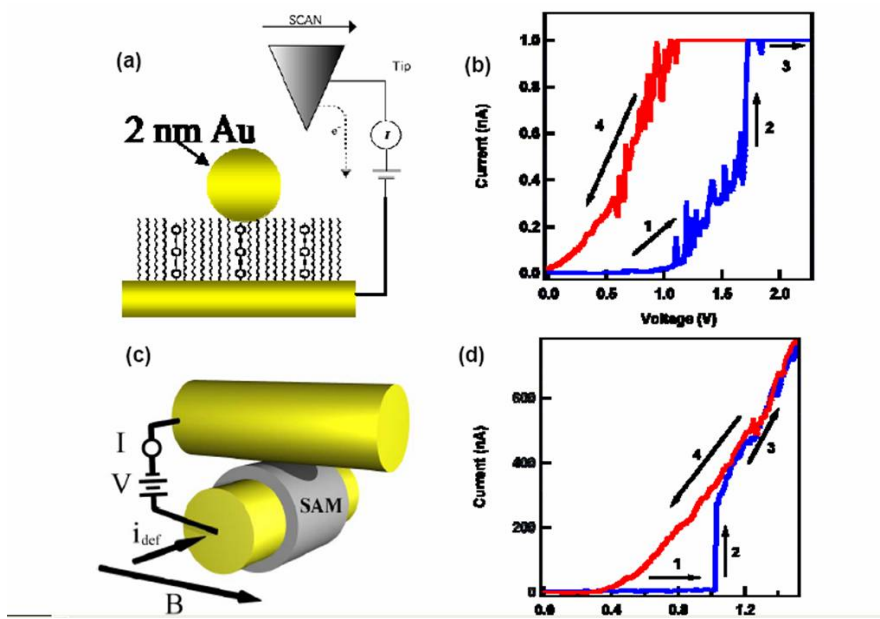

Figure 29. Current-voltage measurements on individual BPDN molecules by STM with BPDN embedded into $C_{11}$ alkane matrix (a) and cross-wire tunneling junction (c) are presented in Figs. (b) and (d) respectively. (From Ref. [74]).

Examples of such behaviors are shown in Figures 26-29. Figures 26[419] and 27[420] show negative differential resistance, while Figs. 28[421] and 29 [74] show hysteresis in different molecular junctions. The molecules involved in these junctions are characterized by the presence of redox centers, i.e. centers that support long-living excess electron states. Such "redox molecules" have been implicated in several other observations of multiple conduction states and non-linear response in molecular junctions operating in a polar (aqueous) environment.[422, 247, 129, 423, 424, 80, 425] This suggests the possibility of polaron formation on the molecule as a possible factor. Indeed, the model (5)-(8) has a positive feedback character: the energy of the resonant level shifts by polaron formation that depends on the electronic occupation of that level. The latter, in turn, depends on the level energy.

Ref. [91] is a study of the nature and possible consequences of this feedback character on the conduction behavior of such junctions, using the reduced one bridge level/ one bridge (primary) oscillator version of the model (5)-(8). This study invokes a mean field approximation akin to the Born-Oppenheimer approximation, which is based on the assumption that the primary vibrational mode is slow relative to the rate at which electrons enter and leave the bridge, i.e. $\omega_0 \ll \Gamma$. In this case the oscillator responds only to the average bridge occupation.[426, 427] The system dynamics is then described by the electronic and oscillator Hamiltonians

$$\hat{H}_{el}(Q) = \varepsilon_0 \hat{d}^\dagger \hat{d} + \sum_{k \in L,R} \varepsilon_k \hat{c}_k^\dagger \hat{c}_k + \sum_{k \in L,R} \left( V_k \hat{c}_k^\dagger \hat{d} + V_k^* \hat{d}^\dagger \hat{c}_k \right) + MQ\hat{d}^\dagger \hat{d} \qquad (86a)$$

$$\hat{H}_{osc} = \omega_0 \hat{a}^\dagger \hat{a} + M\hat{Q}n_0 + \sum_\beta \omega_\beta \hat{b}_\beta^\dagger \hat{b}_\beta + \sum_\beta U_\beta \hat{Q}\hat{Q}_\beta \, ; \qquad n_0 = \left\langle \hat{d}^\dagger \hat{d} \right\rangle \qquad (86b)$$

$$\hat{Q} = \hat{a}^\dagger + \hat{a} \, ; \quad \hat{Q}_\beta = \hat{b}_\beta^\dagger + \hat{b}_\beta \qquad (86c)$$

The steady-state result for oscillator shift coordinate is obtained from (86b)

$$\left\langle \hat{Q} \right\rangle = -\frac{2\omega_0}{\omega_0^2 + (\gamma_{ph}/2)^2} Mn_0 \qquad (87)$$

($\gamma_{ph}$ is the damping rate of the primary oscillator due to its coupling to the phonon bath) and is used for $Q$ in Eq. (86a). This leads to an effective purely electronic Hamiltonian with level energy that depends on its occupation

$$\hat{H}_{el},eff = \bar{\varepsilon}_0(n_0) d^\dagger d + \sum_{k \in L,R} \varepsilon_k \hat{c}_k^\dagger \hat{c}_k + \sum_{k \in L,R} \left( V_k \hat{c}_k^\dagger \hat{d} + V_k^* \hat{d}^\dagger \hat{c}_k \right) \qquad (88a)$$

$$\bar{\varepsilon}_0(n_0) = \varepsilon_0 - 2E_r n_0 \qquad (88b)$$

Here $E_r$ is the reorganization energy (compare Eq. (15))

$$E_r = \frac{M^2 \omega_0}{\omega_0^2 + (\gamma_{ph}/2)^2} \qquad (89)$$

Eq. (88) implies that the steady-state solution for the average electronic population in the bridge level is given by the equation

$$n_0 = \int_\infty^\infty \frac{dE}{2\pi} \frac{f_L(E)\Gamma_L + f_R(E)\Gamma_R}{[E - \bar{\varepsilon}_0(n_0)]^2 + [\Gamma/2]^2} \, ; \quad \Gamma = \Gamma_L + \Gamma_R \qquad (90)$$

whose non-linear form allows for multiple solution and multistability properties. Similar models were recently discussed by several authors.[257, 270, 269, 92] The dynamical consequences of this multistability are still under discussion. Whether they can lead to hysteresis behavior and memory effect as suggested in [91] or to intermittent noise associated with transitions between two locally stable states as discussed in [270, 92] is an issue of relative timescales – the observation time vs. the rate of transitions between locally stable states. An interesting possibility that such a mechanism can be the cause of observed negative differential conduction phenomena has also been pointed out,[91] and may again depend on relative timescales.[90] On the other hand, a recent experimental study of hysteretic conductance in gated molecular junctions based on the redox molecule bipyridyl-dinitro oligophenylene-ethynylene dithiols (BPDN-DT)[86] indicates that the observed behavior is not sensitive to the gate potential (in contrast to the $\varepsilon_0$ dependence in (90)), suggesting that at least in this system the actual mechanism may go beyond the simple picture described above.

## 9. Heating and heat conduction

Localized Joule heating poses a crucial question for the functionality and reliability of molecular devices. The combination of small molecular heat capacity and inefficient heat transfer away from it might cause a large temperature increase that would affect the stability and integrity of molecular junctions. The rates at which heat is deposited in and transported away from the conducting junction are therefore crucial to the successful realization of nano electronic devices.

### 9a. General considerations

In insulators heat is conducted by atomic vibrations, while in metals electrons are the dominant carriers. For a molecular system connecting between two metal electrodes both carrier types exist and mutually interact. A unified description of their dynamics[133] starts again from the Hamiltonian (5)-(8), focusing now on the problems of heat generation and transport. A general framework for discussing these issues is again provided by the non-equilibrium Green function (NEGF) formalism, pioneered for this application by Datta and coworkers[191, 192] and further advanced recently by several groups. [193-195] Fluxes in this formalism are expressed in terms of the Keldysh Green functions (GFs) for electrons and phonons

$$G_{ij}(\tau,\tau') = -i\left\langle T_c \hat{d}_i(\tau)\hat{d}_j^\dagger(\tau')\right\rangle \tag{91}$$

$$D_{\alpha\alpha'}(\tau,\tau') = -i\left\langle T_c \hat{Q}_\alpha(\tau)\hat{Q}_{\alpha'}^\dagger(\tau')\right\rangle \tag{92}$$

(and by their real time projections, $G^a, G^r, G^>, G^<$, and same for $D$) and by the corresponding self energies (SEs). At steady state the net electronic fluxes into the junction at each contact, $K = L, R$, are given by

$$I_K = \frac{1}{\hbar}\int_{-\infty}^{+\infty}\frac{dE}{2\pi}i_K(E) \quad ; \quad i_K(E) = i_K^{in}(E) - i_K^{out}(E) \tag{93}$$

where

$$\begin{aligned} i_K^{in}(E) &= \text{Tr}\left[\Sigma_{L,R}^<(E) G^>(E)\right] \\ i_K^{out}(E) &= \text{Tr}\left[\Sigma_{L,R}^>(E) G^<(E)\right] \end{aligned} \qquad (94)$$

and where $\Sigma_K^{>,<}$, $K=L,R$, given by Eqs. (28)-(29), are the greater and lesser self energy matrices in the space of the bridge electronic subsystem associated with its electron transfer coupling to the metal electrodes. The corresponding electronic energy fluxes into the junction at each contact are

$$J_{E,K}^{el} = \int_{-\infty}^{+\infty} \frac{dE}{2\pi} E\, i_K(E) \qquad (95)$$

In the absence of particle and energy sources and sinks $I_L = -I_R$ and $J_{E,L}^{el} = -J_{E,R}^{el}$. In the presence of electron-phonon interactions in the junction, the primary phonons effectively enter as source/sink to the electronic energy balance. The rate of energy transfer between the electron and phonon subsystem on the molecular bridge is therefore given by

$$J_{\Delta E}^{el} = J_{E,L}^{el} + J_{E,R}^{el} \qquad (96)$$

(a positive $J_{\Delta E}^{el}$ indicates energy transfer from electrons to phonons). It is usually assumed that this energy appears as heat in the phonon subsystem, and Eq. (96) provides a starting point for the discussion of heat generation on the junction.[20]

Next consider heat conduction. While our main concern is the conduction of heat out of the junction, a standard heat conduction problem focuses on the heat carried by a system connecting two thermal reservoirs at different temperatures. The heat carried by the electronic current through the interface $K$ is given by[150]

$$J_{Q,K}^{el} = -\int_{-\infty}^{+\infty} \frac{dE}{2\pi} (E - \mu_K) i_K(E) \qquad (97)$$

In a biased junction this represents mostly Joule heating in the leads. For a molecular bridge connecting two reservoirs at different temperatures and without potential bias this is the heat carried by the thermoelectric (Seebeck) current. The latter contribution to the heat conduction in unbiased junctions is usually much smaller than that due to phonons.

---

[20] In the absence of coupling to phonons on the bridge heat is generated only in the leads. (One still assumes the existence of a dissipation mechanism that keeps the leads in their corresponding equilibrium states). The heat generation rate in the lead $K=L,R$ by the electronic current, is given by Eq. (97).

Consider now the phononic heat transport.[21] A general quantum expression for the phonon thermal flux within the NEGF formalism can be obtained for the model represented by Eqs. (5), (6) and (8)[22] [193-195] The energy/heat flux from the phonon thermal bath $K$ into the junction is given by

$$J_K^{ph} = -\int_0^\infty \frac{d\omega}{2\pi} \omega \, \text{Tr}\left[\Pi_K^{ph,<}(\omega) D^>(\omega) - \Pi_K^{ph,>}(\omega) D^<(\omega)\right] \qquad (98)$$

where Tr stands for summing over all the bridge (primary) vibrations, $\Pi_K^{ph,>}$ and $\Pi_K^{ph,<}$ are the greater and lesser self energy matrices of these vibrations due to their coupling to the bath

$$\begin{aligned}
\left[\Pi_K^{ph,<}(\omega)\right]_{mm'} &= -i\,\Omega_{mm'}^K(\omega) F_K(\omega) \\
\left[\Pi_K^{ph,>}(\omega)\right]_{mm'} &= -i\,\Omega_{mm'}^K(\omega) F_K(-\omega)
\end{aligned} \qquad (99)$$

where

$$\begin{aligned}
F_K(\omega) &= \begin{cases} N_K(\omega) & \omega > 0 \\ 1 + N_K(|\omega|) & \omega < 0 \end{cases} \\
\Omega_{mm'}^K(\omega) &= 2\pi \sum_\beta U_{m\beta} U_{\beta m'} \, \delta(\omega - \omega_\beta)
\end{aligned} \qquad (100)$$

and

$$N_K(\omega) = N_{eq}(\omega, T_K) \equiv \left[\exp(\omega/k_B T_K) - 1\right]^{-1} \qquad (101)$$

is the Bose-Einstein distribution in the contact $K$. A simpler expression can be obtained from (98) for the harmonic bridge model (7) in the case where the electron-phonon interaction does not cause energy exchange between these subsystems *on the bridge.* In this (artificial) situation, phonons may cause decoherence of electronic motion on the bridge but do not exchange energy with electrons on the bridge, so that at steady-state their flux is the same throughout the junction including the $L$ and $R$ interfaces. If, in addition, the matrices $\Omega^L(\omega)$ and $\Omega^R(\omega)$ are proportional to each other, i.e. $\Omega^L(\omega) = c\,\Omega^R(\omega)$ with $c$=constant, then Eq. (98) leads to[195]

---

[21] For transport by phonons the energy and heat fluxes are equivalent, because in the absence of particle conservation there is no chemical potential for phonons.

[22] The simplified characteristics of the bridge model, Eq. (7), are not needed here: Eq. (98) can be derived for a general molecular Hamiltonian (including, e.g., anharmonic interactions) provided that the interaction with the external (free phonon) bath(s) is bilinear. The derivation follows the steps of an analogous development for the electronic current[226], [225] and relies on the non-crossing approximation[393] which in the present context amounts to assuming that the interactions of the 'system' with different 'bath' environments are independent of each other.

$$J^{ph} = \frac{1}{\hbar} \int_0^\infty \frac{d\omega}{2\pi} \omega \, \text{Tr}\left[\Omega^L(\omega) D^r(\omega) \Omega^R(\omega) D^a(\omega)\right] (N_L(\omega) - N_R(\omega))$$

$$+ \frac{1}{\hbar} \int_0^\infty \frac{d\omega}{2\pi} \omega \, \text{Tr}\left[\frac{\Omega^L(\omega) \Omega^R(\omega)}{\Omega(\omega)} D^r(\omega) \Omega_{el}(\omega) D^a(\omega)\right] (N_L(\omega) - N_R(\omega)) \quad (102)$$

where $\Omega(\omega) = \Omega^L(\omega) + \Omega^R(\omega)$ and $\Omega_{el}(\omega) = -2\,\text{Im}\left[\Pi_{el}^r(\omega)\right]$ is the imaginary part of the retarded projection of the primary phonons SE due to their coupling to the electronic subsystem on the bridge. The result (102) contains additively the heat conduction by the pure harmonic bridge and a correction term associated with the electron-phonon interaction.[23] We note that the latter term is responsible for the lifetime broadening of these phonons due to their coupling (induced by their interaction with the bridge electronic system) to electron-hole excitations in the leads, that was argued to dominate the broadening of vibrational features in inelastic electron tunneling spectroscopy at low temperatures[234]

In the absence of electronic conduction, e.g. for a harmonic bridge connecting between dielectric thermal baths, Eq. (102) yields the pure phononic heat flux between two thermal phonon reservoirs connected by a harmonic bridge[24]

$$J^{ph} = \frac{1}{\hbar} \int_0^\infty \frac{d\omega}{2\pi} \omega \, \text{Tr}\left[\Omega^L(\omega) D^r(\omega) \Omega^R(\omega) D^a(\omega)\right] (N_L(\omega) - N_R(\omega)) \quad (103)$$

which was obtained in different ways before.[428-430] In the most general case, where both electrons and phonons are transported and exchange energy in the junction, the general expressions (97) and (98) have to be used, although the phononic contribution (98) is expected to dominate in molecular junctions.

Junction heating is determined by the balance between the rate at which heat is deposited in the junction and the rate at which it is conducted away. Eqs. (91)-(103) present a general formalism for treating this problem, however the application to realistic junction models is prohibitively complex (some simple model results are presented below). Here we review earlier approaches to these problems that can be applied to complex system at the price of disregarding the (presumably small) electronic contribution to the heat conduction. Such considerations usually address separately the heat deposit and conduction processes.

---

[23] It is interesting to note that the same formal form, Eq. (102) is obtained also in the more general case where anharmonic interactions exist between bridge phonons, except that $\Omega_{el}(\omega)$ is now replaced by a more general term that includes also the effects of these interactions. We will not discuss this issue further in this review. (For application of the NEGF formalism to anharmonic effects in molecular heat conduction see Ref. [194])

[24] Note that while (102) relies on the equality $\Omega^L(\omega) = c\,\Omega^R(\omega)$, Eq. (103) can be obtained from (98) without this restriction.

## 9b. Heat generation

When a current $I$ traverses a wire under potential bias $\Phi$, the power converted into heat is $W = I\Phi$. In nanojunctions with sizes small relative to the electron mean free path most of this power, $(1-\eta)W$ with $\eta \ll 1$, is dissipated in the leads. However, as discussed in Sect. 4, conduction can take place also by electron activation and hopping. In this incoherent transport limit $\eta = 1$.

We are concerned with the fraction $\eta$ of the power that is converted into heat in the bridge region. To emphasize the importance of this issue note that in a junction carrying 1nA under a bias of 1V, the dissipated power is $W = 10^{10}$ eV/s, while 10eV deposited locally on the bridge are more than enough to destroy a molecule. The magnitude of the fraction $\eta$ is therefore of utmost importance as is the rate, discussed below, at which heat is conducted away from the junction.

As stated above, heat generation in current-carrying molecular junctions is defined as the process of energy transfer from the molecular electronic subsystem to the underlying nuclear motion. In doing so we tacitly assume that the energy transferred appears in the nuclear subsystem as heat, i.e. randomized motion.[25] Neither the above definition nor the assumption is obvious. For example, in metallic current carrying systems, electrons move systematically on top of a distribution, assumed thermal, that can be used to define their energy. Conversion of current energy to heat amounts to destroying the systematic part of this motion, transforming the associated kinetic energy into thermal motion expressed as a temperature rise. This randomization of electronic motion is caused by scattering off impurity centers, phonons or other electrons. A local equilibrium assumption is often invoked, with the electron and phonon distributions assumed to have the same temperature.[26] It is only in molecular wires, where conduction is often described as a succession of single carrier (electron or hole) transfer events, that heating is naturally described as energy transfer from these carriers to the phonon subsystem.

---

[25] Exceptions are known. Indeed, in shuttle conductance (see Section 10) some of this energy appears as coherent oscillations of a nuclear coordinate.

[26] Such a picture may break down in processes involving ultrafast optical excitation of metal electrons.

Experimental manifestations of this energy transfer appear mostly as current induced conformational and chemical changes in the molecular bridge[190, 288] (see Section 10). Current-induced heating was suggested as the reason for the observation that a hysteresis loop in the I/V response of a metallic nanojunction that undergoes voltage dependent configurational changes shrinks with increasing current and is eventually replaced by 2-level fluctuations between the two configuration/conduction states.[431] It was also implicated in the voltage dependence of the most probable breakdown force in an octanedithiol-gold conducting AFM breakjunction immersed in toluene, where the activated nature of the breakdown process was used to estimate the junction temperature.[432] The latter work has led to an estimate of the current-induced junction heating in that system, placing it at $\sim 30K$ increase above room temperature at a bias of 1V.

Theoretical aspects of this problem were discussed by several groups. In the NEGF approach of Lake and Datta[191, 192] the model includes a single electron Hamiltonian and a set of localized phonons (phonon $m$ localized at position $\mathbf{r}_m$), kept at thermal equilibrium

$$\hat{H}_0 = \frac{\hat{\mathbf{p}}^2}{2m} + V(\mathbf{r}) + \sum_m \hbar \omega_m \left( \hat{a}_m^\dagger \hat{a}_m + \frac{1}{2} \right) \quad (104)$$

and a local interaction between them

$$\hat{H}' = \sum_m U \delta(\mathbf{r} - \mathbf{r}_m) \left( \hat{a}_m^\dagger + \hat{a}_m \right) \quad (105)$$

For a 1-dimensional conductor these authors calculate the particle current density $J_N(z;E)$ per unit energy, which is related to the total (position independent) particle current, $I_N = \int dE\, J_N(z;E)$ and to the energy current

$$I_E(z) = \int dE\, E\, J_N(z;E) \quad (106)$$

In this picture the power transferred to phonons, i.e. deposited as heat, can be described locally

$$P(z) = -\frac{d}{dz} I_E(z) \quad (107)$$

In the low bias limit these authors find that this power contains two terms. One, quadratic in the applied bias, is identified with the Joule heat. The other, linear in the applied bias, is a manifestation of the Thomson thermoelectric effect in this system.

Another important result of this work is the strong enhancement of heat generation observed near resonance in a double barrier tunnel structure.

Segal and Nitzan[151] have studied this problem using the model (18)-(19) – the same model that was used[168] to describe the crossover from coherent tunneling to activated hopping transport (see Section 4). As discussed in Section 3d, the quantum master equation derived for this model[170, 171] can be used to evaluate differential transmission coefficients $\mathcal{T}_{L \to R}(E, E_0)$ and $\mathcal{T}_{R \to L}(E, E_0)$ for an inelastic transmission process in which an electron entering the lead with energy $E_0$ is scattered out with energy $E$. Approximate expressions for the total particle current and the power left on the bridge are then given by

$$I_N = \frac{1}{\pi\hbar} \int_{-\infty}^{\infty} dE_0 \int_{-\infty}^{\infty} dE \big[ \mathcal{T}_{L \to R}(E_0, E, \Phi) f(E_0)(1 - f(E + e\Phi)) \\ - \mathcal{T}_{R \to L}(E_0, E, \Phi) f(E_0 + e\Phi)(1 - f(E)) \big] \quad (108)$$

$$I_E = -\frac{1}{\pi\hbar} \int_{-\infty}^{\infty} dE_0 \int_{-\infty}^{\infty} dE \big[ \mathcal{T}_{L \to R}(E_0, E, \Phi) f(E_0)(1 - f(E + e\Phi)) \\ + \mathcal{T}_{R \to L}(E_0, E, \Phi) f(E_0 + e\Phi)(1 - f(E)) \big] (E - E_0) \quad (109)$$

Note that (109) is the equivalent of $\int dz P(z)$ where $P(z)$ is the local power dissipation, Eq. (107). It can be shown[151] that in the low bias limit, $\Phi \to 0$, it is proportional to $\Phi^2$, as expected for the Joule heat, however no component linear in $\Phi$ is obtained in this approximation. Estimating the electron-phonon coupling strength from the order of magnitude of reorganization energies in organic systems, Segal and Nitzan estimate the order of magnitude fraction $\eta$ of the available power $eI_N\Phi$ that remains on the bridge to be of order ~0.1, increasing with bridge length. As in the theory of Lake and Datta, it increases strongly when resonance transmission is approached.

Todorov and coworkers[214, 215, 153] have addressed this problem also within the tight binding model (18b) for the bridge, modeling the deviations $\mathbf{u}_n$ of the underlying ions from their equilibrium positions as independent harmonic oscillators and representing the electron-phonon interaction by the lowest order expansion of the tight binding elements $(H_M)_{n,n'}$ in these deviations

$$H_{MB} = \sum_n \sum_{\nu=1,2,3} F_{n\nu} u_{n\nu} \sum_{n'} (|n\rangle\langle n'| + |n'\rangle\langle n|) (\partial H_{n,n'} / \partial R_{n\nu})_{eq} \quad (110)$$

Here $\mathbf{R}_n$ are ionic positions, $v$ goes over the Cartesian directions and the derivatives with respect to ion positions are evaluated at the equilibrium configuration. The energy transfer rate is evaluated by low order quantum perturbation theory in the basis of delocalized electronic states $\{\alpha\}$.[27] When these are taken as Bloch states of a uniform 1-dimensional chain, the net energy transfer rate into a single ion-oscillator $n$ of mass $M$ and frequency $\omega$ is obtained in the form (cf Eq. (11) of Ref. [153])

$$P_n \approx \frac{2\pi\hbar}{M} \frac{1}{\pi^2} \frac{H'^2}{H^2} \left[ \left(|e\Phi| - \hbar\omega\right)\Theta\left(|e\Phi| - \hbar\omega\right) - 2N(\omega)\hbar\omega \right] \quad (111)$$

where $H$ and $H'$ stand for the equilibrium nearest-neighbor hopping element ($V_{n,n+1}$ of Eq. (18b)) and its derivative with distance, $\Theta$ is the step function and $N(\omega) = \left[\exp(\hbar\omega/k_B T) - 1\right]^{-1}$ ($T$ is the effective bridge temperature) is the oscillator thermal population. Eq. (111) implies that when this temperature is low only energy deposition into the oscillator is possible (provided $|e\Phi| > \hbar\omega$), while in general the direction of energy flow is determined by the balance between current-induced heating and cooling. An interesting outcome from this analysis is the observation[153] that in the high bias limit Eq. (111) can be approximated by the result of a classical analysis

$$P_n = 4\frac{m}{M} Ie|\Phi| \quad (112)$$

where $m$ and $M$ are electron and ion masses, respectively and $I$ is the electron current (so that $Ie\Phi$ is the available power). Accordingly, the fraction of the available energy converted to heat is determined by the mass ratio. Multiplying by the number of ions brings this perturbation theory-based estimate to the order of 1%.

Di Ventra and coworkers have combined a similar quantum perturbation methodology with ab-initio calculations of electron-vibrational coupling in realistic models of molecular junctions.[212, 213, 433] The net heat transfer between the electronic and nuclear subsystems is evaluated from a suitable generalization of Eq. (111). In conjunction with an estimate of heat conduction out of the junction vibrational subsystem, estimates of the expected temperature rise (a few tens of degrees in various alkanes[213]) could be made.

Finally, while not addressing the issue of power dissipation, several workers have investigated coupled electron and nuclear dynamics in a current carrying junction.[135, 275, 137, 241, 242, 276, 277, 188, 279] Such studies are usually limited to simple models, addressing one oscillator as part of the molecular system (the rest of the nuclear environment can be regarded as a thermal bath) and are based on a kinetic description (expected to be valid in the weak molecule-leads coupling – the Coulomb blockade limit) of the electron hopping in and out of the bridge and coupled to the oscillator motion. We have discussed examples of such works in Section 3d. Thus, within its range of validity and for the simple model considered, Eq. (33) accounts for the balance between the heating and cooling processes in the current carrying junctions. However, in realistic situations energy loss by vibrational relaxation depends on junction geometry and on the molecular structure – the same factors that determine also an independent junction transport attribute – its heat conduction property. We turn to this issue next.

---

[27] A similar methodology was used by Galperin and Nitzan in a study of inelastic energy transfer to vibrations during electron tunneling transmission through water.[154]

## 9c. Heat conduction

While heat conduction is an essential ingredient in the balance of processes that determine junction heating, it is easier to study and analyze as an independent process. To this end we may consider a molecular wire suspended between two heat reservoirs characterized by different temperatures. When these reservoirs are insulators heat is carried by nuclear motions, i.e. phonons. When they are metals, electrons contribute as well, and may dominate the heat transport. In the latter case cross transport (thermoelectric) phenomena are also encountered.

Here we focus on phononic heat transfer. Theoretical interest in this issue goes back to Peierls' early work.[434] Recently it was found that thermal transport properties of nanowires can be very different from the corresponding bulk properties. For example, Rego and Kirczenow[428] have shown theoretically that in the low temperature ballistic regime, the phonon thermal conductance of a 1 dimensional quantum wire is quantized, and have obtained $g = \pi^2 k_B^2 T / 3h$ as the universal quantum heat conductance unit, where $k_B$ and $h$ are the Boltzmann and Planck constants, respectively, and $T$ is the temperature. Also of considerable interest are attempts to derive the macroscopic Fourier law of heat conduction in 1-dimensional systems from microscopic considerations. The Fourier law is a relationship between the heat current $J$ per unit area $\mathcal{A}$ and the temperature gradient $\nabla T$

$$J / \mathcal{A} = -\tilde{K} \nabla T \qquad (113)$$

where $\mathcal{A}$ is the cross-section area normal to the direction of heat propagation and $\tilde{K}$ is the thermal conductivity (the thermal conductance $K$ is defined as $K = J/\Delta T$). Perfect harmonic chains were theoretically investigated by Rieder and Lebowitz[435] and by Zürcher and Talkner[436] who found that heat flux in these systems is proportional to the temperature difference and not to the temperature gradient. Consequently, the thermal conductivity diverges with increasing chain length. Anomalous heat conduction was also found in 1-dimensional models of colliding hard particles.[437, 438] Different models that potentially avoid this divergence and yield Fourier law conduction were discussed. Some invoke impurities and disorder[439, 440], others[441, 442] consider anharmonicity as the source of normal heat conduction. Numerical simulations for chains with a random potential were performed by Mokross,[443] and the role of phonon-lattice interaction was studies by Hu et al.[444] Still, there is yet no convincing and conclusive result about the validity of Fourier law in 1D systems. Another aspect that was the subject of recent discussion is the possible asymmetry in the directionality of heat transfer, and several model nanojunction systems that show heat rectification behavior were discussed.[445-450] Rectification was associated with nonlinear (anharmonic) response, both in classical and quantum models. Strictly quantum effects, e.g. interference and quantum statistics, in heat and energy transport were also investigated.[451, 452]

Experimentally, remarkable progress has been achieved in the last decade in nanoscale thermometry, and measurements on the scale of the mean free path of phonons and electrons are possible. Using scanning thermal microscopy methods one can obtain the spatial temperature distribution of the sample surface, study local thermal properties of materials, and perform calorimetry at nanometric scale.[453, 454] The thermal conductivity and thermoelectric power of single carbon nanotubes were studied both experimentally[455-457] and theoretically[458, 459] [460, 461]. In a different experiment, Schwab et al[462] have observed the quantum thermal conductance in a nano fabricated 1D structure, which behaves essentially like a phonon waveguide. Their results agree with the theoretical predictions.[428] These and other experimental and theoretical developments in this field have been recently reviewed.[463]

In the absence of electronic conduction and of electron-phonon coupling, and in the harmonic approximation, the heat flux through a molecular bridge connecting two thermal phonon reservoirs is given by Eq. (103). Segal et al [430] have evaluated this flux explicitly for a harmonic molecule characterized by a set of normal modes and coupled through its end atoms to harmonic heat reservoirs. They have also performed classical mechanics simulations in order to assess the role played by anharmonicity. Application to the heat transport properties of alkane molecules has yielded several conclusions of general nature:

(a) At room temperature and below, molecular anharmonicity is not an important factor in the heat transport properties of alkanes of length up to several tens of carbon atoms.

(b) At room temperature, the efficiency of heat transport by alkane chains decreases with chain size above 3-4 carbons, then saturates and becomes length independent for moderate sizes of up to a few tens of carbon atoms. This observation agrees with a recent experimental observation of vibrational energy transfer in alkane chains [464] and of heat conduction in alkanedithiol SAMs.[465]

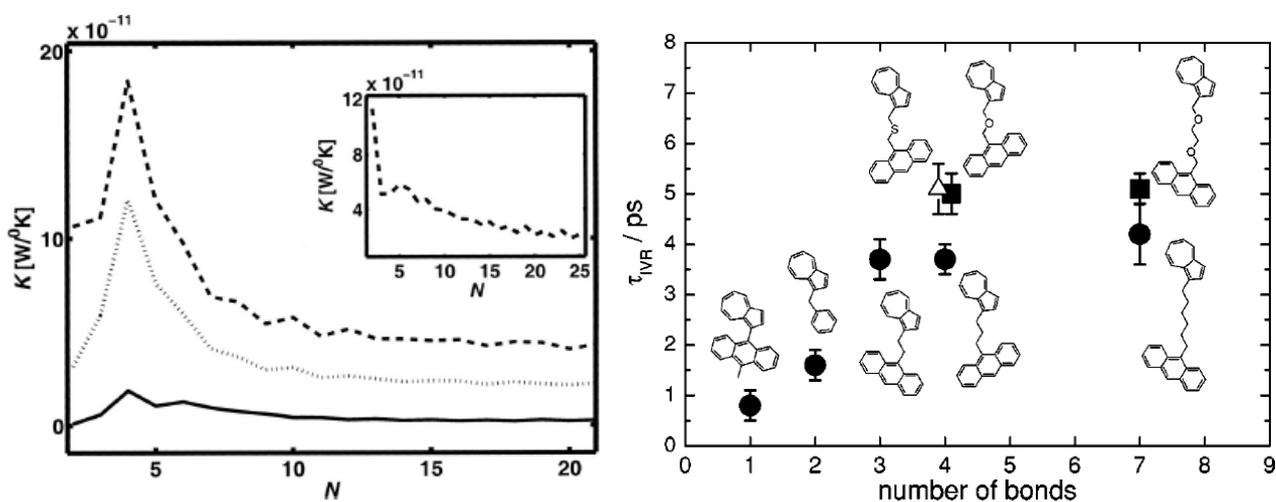

Fig. 30. Left: The heat transport coefficient (heat flux per unit T difference between hot and cold bath) displayed as a function of alkane bridge length, for a particular model of molecule-heat baths coupling (Fig. 2 of Ref. [430]; see there for details) at 50K (full line), 300K (dotted line) and 1000K (dashed line). The inset shows the T=1000K result for a molecule-bath coupling which is 15 times stronger. Right: Vibrational energy transfer times $\tau_{IVR}$ between Azulene and Anthracene species connected by alkane bridges of varying lengths, displayed against these lengths (From Ref. [464]).

(c) At low temperature, the heat transport efficiency increases with chain size. This is a quantum effect: at low temperatures only low frequency modes can be populated and contribute to phonon transport, however such modes are not supported by short molecules and become available only in longer ones.

Theoretical results demonstrating points (b) and (c) are shown in the left panel of Fig. 30. The experimental dependence of vibrational energy transfer along an alkane bridge on its length, showing a similar high temperature trend, is shown in the right panel of that Figure.

While vibrational energy transfer and heat conduction in molecular junctions are interesting by themselves, our interest in the present context is in the temperature rise that reflects the balance between electronic energy deposit onto the molecular vibrational subsystem and heat conduction out of the junction region. We discuss this issue next.

**9d. Junction temperature**

The combined effects of energy transfer from electronic to the vibrational degrees of freedom in a conducting junction, and heat conduction out of the junction, lead to energy accumulation in the vibrational (phonon) subspace that may result in molecular decomposition and junction disintegration. An attempt to describe this increasing energy contents as temperature rise, sometimes described locally at different parts of a junction, necessarily requires a proper definition of local temperature in a non-equilibrium system, an obviously ambiguous concept.[463] A common practical definition is to associate this temperature with the average atomic kinetic energy ($k_B T = m \langle v^2 \rangle$) in local regions defined by some coarse graining procedure (a classical procedure valid only when the temperature is high enough) or the energy of local vibrational modes. An alternative method that was shown to be superior[195] uses a fictitious external phonon bath as a "thermometer": It is coupled to any desired mode in the system and the ensuing heat flux is calculated. The temperature of the examined mode is determined to be such, that when assigned to the fictitious bath renders the heat flux between it and the mode zero.

Given such ways to determine a junction temperature, the steady state temperature increase in a current-carrying junction can be examined. An example that demonstrates qualitative aspects of this phenomenon is shown in Figure 31. The main result conveyed by this figure is the existence of two thresholds: A low bias threshold marks the onset of phonon generation at $e\Phi = \hbar\omega_0$ due to inelastic electron tunneling. A higher threshold in the vicinity of $e\Phi = 2\varepsilon_0$ marks the bias at which the molecular electronic level enters the window between the leads Fermi energies in the model used, where the applied bias is taken to distribute evenly at the two metal-molecule contacts.

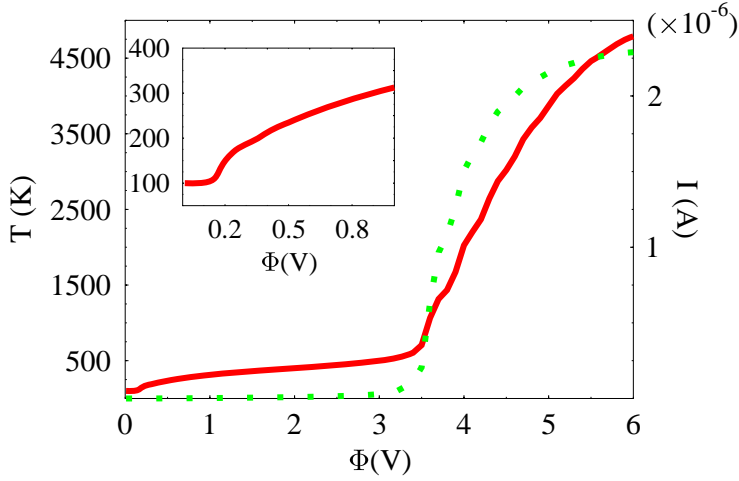

Fig. 31. Temperature increase in a junction where the bridge includes one electronic level ($\varepsilon_0 = 2$eV above the unbiased Fermi energies) coupled to one local vibration ($\omega_0 = 0.2$eV). The electronic couplings to the leads are represented by the electron transfer rates, $\Gamma_L = \Gamma_R = 0.02$eV, the electron-phonon coupling on the bridge is taken $M = 0.2$eV and the leads temperature is $T=100$K. The damping rate of the local vibration due to its coupling to the secondary phonon environment is $\gamma_{ph} = 0.01$eV. The local temperature (full line, red; left vertical axis) is obtained by the measurement technique explained in the text and is plotted against the applied bias. The inset shows the low bias region. Also shown as function of the applied voltage is the current through the junction (dotted line, green; right vertical axis).

Quantitative estimates of the temperature rise in realistic models of molecular junctions where made by Di Ventra and coworkers.[212, 213, 433] These estimates are based on separate calculations of heat generation and dissipation in metallic and molecular wires. Calculations of heat transport in wires connecting between thermal reservoirs have to be supplemented by a relationship between this transport property and the rate of dissipation of heat generated on the junction itself. To this end the authors assume[212] that the steady-state temperature of a bridge connecting thermal baths of temperature $T_L$ and $T_R$ is $(T_L + T_R)/2$. Under this assumption the heat current between a bridge of temperature $T_{BR}$ and the environment of temperature $T_L$ is the same as the current going through the bridge when it connects between reservoirs with temperatures $T_L$ and $T_R$ that satisfies $(T_L + T_R)/2 = T_{BR}$, i.e. $T_R = 2T_{BR} - T_L$. The steady state equality between the rate of heat generation and the rate of heat dissipation based on this estimate gives an equation for the bridge temperature $T_{BR}$. The following observations based on these calculations were made:

(1) Under the same voltage bias, the temperature rise at benzene dithiol junction is considerably smaller than that of a gold wire of similar size because of the larger conduction (therefore higher current) in the latter. In absolute terms, the temperature rise is predicted to be about 15K and 130K above ambient temperatures at a voltage bias of ~1V.[212]

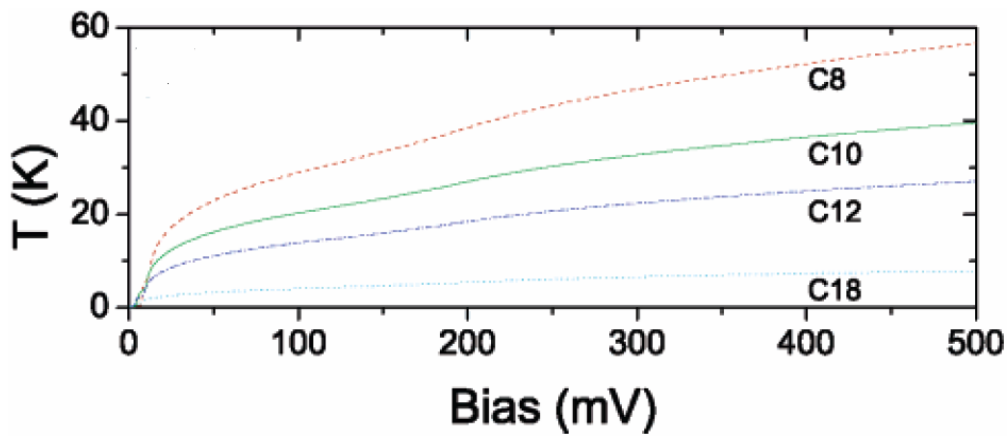

Fig. 32. Estimated junction temperature as a function of bias in alkanethiols junctions of various chain lengths. (From Fig. 1 of Ref. [213]).

(2) In dithiolate alkane chains, estimated temperature rise resulting from the balance between heating and heat conduction is a few tens degrees at 0.5V and depends on chain length (see Fig. 32). The temperature rise is smaller in longer chains characterized by smaller electrical conduction.[213] Decreasing conduction with molecular lengths in these chains overshadows the less efficient heat dissipation in these systems. Recent experimental estimates[432] of the temperature rise are somewhat lower, however these measurements are done in toluene solvent, where more channels to heat dissipation are open.

(3) In contrast to alkanes, in Al wires the temperature rise in current carrying wires is more pronounced for longer chains.[433] In these good conductors the balance between the length effects on conduction and heat dissipation is tipped the opposite way from their molecular counterparts, because length dependence of conduction is relatively weak. Interesting results are obtained vis-à-vis junction stability: Even when the temperature rise is not substantial, junction breakup may be caused by current-induced forces.

It appears that theory has made substantial progress in the study of heating and heat conduction in nanojunctions in general and molecular junctions in particular. Progress in the field seems to depend now on future experimental work.

## 10. Current induced reactions

Heating and heat conduction, discussed in the previous Section, pertain to the issue of junction stability. From this point of view our aim may be to minimize configurational changes induced by charge transport through the molecular bridge. We have seen however (Section 8) that charging-induced configurational changes may be instrumental in affecting junction functionality. A particular example already alluded to is the interesting phenomenon of shuttle transport, [134-141] which is associated with electrostatic feedback between bridge charging and its distance from (i.e. coupling to) the source and drain electrodes.

Taking these ideas a step further, molecular junctions may be studied not as components of electronic devices but as nanoreactors for controlled chemical changes.[187, 466-468, 369, 469-471, 287] From this point of view one is interested in affecting and controlling conformational changes and chemical reactions during the junction operation. Such changes originate from forces exerted on the molecule; short range forces as exerted, for example, by a tip used to push atoms, long range electrostatic forces arising from the imposed potential bias, and forces associated with the transporting current. The latter, current induced forces[306-309, 160, 190, 219, 141, 310, 31, 157, 288] are obviously relevant to the subject of this review. The ultimate result of affecting chemical change depends on the balance between pumping energy into molecular bonds and processes that dissipate excess molecular energy, as already discussed in the previous Section.

Current-induced reactions in molecular junctions, in particular in STM configurations, may become an important tool for nano-fabrication. An extensive discussion of these issues should be a subject of a separate review. Indeed, several such treatments have been published in recent years,[471, 190, 472, 473, 288, 474] and they supplement the present review with regard to this important subject.

## 11. Summary and outlook

Scanning tunneling microscopy, with the associated ability to examine and prepare nanoscale structures, is only 25 years old. The first measurements of actual molecular transport junctions, with the molecule suspended between two electrodes, are only a decade old. The first inelastic tunneling spectra of molecules in such junctions are only two years old. In this short time of development, the preparation, measurement and understanding of molecular transport junctions has progressed very rapidly. In some areas, such as inelastic electron tunneling spectroscopy far from electronic resonances, the situation is very attractive – reproducible experiments can be done, theory can accurately describe the results, and these can be useful for answering questions such as the positions and identities of molecules within the junction, and even the way in which the currents flow through the molecule.

But a general understanding of vibrational effects in nonequilibrium molecular junction transport is still far off. While good measurements are beginning to appear in the Kondo, Coulomb Blockade and near-resonance regimes, our understanding there is far more limited. This is partly because an appropriate general formalism is difficult – The Keldysh nonequilibrium Greens function approach, extensively introduced into molecular transport by Datta and now standard in the field, is very difficult for problems such as simultaneous electron/electron interaction and correlation, electron/phonon coupling and phonon anharmonieity. These are all encountered in such recent measurements as the observation of inelastic co-tunneling and Kondo lines in the Coulomb Blockade regime.

This overview has focused on a general description of the major problems in the field, with brief remarks on their experimental observation and an outline of different theoretical and computational approaches taken. Because the field is vast, our focus has been resolutely on vibrational effects, and even then we had to be very schematic in some areas, to present the rough outlines of the theory and modeling approaches.

With the constant improvement of experimental capabilities, the field is becoming far more sophisticated and very challenging. Attention thus far has focused on relatively simple molecular structures and geometries. But the intrinsic molecular features of stereochemical change (the ability of a given molecular structure to occupy different points on the potential energy surface) and of vibrational reorganization (molecules and their ions are generally different in their geometries), coupled with a very strong electronic polarizability of almost all molecules (that result in the molecular orbital energies being different for anions, cations, and neutrals) substantially complicate the quantitative, and even qualitative, understanding of how current is transported in molecular junctions, and certainly of the vibrational effects on such currents.

The last three chapters of this overview discuss briefly some of the newest areas – utilizing current in molecular junctions to effect chemical transformation and bond breaking, heat and thermal transport in molecular junctions and the effects of strong electron correlations. We've not discussed issues such as the interplay between molecular stereochemical change and applied voltage/current or the reactivity of transmitting junctions, because there is only very fragmentary (if any) experimental information available. Clearly these will be among the interesting fields of the future, as will transport, vibrational and heating effects in large biological entities.

The similar statistics of photons and phonons implies that much of the work discussed here can also be used to approach problems of photoexcitation and photoemission in molecular junctions. That field has some very interesting practical applications, as well as fundamental challenges.

In the first six chapters, we have encountered many situations in which theories simply are not adequate to explain the phenomena, or where taking the theory beyond a minimalist model towards actual computation is exceedingly difficult. Progress is being made on the Coulomb Blockade regime, where interesting effects such as frequency softening with partial charging on the diamond edges, and roles of molecular reorganization and polarization in such important phenomena as junction hysteresis and negative differential resistance can be investigated in a more quantitative fashion. Indeed, very early results are beginning to appear utilizing inelastic electron tunneling spectroscopy to examine the behavior in hysteretic ranges of the conductance spectrum – recent work from IBM has demonstrated the value of such structures acting as single molecule switches and memories, so that these issues are important for both applied and fundamental reasons.

Strong interaction effects with vibrations, manifested both in the Franck Condon blockade and in unusual shapes and "fuzziness" in Coulomb diamonds for molecular junctions, as opposed to traditional quantum dot junctions, are puzzles that should be approached fairly soon. The full understanding of the roles of electronic and phonon excitations and flows in nonequilibrium transport junctions represents an important component of the major, fundamental challenge involved in describing molecular systems in highly nonequilibrium situations.

## Acknowledgements.

MG and MR are grateful to the DARPA MoleApps program, to the International Division of NSF and to the MURI program of DoD for support. The research of AN is supported by the Israel Science Foundation, the US-Israel Binational Science Foundation and the Germany-Israel Foundation. AN thanks Prof. Jan van Ruitenbeek for an helpful exchange of ideas and information.

# References

[1]  Mirkin C A and Ratner M A 1992 Molecular Electronics *Ann. Rev. Phys. Chem.* **43** 719-54
[2]  Aviram A ed 1992 *Molecular Electronics : Science and Technology* (College Park, MD: American Institute of Physics)
[3]  Petty M C, Bryce M R and Bloor D eds 1995 *An Introduction to Molecular Electronics* (Oxford: Oxford University Press)
[4]  Jortner J and Ratner M eds 1997 *Molecular electronics* (Oxford: Blackwell Science)
[5]  Aviram A, Ratner M and Mujica V eds 2002 *Molecular Electronics II* (New York: New York academy of Sciences)
[6]  Ratner M A and Reed M A 2002 *Encyclopedia of Physical Science and Technology*: Academic Press)
[7]  Nitzan A 2001 Electron transmission through molecules and molecular interfaces *Ann. Rev. Phys. Chem.* **52** 681- 750
[8]  Mujica V and Ratner M A 2002 *Handbook of Nanoscience, Engineering and Technology,* ed W A G III*, et al.* (Boca Raton, FL: CRC Press)
[9]  Carroll R L and Gorman C B 2002 *Angew. Chem. Int. Edn.* **41** 4378–400
[10] Reed M A and Lee T eds 2003 *Molecular Nanoelectronics* (Stevenson Ranch, CA: American Scientific Publishers)
[11] Tour J M 2003 *Molecular Electronics: Commercial Insights, Chemistry, Devices, Architectures and Programming* (River Edge, N.J.: World Scientific)
[12] Heath J R and Ratner M A 2003 Molecular Electronics *Physics Today* **56(5)** 43
[13] Nitzan A and Ratner M 2003 Electron transport in molecular wire junctions: Models and Mechanisms *Science* **300** 1384-9
[14] Kagan C R and Ratner M A eds 2004  vol 29 (New York: Material Research Society)
[15] McCreery L 2004 *Chem. Materials* **16** 4477-96
[16] Cuniberti G, Fagas G and Richter K eds 2005 *Introducing Molecular Electronics* (Berlin: Springer)
[17] Joachim C and Ratner M A 2005 *Proc. Nat. Acad. Sci.* **102** 8800
[18] Selzer Y and Allara D L 2006 Single-molecule electrical junctions *Ann. Rev. Phys. Chem.* **57** 593-623
[19] *Faraday Discussions 231, 2006: Molecular Wires and Nanoscale Conductors*
[20] Marcus R A 1956 On the Theory of Oxidation-Reduction Reactions Involving Electron Transfer. I. *J. Chem. Phys.* **24** 966
[21] Marcus R A and Sutin N 1985 Electron transfer in chemistry and biology *Biochem. Biophys. Acta* **811** 265
[22] Nitzan A 2006 *Chemical Dynamics in Condensed Phases* (Oxford: Oxford University Press)
[23] Nitzan A 2001 A relationship between electron transfer rates and molecular conduction. *J. Phys. Chem. A* **105** 2677-9
[24] Medvedev E S and Stuchebrukhov A A 1997 Inelastic tunneling in long-distance electron transfer reactions *J. Chem. Phys.* **107** 3821-31

[25]   Brisker D and Peskin U 2006 Vibrational anharmonicity effects in electronic tunneling through molecular bridges *The Journal of Chemical Physics* **125** 111103-4
[26]   Caspary M and Peskin U 2006 Site-directed electronic tunneling through a vibrating molecular network *The Journal of Chemical Physics* **125** 184703-13
[27]   Yeganeh S and Ratner M A to be published
[28]   Kubatkin S, Danilov A, Hjort M, Cornil J, Bredas J-L, Stuhr-Hansen N, Hedegard P and Bjornholm T 2003 Single-electron transistor of a single organic molecule with access to several redox states *Nature* **425** 698 - 701
[29]   Jaklevic R C and Lambe J 1966 *Physical Review Letters* **17** 1139
[30]   Kirtley J, Scalapino D J and Hansma P K 1976 *Phys. Rev. B-Condens Matter* **14** 3177
[31]   Jorn R and Seideman T 2006 Theory of current-induced dynamics in molecular-scale devices *The Journal of Chemical Physics* **124** 084703
[32]   Stipe B C, Rezaei M A and Ho W 1999 A variable-temperature scanning tunneling microscope capable of single-molecule vibrational spectroscopy *Review of Scientific Instruments* **70** 137-43
[33]   Gaudioso J, Laudon J L and Ho W 2000 Vibrationally mediated negative differential resistance in a single molecule *Phys. Rev. Letters* **85** 1918-21
[34]   Wang W, Lee T, Kretzschmar I and Reed M A 2004 Inelastic Electron Tunneling Spectroscopy of Alkanedithiol Self-Assembled Monolayers *Nano Letters* **4** 643 - 6
[35]   Kushmerick J G, Lazorcik J, Patterson C H, Shashidhar R, Seferos D S and Bazan G C 2004 Vibronic Contributions to Charge Transport Across Molecular Junctions *Nano Letters* **4** 639 - 42
[36]   Kushmerick J G, Blum A S and Long D P 2006 *Analytica Chimica Acta* **568** 20-7
[37]   Long D P, Lazorcik J L, Mantooth B A, Moore M H, Ratner M A, Troisi A, Yao Y, Ciszek J W, Tour J M and Shashidar R in press Effect of hydration on molecular junction transport *Nature Materials*
[38]   Troisi A, Beebe J M and et.al. *Nature, submitted*
[39]   Smit R H M, Noat Y, Untiedt C, Lang N D, Hemert M C V and Ruitenbeek J M V 2002 Measurement of the conductance of a hydrogen molecule *Nature* **419** 906 - 9
[40]   Ho W 2002 Single-molecule chemistry *The Journal of Chemical Physics* **117** 11033-61
[41]   Troisi A and Ratner M A 2005 Modeling the inelastic electron tunneling spectra of molecular wire junctions *Phys. Rev. B-Condens Matter* **72** 033408
[42]   Troisi A and Ratner M A 2006 Molecular signatures in the transport properties of molecular wire junctions: What makes a junction "molecular"? *SMALL* **2** 172-81
[43]   Troisi A and Ratner M A 2004 Conformational Molecular Rectifiers *Nano Letters* **4** 591-5
[44]   Troisi A and Ratner M A 2006 Molecular Transport Junctions: Propensity Rules for Inelastic Electron Tunneling Spectra *Nano Lett.* **6** 1784-8
[45]   Troisi A and Ratner M A **J. Chem. Phys., in press**
[46]   Lauhon L J and Ho W 2000 Control and Characterization of a Multistep Unimolecular Reaction *Phys. Rev. Lett.* **84** 1527-30
[47]   Donhauser Z J, Mantooth B A, Kelly K F, Bumm L A, Monnell J D, Stapleton J J, Price Jr. D W, Rawlett A M, Allara D L, Tour J M and Weiss P S 2001

Conductance Switching in Single Molecules Through Conformational Changes *Science* **292** 2303-7

[48] Donhauser Z, Mantooth B A, Pearl T P, Kelly K F, Nanayakkara S U and Weiss P S 2002 Matrix-mediated control of stochastic single molecule conductance switching *2001 International Conference on Solid State Devices and Materials (SSDM 2001). Tokyo, Japan. 26-28 Sept. 2001.* **41** 4871-7

[49] Ramachandran G K, Hopson T J, Rawlett A M, Nagahara L A, Primak A and Lindsay S M 2003 A Bond-Fluctuation Mechanism for Stochastic Switching in Wired Molecules *Science* **300** 1413-6

[50] Haiss W, Zalinge H v, Bethell D, Ulstrup J, Schiffrin D J and Nichols R J 2006 Thermal gating of the single molecule conductance of alkanedithiols *Faraday Discussions* **131** 253 - 64

[51] Venkataraman L, Klare J E, Nuckolls C, Hybertsen M S and Steigerwald M L 2006 Dependence of single-molecule junction conductance on molecular conformation *Nature* **442** 904-7

[52] Olson M, Mao Y, Windus T, Kemp M, Ratner M, Leon N and Mujica V 1998 A conformational study of the influence of vibrations on conduction in molecular wires *Journal of Physical Chemistry B* **102** 941-7

[53] Emberly E G and Kirczenow G 2001 Current-driven conformational changes, charging, and negative differential resistance in molecular wires. *Phys. Rev. B-Condens Matter* **64** 125318/1-5.

[54] Emberly E G and Kirczenow G 2003 The smallest molecular switch. *Phys. Rev. Letters* **91** 188301/1-4.

[55] Taylor J, Brandbyge M and Stokbro K 2003 Conductance switching in a molecular device: the role of side groups and intermolecular interactions. *Phys. Rev. B-Condens Matter* **68** 121101-1-4

[56] Pati R and Karna S 2004 Current switching by conformational change in a pi - sigma - pi molecular wire. *Phys. Rev. B-Condens Matter* **69** 155419-1-5

[57] Pecchia A, Gheorghe M, Latessa L, Di Carlo A and Lugli P 2004 The influence of thermal fluctuations on the electronic transport of alkeno-thiolates *IEEE Trans. on Nanotechnology* **3** 353-7

[58] Cizek M, Thoss M and Domcke W 2005 Charge transport through a flexible molecular junction *Czech. J. Phys.* **55** 189-202

[59] Berthe M, Urbieta A, Perdigao L, Grandidier B, Deresmes D, Delerue C, Stievenard D, Rurali R, Lorente N, Magaud L and Ordejon P 2006 Electron Transport via Local Polarons at Interface Atoms *Physical Review Letters* **97** 206801-4

[60] Moore A M, Dameron A A, Mantooth B A, Smith R K, Fuchs D J, Ciszek J W, Maya F, Yao Y, Tour J M and Weiss P S 2006 Molecular Engineering and Measurements To Test Hypothesized Mechanisms in Single Molecule Conductance Switching *J. Am. Chem. Soc.* **128** 1959-67

[61] Lewis P A, Inman C E, Yao Y, Tour J M, Hutchison J E and Weiss P S 2004 Mediating Stochastic Switching of Single Molecules Using Chemical Functionality *J. Am. Chem. Soc.* **126** 12214-5

[62] Kulzer F and Orrit M 2004 *Ann. Revs. Phys. Chem.* **55** 585-611

[63] Basch H, Cohen R and Ratner M A 2005 Interface geometry and molecular junction conductance: Geometric fluctuation and stochastic switching *NANO LETTERS* **5** 1668-75

[64] Xue Y and Ratner M 2003 Microscopic study of electrical transport through individual molecules with metallic contacts. I. Band lineup, voltage drop, and high-field transport *Phys. Rev. B-Condens Matter* **68** 115406

[65] Nara J, Geng W T, Kino H, Kobayashi N and Ohno T 2004 Theoretical investigation on electron transport through an organic molecule: Effect of the contact structure *The Journal of Chemical Physics* **121** 6485-92

[66] Dalgleish H and Kirczenow G 2006 A New Approach to the Realization and Control of Negative Differential Resistance in Single-Molecule Nanoelectronic Devices: Designer Transition Metal-Thiol Interface States *Nano Letters* **6** 1274-8

[67] Basch H and Ratner M A 2005 Binding at molecule/gold transport interfaces. V. Comparison of different metals and molecular bridges *The Journal of Chemical Physics* **123** 234704

[68] Moresco F, Meyer G, Rieder K-H, Hao T, Gourdon A and Joachim C 2000 Conformational changes of single molecules induced by scanning tunneling microscopy manipulation: a route to molecular switching. *Phys. Rev. Lett.* **86** 672-5

[69] Solak A O, Ranganathan S, Itoh T and McCreery R L 2002 A mechanism for conductance switching in carbon-based molecular electronic junctions *Electrochemical & Solid State Letters* **5** E43-6

[70] McCreery R, Dieringer J, Solak A O, Snyder B, Nowak A, McGovern W and DuVall S 2003 Molecular Rectification and Conductance Switching in Carbon-Based Molecular Junctions by Structural Rearrangement Accompanying Electron Injection *J. Am. Chem. Soc.* **125** 10748-58

[71] Dulic D, van der Molen S, Kudernac T, Jonkman H, de Jong J, Bowden T, van Esch J, Feringa B and van Wees B 2003 One-way optoelectronic switching of photochromic molecules on gold. *Phys. Rev. Letters* **91**

[72] Loppacher C, Guggisberg M, Pfeiffer O, Meyer E, Bammerlin M, Luthi R, Schlittler R, Gimzewski J K, Tang H and Joachim C 2003 Direct determination of the energy required to operate a single molecule switch *Phys. Rev. Letters* **90** 1-4

[73] Li J, Speyer G and Sankey O F 2004 Conduction Switching of Photochromic Molecules *Physical Review Letters* **93** 248302

[74] Blum A S, Kushmerick J G, Long D P, Patterson C H, Yang J C, Henderson J C, Yao Y, Tour J M, Shashidhar R and Ratna B R 2005 Molecularly inherent voltage-controlled conductance switching *Nature Materials* **4** 167–72

[75] Cai L T, Cabassi M A, Yoon H, Cabarcos O M, McGuiness C L, Flatt A K, Allara D L, Tour J M and Mayer A S 2005 Reversible bistable switching in nanoscale thiol-substituted oligoaniline molecular junctions *NANO LETTERS* **5** 2365-72

[76] He J, Chen F, Liddell P A, Andreasson J, Straight S D, Gust D, Moore T A, Moore A L, Li J, Sankey O F and Lindsay S M 2005 Switching of a photochromic molecule on gold electrodes: single-molecule measurements *Nanotechnology* **16** 695-702

[77] Haiss W, vanZalinge H, Higgins S J, Bethell D, Hobenreich H, Schiffrin D J and Nichols R J 2003 Redox State Dependence of Single Molecule Conductivity *J. Am. Chem. Soc.* **125** 15294-5

[78] He H, Li X L, Tao N J, Nagahara L A, Amlani I and Tsui R 2003 Discrete conductance switching in conducting polymer wires *Phys. Rev. B-Condens Matter* **68** 45302-1

[79] Tao N J 2005 Measurement and control of single molecule conductance *J. Materials Chemistry* **15** 3260-3

[80] Li X, Xu B, Xiao X, Yang X, Zang L and Tao N 2006 Controlling charge transport in single molecules using electrochemical gate *Faraday Discussions* **131** 111-20

[81] Tran E, Duati M, Whitesides G M and Rampi M A 2006 Gating current flowing through molecules in metal–molecules–metal junctions *Faraday Discussions* **131** 197-203

[82] Mantooth B A, Donhauser Z J, Kelly K F and Weiss P S 2002 Cross-correlation image tracking for drift correction and adsorbate analysis *Review of Scientific Instruments* **73** 313-7

[83] Park J, Pasupathy A N, Goldsmith J I, Chang C C, Yaish Y, Petta J R, Rinkoski M, Sethna J P, Abruna H D, McEuen P L and Ralph D L 2002 **Coulomb blockade and the Kondo effect in single-atom transistors** *Nature* **417** 722-5

[84] Liang W, Shores M P, Bockrath M, Long J R and Park H 2002 **Kondo resonance in a single-molecule transistor** *Nature* **417** 725-9

[85] Chen J, Wang W, Klemic J, Reed M A, Axelrod B W, Kaschak D M, Rawlett A M, Price D W, Dirk S M, Tour J M, Grubisha D S and Bennett D W 2002 Molecular Wires, Switches, and Memories *Ann NY Acad Sci* **960** 69-99

[86] Keane Z K, Ciszek J W, Tour J M and Natelson D 2006 Three-terminal devices to examine single-molecule conductance switching *Nano Letters* **6** 1518-21

[87] Poot M, Osorio E, O'Neill K, Thijssen J M, Vanmaekelbergh D, van Walree C A, Jenneskens L W and van der Zant H S J 2006 Temperature dependence of three-terminal molecular junctions with sulfur end-functionalized tercyclohexylidenes *Nano Letters* **6** 1031-5

[88] Natelson D 2006 *Handbook of Organic Electronics and Photonics,* ed N S Nalwa: American Scientific Publishers )

[89] Xu B, Zhang P, Li X and Tao N 2004 Direct Conductance Measurement of Single DNA Molecules in Aqueous Solution *Nano Letters* **4** 1105-8

[90] Kiehl R A, Le J D, Candra P, Hoye R C and Hoye T R 2006 Charge storage model for hysteretic negative-differential resistance in metal-molecule-metal junctions *Applied Physics Letters* **88** 172102

[91] Galperin M, Ratner M A and Nitzan A 2004 Hysteresis,Switching,and Negative Differential Resistance in Molecular Junctions:A Polaron Model *Nano Lett.* **5** 125-30

[92] Mozyrsky D, Hastings M B and Martin I 2006 Intermittent polaron dynamics: Born-Oppenheimer approximation out of equilibrium *Phys. Rev. B-Condens Matter* **73** 035104

[93] Tersoff J and Hamann D R 1983 Theory and application for the scanning tunneling microscope *Phys. Rev. Letters* **50** 1998

[94] Tersoff J and Hamann D R 1985 Theory of scanning tunneling microscope *Phys. Rev. B-Condens Matter* **31** 805-13

[95] Meyer E, Hug H J and Bennewitz R 2004 *Scanning probe microscopy : the lab on a tip* (Berlin: Springer)

[96] Qiu X H, Nazin G V and Ho W 2003 Vibrationally Resolved Fluorescence Excited with Submolecular Precision *Science* **299** 542-6

[97] Hahn J R and Ho W 2006 Imaging and vibrational spectroscopy of single pyridine molecules on Ag(110) using a low-temperature scanning tunneling microscope *The Journal of Chemical Physics* **124** 204708

[98] Maddox J B, Harbola U, Liu N, Silien C, Ho W, Bazan G C and Mukamel S 2006 Simulation of single molecule inelastic electron tunneling signals in paraphenylene-vinylene oligomers and distyrylbenzene[2.2]paracyclophanes *JOURNAL OF PHYSICAL CHEMISTRY A* **110** 6329-38

[99] Lee H J, Lee J H and Ho W 2005 Vibronic Transitions in Single Metalloporphyrins *ChemPhysChem* **6** 971-5

[100] Qiu X H, Nazin G V and Ho W 2004 Vibronic States in Single Molecule Electron Transport *Phys. Rev. Lett.* **92** 206102

[101] Hersam M and Reifenberger R 2004 Charge transport through molecular junctions *MRS Bulletin* **29** 385-90

[102] Hong S, Reifenberger R, Tian W, Datta S, Henderson J and Kubiak C P 2000 *Superlattices & Microstructures*: Academic Press) pp 289-303

[103] Guisinger N P, Yoder N L and Hersam M C 2005 Probing charge transport at the single-molecule level on silicon by using cryogenic ultra-high vacuum scanning tunneling microscopy *PNAS* **102** 8838-43

[104] Guisinger N P, Greene M E, Basu R, Baluch A S and Hersam M C 2004 Room Temperature Negative Differential Resistance through Individual Organic Molecules on Silicon Surfaces *Nano Letters* **4** 55-9

[105] Cao X and Hamers R J 2001 Silicon Surfaces as Electron Acceptors: Dative Bonding of Amines with Si(001) and Si(111) Surfaces *J. Am. Chem. Soc.* **123** 10988-96

[106] Piva P G, DiLabio G A, Pitters J L, Zikovsky J, Rezeq M d, Dogel S, Hofer W A and Wolkow R A 2005 Field regulation of single-molecule conductivity by a charged surface atom *Nature* **435** 658-61

[107] D. Janes et al., personal communciation

[108] McCreery R L, Viswanathan U, Kalakodimi R P and Nowak A M 2006 Carbon/molecule/metal molecular electronic junctions: the importance of contacts *Faraday Discussions* **131** 33 - 43

[109] Anariba F, Steach J K and McCreery R L 2005 Strong Effects of Molecular Structure on Electron Transport in Carbon/Molecule/Copper Electronic Junctions *The Journal of Physical Chemistry B* **109** 11163-72

[110] Guo X, Small J P, Klare J E, Wang Y, Purewal M S, Tam I W, Hong B H, Caldwell R, Huang L, O'Brien S, Yan J, Breslow R, Wind S J, Hone J, Kim P and Nuckolls C 2006 Covalently Bridging Gaps in Single-Walled Carbon Nanotubes with Conducting Molecules *Science* **311** 356-9

[111] Lodha S and Janes D B 2006 Metal/molecule/p-type GaAs heterostructure devices *Journal of Applied Physics* **100** 024503-8

[112] Rakshit T, Liang G C, Ghosh A W, Hersam M C and Datta S 2005 Molecules on silicon: Self-consistent first-principles theory and calibration to experiments *Physical Review B* **72** 125305

[113] Rakshit T, Liang G-C, Ghosh A W and Datta S 2004 Silicon-based molecular electronics *Nano Letters* **4** 1803 -7

[114] Mujica V and Ratner M A 2006 Semiconductor/molecule transport junctions: An analytic form for the self-energies *CHEMICAL PHYSICS* **326** 197-203

[115] Chen J and Reed M A 2002 Electronic transport of molecular systems *Chemical Physics* **281** 127-45

[116] Selzer Y, Salomon A and Cahen D 2002 The Importance of Chemical Bonding to the Contact for Tunneling through Alkyl Chains *J. Phys. Chem. B* **106** 10432-9

[117] Lindsay S 2006 Molecular wires and devices: Advances and issues *Faraday Discussions* **131** 403 - 9

[118] Metzger R M, Xu T and Peterson I R 2001 Electrical Rectification by a Monolayer of Hexadecylquinolinium Tricyanoquinodimethanide Measured between Macroscopic Gold Electrodes *J. Phys. Chem. B* **105** 7280-90

[119] Salomon A, Cahen D, Lindsay S M, Tomfohr J, Engelkes V B and Frisbie C D 2003 Comparison of Electronic Transport Measurements on Organic Molecules *Advanced Materials* **15** 1881-90

[120] Cai L T, Skulason H, Kushmerick J G, Pollack S K, Naciri J, Shashidhar R, Allara D L, Mallouk T E and Mayer T S 2004 Nanowire-Based Molecular Monolayer Junctions: Synthesis, Assembly, and Electrical Characterization *Journal of Physcial Chemistry B* **108** 2827-32

[121] Wang W, Lee T and Reed M A 2005 Electron tunnelling in self-assembled monolayers *Reports on Progress in Physics* **68** 523-44

[122] Mbindyo J K N, Mallouk T E, Mattzela J B, Kratochvilova I, Razavi B, Jackson T N and Mayer T S 2002 Template Synthesis of Metal Nanowires Containing Monolayer Molecular Junctions *J. Am. Chem. Soc.* **124** 4020-6

[123] Yaliraki S N and Ratner M A 1998 Molecule-interface coupling effects on electronic transport in molecular wires *J. Chem. Phys.* **109** 5036-43

[124] Blum A S, Kushmerick J G, Pollack S K, Yang J C, Moore M, Naciri J, Shashidhar R and Ratna B R 2004 Charge Transport and Scaling in Molecular Wires *The Journal of Physical Chemistry B* **108** 18124-8

[125] Selzer Y, Cai L, Cabassi M A, Yao Y, Tour J M, Mayer T S and Allara D L 2005 Effect of Local Environment on Molecular Conduction: Isolated Molecule versus Self-Assembled Monolayer *Nano Letters* **5** 61-5

[126] Adams D M, Brus L, Chidsey C E D, Creager S, Kagan C R, Kamat P V, Lieberman M, Marcus R A, Metzger R M, Michel-Beyerle M E, Newton M D, Rolison D R, Sankey O, Schanze K S and Zhu X 2003 Charge Transfer on the Nanoscale: Current Status *J. Phys. Chem. B* **107** 6668-97

[127] Venkataraman L, Klare J E, Tam I W, Nuckolls C, Hybertsen M S and Steigerwald M L 2006 Single-Molecule Circuits with Well-Defined Molecular Conductance *Nano Letters* **6** 458-62

[128] Haiss W, Nichols R J, Zalinge H v, Higgins S J, Bethell D and Schiffrin D J 2004 Measurement of single molecule conductivity using the spontaneous formation of molecular wires *PCCP* **6** 4330 - 7

[129] Xiao X Y, Brune D, He J, Lindsay S, Gorman C B and Tao N J 2006 Redox-gated electron transport in electrically wired ferrocene molecules *CHEMICAL PHYSICS* **326** 138-43

[130] Choi J P and Murray R W 2006 Electron Self-Exchange between $Au_{140}^{+/0}$ Nanoparticles Is Faster Than That between $Au_{38}^{+/0}$ in Solid-State, Mixed-Valent Films *J. Am. Chem. Soc.* **128** 10496-502

[131] Leopold M C and Donkers R L 2004 *Faraday Disc.* **125** 63-76

[132] J. Liao L B M L C S M C 2006 Reversible Formation of Molecular Junctions in 2D Nanoparticle Arrays *Advanced Materials* **18** 2444-7

[133] Galperin M, Ratner M and Nitzan A cond-mat/0611169 Heat conduction in molecular transport junctions

[134] Gorelik L Y, Isacsson A, Jonson M, Kasemo B, Shekhter R I and Voinova M V 1998 Micro-mechanical charge transfer mechanism in soft Coulomb blockade nanostructures *Physica B* **251** 197-200

[135] Gorelik L Y, Isacsson A, Voinova M V, Kasemo B, Shekhter R I and Jonson M 1998 Shuttle Mechanism for Charge Transfer in Coulomb Blockade Nanostructures *Phys. Rev. Lett.* **80** 4526–9

[136] Armour A D and MacKinnon A 2002 Transport via a quantum shuttle *Phys. Rev. B-Condens Matter* **66** 035333

[137] McCarthy K D, Prokof'ev N and Tuominen M T 2003 Incoherent dynamics of vibrating single-molecule transistors *Phys. Rev. B-Condens Matter* **67** 245415

[138] Fedorets D, Gorelik L Y, Shekhter R I and Jonson M 2004 Quantum Shuttle Phenomena in a Nanoelectromechanical Single-Electron Transistor *Phys. Rev. Letters* **92** 166801

[139] Smirnov A Y, Mourokh L G and Horing N J M 2004 Temperature dependence of electron transport through a quantum shuttle *Phys. Rev. B-Condens Matter* **69** 155310

[140] Pistolesi F and Fazio R 2005 Charge Shuttle as a Nanomechanical Rectifier *Phys. Rev. Letters* **94** 036806

[141] Kaun C-C and Seideman T 2005 Current-Driven Oscillations and Time-Dependent Transport in Nanojunctions *Phys. Rev. Letters* **94** 226801

[142] Buttiker M and Landauer R 1982 *Phys. Rev. Lett.* **49** 1739-42

[143] Nitzan A, Jortner J, Wilkie J, Burin A L and Ratner M A 2000 *Tunneling Time for Electron Transfer Reactions J. Phys. Chem. B* **104** 5661-5

[144] Galperin M, Nitzan A and Peskin U 2001 Traversal times for electron tunneling through water *J. Chem. Phys.* **114** 9205-8

[145] Marcus R A 1956 Electrostatic Free Energy and Other Properties of States Having Nonequilibrium Polarization. I. *J. Chem. Phys.* **24** 979

[146] Marcus R A 1993 Electron transfer reactions in chemistry: Theory and experiments *Rev. Mod. Phys.* **65** 599

[147] Marcus R A 1965 *J. Chem. Phys.* **43** 679

[148] Migdal A B 1958 *Sov. Phys. JETP* **7** 996

[149] Eliashberg G M 1960 *Soviet Physics JETP* **11** 696

[150] Mahan G D 2000 *Many-particle physics* (New York: Plenum press)

[151] Segal D and Nitzan A 2002 Heating in current carrying molecular junctions *J. Chem. Phys.* **117** 3915-27

[152] Krishna V and Tully J C 2006 Vibrational lifetimes of molecular adsorbates on metal surfaces *The Journal of Chemical Physics* **125** 054706, and references therein.

[153] Horsfield A P, Bowler D R, Fisher A J, Todorov T N and Montgomery M J 2004 Power dissipation in nanoscale conductors: classical, semi-classical and quantum dynamics *Journal of Physics: Condensed Matter* **16** 3609-22

[154] Galperin M and Nitzan A 2001 Inelastic effects in electron tunneling through water layers *J. Chem. Phys.* **115** 2681-94

[155] Troisi A, Ratner M A and Nitzan A 2003 Vibronic effects in off-resonant molecular wire conduction *J. Chem Phys.* **118** 6072-82

[156] Chen Y-C, Zwolak M and Di Ventra M 2004 Inelastic Current-Voltage Characteristics of Atomic and Molecular Junctions *Nano Letters* **4** 1709-12

[157] Leeuwen R v 2004 First-principles approach to the electron-phonon interaction *Phys. Rev. B-Condens Matter* **69** 115110

[158] Lorente N, Persson M, Lauhon L J and Ho W 2001 Symmetry Selection Rules for Vibrationally Inelastic Tunneling *Phys. Rev. Lett.* **86** 2593 - 6

[159] Seminario J M and Cordova L E 2004 Theoretical interpretation of intrinsic line widths observed in inelastic electron tunneling scattering experiments *J. Phys. Chem. A* **108** 5142-4

[160] Sergueev N, Roubtsov D and Guo H 2005 Ab Initio Analysis of Electron-Phonon Coupling in Molecular Devices *Physical Review Letters* **95** 146803-4

[161] Paulsson M, Frederiksen T and Brandbyge M 2005 Modeling inelastic phonon scattering in atomic- and molecular-wire junctions *Phys. Rev. B-Condens Matter* **72** 201101

[162] Grobis M, Khoo K H, Yamachika R, Lu X, Nagaoka K, Louie S G, Crommie M F, Kato H and Shinohara H 2005 Spatially Dependent Inelastic Tunneling in a Single Metallofullerene *Phys. Rev. Letters* **94** 136802

[163] Jiang J, Kula M, Lu W and Luo Y 2005 First-Principles Simulations of Inelastic Electron Tunneling Spectroscopy of Molecular Electronic Devices *Nano Letters* **5** 1551-5

[164] Bocquet M-L, Lesnard H and Lorente N 2006 Inelastic Spectroscopy Identification of STM-Induced Benzene Dehydrogenation *Phys. Rev. Letters* **96** 096101

[165] Frederiksen T, Brandbyge M, Lorente N and Jauho A-P 2004 Inelastic Scattering and Local Heating in Atomic Gold Wires *Phys. Rev. Letters* **93** 256601

[166] Paulsson M, Frederiksen T and Brandbyge M 2006 Inelastic Transport through Molecules: Comparing First-Principles Calculations to Experiments *Nano Letters* **6** 258-62

[167] Kula M, Jiang J and Luo Y 2006 Probing Molecule-Metal Bonding in Molecular Junctions by Inelastic Electron Tunneling Spectroscopy *Nano Letters* **6** 1693 -8

[168] Segal D, Nitzan A, Davis W B, Wasilewski M R and Ratner M A 2000 Electron transfer rates in bridged molecular systems 2: A steady state analysis of coherent tunneling and thermal transitions *J. Phys. Chem. B* **104** 3817

[169] Segal D, Nitzan A, Ratner M A and Davis W B 2000 Activated Conduction in Microscopic Molecular Junctions *J. Phys. Chem.* **104** 2790

[170] Segal D and Nitzan A 2001 Steady state quantum mechanics of thermally relaxing systems *Chem. Phys.* **268** 315-35

[171] Segal D and Nitzan A 2002 Conduction in molecular junctions: inelastic effects *Chem. Phys.* **281** 235-56

[172] May V 2002 Electron transfer through a molecular wire: consideration of electron-vibrational coupling within the Liouville space pathway technique *Phys. Rev. B-Condens Matter* **66** 245411

[173] Petrov E G, Zelinskyy Y R and May V 2002 Bridge Mediated Electron Transfer: A Unified Description of the Thermally Activated and Superexchange Mechanisms *J. Phys. Chem. B* **106** 3092 -102

[174] Petrov E G, May V and Hanggi P 2003 Spin-boson description of electron transmission through a molecular wire *Chem. Phys.* **296** 251

[175] Petrov E G 2006 Towards a many-body theory for the combined elastic and inelastic transmission through a single molecule *Chemical Physics* **326** 151-75

[176] Li X-Q, Luo J-Y, Yang Y-G, Cui P and Yan Y-J 2005 Quantum master-equation approach to quantum transport through mesoscopic systems *Phys. Rev. B-Condens Matter* **71** 205304

[177] Gebauer R and Car R 2004 Current in Open Quantum Systems *Physical Review Letters* **93** 160404

[178] Redfield A G 1957 *IBM J. Res. Develop* **1** 19

[179] Redfield A G 1965 *Adv. Magen. Reson.* **1** 1

[180] May V and Kühn O 2000 *Charge and energy transfer dynamics in molecular systems* (Berlin: Wiley-VCH), Chapter 3.

[181] Datta S 1990 A Simple Kinetic Equation for Steady-state Quantum Transport *J. Phys.: Cond. Matter* **2** 8023-52

[182] Neofotistos G, Lake R and Datta S 1991 Inelastic Scattering Effects on Single Barrier Tunneling *Phys. Rev. B-Condens Matter* **43** 2442-5

[183] McLennan M J, Lee Y and Datta S 1991 Voltage Drop in Mesoscopic Systems: A Numerical Study using a Quantum Kinetic Equation *Phys. Rev. B-Condens Matter* **43** 13846-84

[184] D'Amato J L and Pastawski H M 1990 Conductance of a disordered linear chain including inelastic scattering events *Phys. Rev. B-Condens Matter* **41** 7411-20

[185] Pastawski H M, Torres L E F F and Medina E 2002 Electron–phonon interaction and electronic decoherence in molecular conductors *Chemical Physics* **281** 257-78

[186] Buttiker M 1988 Coherent and sequential tunneling in series barriers *IBM J. Res. Develop.* **32** 63-75

[187] Walkup R E, Newns D M and Avouris P 1993 *Atomic and Nanometer Scale Modification of Materials: Fundamentals and Applications,* ed P Avouris (Amsterdam: Kluwer Academic Publishers) pp 97-100

[188] Koch J, Semmelhack M, Oppen F v and Nitzan A 2006 Current-induced nonequilibrium vibrations in single-molecule devices *Phys. Rev. B-Condens Matter* **73** 155306

[189] May V and Kuhn O 2006 IV characteristics of molecular wires in the presence of intramolecular vibrational energy redistribution *CHEMICAL PHYSICS LETTERS* **420** 192-8

[190] Seideman T 2003 Current-driven dynamics in molecular-scale devices *J. Phys.-Cond. Mat.* **15** R521-49

[191] Lake R K and Datta S 1992 Energy ballance and heat exchange in mesoscopic systems *Phys. Rev. B-Condens Matter* **46** 4757-63

[192] Lake R K and Datta S 1992 Nonequilibrium Green's-function method applied to double-barrier resonant-tunneling diodes *Phys. Rev. B-Condens Matter* **45** 6670

[193] Wang J-S, Wang J and Zeng N 2006 Nonequilibrium Green's function approach to mesoscopic thermal transport *Phys. Rev. B-Condens Matter* **74** 033408

[194] Mingo N 2006 Anharmonic phonon flow through molecular-sized junctions *Physical Review B (Condensed Matter and Materials Physics)* **74** 125402-13

[195] Galperin M, Ratner M and Nitzan A to be published Inelastic effects in molecular junctions in the Coulomb and Kondo regimes: Nonequilibrium equation-of-motion approach.

[196] Landauer R 1957 *IBM J. Res. Dev.* **1** 223

[197] Landauer R 1970 Electrical resistance of disordered 1-dimensional lattices *Phil. mag.* **21** 863-7

[198] Imry Y 1986 *Directions in Condensed Matter Physics,* ed G Grinstein and G Mazenko (Singapore: World Scientific) p 101

[199] Datta S 1995 *Electric transport in Mesoscopic Systems* (Cambridge: Cambridge University Press)

[200] Domcke W and Cederbaum L S 1977 A simple formula for the vibrational structure of resonances in electron-molecule scattering *J. Phys. B: At. Mol. Phys.* **10** L47-L52

[201] Domcke W and Cederbaum L S 1980 Vibration-induced narrowing of electron scattering resonance near threshold *J. Phys. B: Atom. Molec. Phys.* **13** 2829-38

[202] Wingreen N S, Jacobsen K W and Wilkins J W 1988 Resonant Tunneling with Electron-Phonon Interaction: An Exactly Solvable Model *Phys. Rev. Lett.* **61** 1396-99

[203] Wingreen N S, Jacobsen K W and Wilkins J W 1989 Inelastic scattering in resonant tunneling *Phys. Rev. B-Condens Matter* **40** 11834-50

[204] Bonca J and Trugman S A 1995 Effect of inelastic processes on tunneling *Phys. Rev. Letters* **75** 2566-9

[205] Bonca J and Trugman S A 1997 Inelastic quantum transport *Phys. Rev. Letters* **79** 4874-7
[206] Haule K and Bonca J 1999 Inelastic tunneling through mesoscopic structures *Phys. Rev. B-Condens Matter* **59** 13087-93
[207] Ness H and Fisher A J 1999 Quantum inelastic conductance through molecular wires *Phys. Rev. Lett.* **83** 452-5
[208] Ness H, Shevlin S A and Fisher A J 2001 Coherent electron- phonon coupling and polaronlike transport in molecular wires *Phys. Rev. B-Condens Matter* **63** 125422
[209] Ness H and Fisher A J 2002 Coherent electron injection and transport in molecular wires: inelastic tunneling and electron-phonon interactions *Chemical Physics* **281** 279-92
[210] Ness H and Fisher A J 2003 *Eur. Phys. J.* **D24** 409-12
[211] Ness H and Fisher A J 2005 Molecular Electronics Special Feature: Vibrational inelastic scattering effects in molecular electronics *PNAS* **102** 8826-31
[212] Chen Y C, Zwolak M and Di Ventra M 2003 Local Heating in Nanoscale Conductors *Nano Lett.* **3** 1691-4
[213] Chen Y-C, Zwolak M and Di Ventra M 2005 Inelastic Effects on the Transport Properties of Alkanethiols *Nano Letters* **5** 621-4
[214] Todorov T N 1998 Local heating in ballistic atomic-scale contracts *Phil. Mag. B* **77** 965-73
[215] Montgomery M J, Todorov T N and Sutton A P 2002 Power dissipation in nanoscale conductors *J Phys.: Cond. Matter* **14** 5377-89
[216] Montgomery M J and Todorov T N 2003 Electron–phonon interaction in atomic-scale conductors: Einstein oscillators versus full phonon modes *J Phys.: Cond. Matter* **15** 8781-95
[217] Montgomery M J, Hoekstra J, Todorov T N and Sutton A P 2003 Inelastic current–voltage spectroscopy of atomic wires *J Phys.: Cond. Matter* **15** 731-42
[218] Alavi S and Seideman T 2001 Reaction induced by a scanning tunneling microscope: Theory and application *The Journal of Chemical Physics* **115** 1882-90
[219] Alavi S, Larade B, Taylor J, Guo H and Seideman T 2002 Current-triggered vibrational excitation in single-molecule transistors *Chemical Physics* **281** 293-303
[220] Cizek M, Thoss M and Domcke W 2004 Theory of vibrationally inelastic electron transport through molecular bridges *Phys. Rev. B-Condens Matter* **70** 125406
[221] Keldysh L V 1965 *Sov. Phys. JETP* **20** 1018
[222] Kadanoff L P and Baym G 1962 *Quantum Statistical Mechanics. Green's Function Methods in Equilibrium and Nonequilibrium Problems* (Reading, MA.: Benjamin)
[223] Wagner M 1991 Expansions of non-equilibrium Green's functions *Phys. Rev. B* **44** 6104-17
[224] Datta. S 2005 *Quantum Transport: Atom to Transistor* (Cambridge: Cambridge University Press)
[225] Haug H and Jauho A-P 1996 *Quantum Kinetics in Transport and Optics of Semiconductors* vol 123 (Berlin: Springer)
[226] Meir Y and Wingreen N S 1992 Landauer formula for the current through an interacting electron region *Phys. Rev. Lett.* **68** 2512-5

[227] Jauho A P, Wingreen N S and Meir Y 1994 Time-dependent transport in interacting and noninteracting resonant-tunneling systems *Phys. Rev. B-Condens Matter* **50** 5528–44

[228] Caroli C, Combescot R, Lederer D, Nozieres P and Saint-James D 1971 A direct calculation of the tunneling current IV. Electron-phonon interaction effects *J. Phys. C: Solid State* **5** 21-42

[229] Datta S and Lake R K 1991 Voltage probes and inelastic scattering *Phys. Rev. B-Condens Matter* **44** 6538–41

[230] Lorente N and Persson M 2000 Theory of Single Molecule Vibrational Spectroscopy and Microscopy *Phys. Rev. Lett.* **85** 2997-3000

[231] Tikhodeev S, Natario M, Makoshi K, Mii T and Ueba H 2001 Contribution to a theory of vibrational scanning tunneling spectroscopy of adsorbates: Nonequilibrium Green's function approach *Surf. Sci.* **493** 63-70

[232] Mii T, Tikhodeev S and Ueba H 2002 Theory of vibrational tunneling spectroscopy of adsorbates on metal surfaces *Surface Science* **502-503** 26-33

[233] Mii T, Tikhodeev S G and Ueba H 2003 Spectral features of inelastic electron transport via a localized state *Phys. Rev. B-Condens Matter* **68** 205406

[234] Galperin M, Ratner M A and Nitzan A 2004 On the Line Widths of Vibrational Features in Inelastic Electron Tunneling Spectroscopy *Nano Letters* **4** 1605-11

[235] Galperin M, Ratner M A and Nitzan A 2004 Inelastic electron tunneling spectroscopy in molecular junctions: Peaks and dips *J. Chem. Phys.* 11965-79

[236] Asai Y 2004 Theory of Inelastic Electric Current through Single Molecules *Phys. Rev. Letters* **93** 246102

[237] Pecchia A, Di Carlo A, Gagliardi A, Sanna S, Frauenheim T and Gutierrez R 2004 Incoherent Electron-Phonon Scattering in Octanethiols *Nano Letters* **4** 2109-14

[238] Solomon G C, Gagliardi A, Pecchia A, Frauenheim T, Di Carlo A, Reimers J R and Hush N S 2006 Understanding the inelastic electron-tunneling spectra of alkanedithiols on gold *JOURNAL OF CHEMICAL PHYSICS* **124** 094704

[239] Hod O, Baer R and Rabani E in press Inelastic effects in Aharonov-Bohm molecular interferometers *Phys. Rev. Lett.*

[240] Viljas J K, Cuevas J C, Pauly F and Hafner M 2005 Electron-vibration interaction in transport through atomic gold wires *Phys. Rev. B-Condens Matter* **72** 245415

[241] Blanter Y M, Usmani O and Nazarov Y V 2004 Single-Electron Tunneling with Strong Mechanical Feedback *Phys. Rev. Letters* **93** 136802

[242] Chtchelkatchev N M, Belzig W and Bruder C 2004 Charge transport through a single-electron transistor with a mechanically oscillating island *Phys. Rev. B-Condens Matter* **70** 193305

[243] Weig E M, Blick R H, Brandes T, Kirschbaum J, Wegscheider W, Bichler M and Kotthaus J P 2004 Single-Electron-Phonon Interaction in a Suspended Quantum Dot Phonon Cavity *Phys. Rev. Letters* **92** 046804-4

[244] Yu L H, Keane Z K, Ciszek J W, Cheng L, Stewart M P, Tour J M and Natelson D 2004 Inelastic Electron Tunneling via Molecular Vibrations in Single-Molecule Transistors *Phys. Rev. Let.* **93** 266802

[245] Cornaglia P S and Grempel D R 2005 Magnetoconductance through a vibrating molecule in the Kondo regime *Phys. Rev. B-Condens Matter* **71** 245326

[246] Mravlje J, Ramsak A and Rejec T 2005 Conductance of deformable molecules with interaction *Phys. Rev. B-Condens Matter* **72** 121403-4

[247] Visoly-Fisher I, Daie K, Terazono Y, Herrero C, Fungo F, Otero L, Durantini E, Silber J J, Sereno L, Gust D, Moore T A, Moore A L and Lindsay S M 2006

Conductance of a biomolecular wire *PROCEEDINGS OF THE NATIONAL ACADEMY OF SCIENCES OF THE UNITED STATES OF AMERICA* **103** 8686-90

[248] Yan L 2006 Inelastic Electron Tunneling Spectroscopy and Vibrational Coupling *J. Phys. Chem. A*

[249] Lundin U and McKenzie R H 2002 Temperature dependence of polaronic transport through single molecules and quantum dots *Phys. Rev. B-Condens Matter* **66** 075303

[250] Zhu J-X and Balatsky A V 2003 Theory of current and shot-noise spectroscopy in single-molecular quantum dots with a phonon mode *Phys. Rev. B-Condens Matter* **67** 165326

[251] Alexandrov A, Bratkovsky A M and Williams R S 2003 Bistable tunneling current through a molecular quantum dot *Phys. Rev. B-Condens Matter* **67** 075301

[252] Alexandrov A and Bratkovsky A M 2003 Memory effect in a molecular quantum dot with strong electron-vibron interaction *Phys. Rev. B-Condens Matter* **67** 235312

[253] Král P 1997 Nonequilibrium linked cluster expansion for steady-state quantum transport *Phys. Rev. B-Condens Matter* **56** 7293–303

[254] Flensberg K 2003 Tunneling broadening of vibrational sidebands in molecular transistors. *Phys. Rev. B-Condens Matter* **68** 205323-1-7

[255] Galperin M, Nitzan A and Ratner M A 2006 Resonant inelastic tunneling in molecular junctions *Phys. Rev. B-Condens Matter* **73** 045314

[256] Hyldgaard P, Hershfield S, Davies J H and Wilkins J W 1994 Resonant Tunneling with an Electron-Phonon Interaction *Annals of Physics* **236** 1-42

[257] Mitra A, Aleiner I and Millis A 2004 Phonon effects in molecular transistors: quantal and classical treatment *Phys. Rev. B-Condens Matter* **69** 245302

[258] Ryndyk D A, Hartung M and Cuniberti G 2006 Nonequilibrium molecular vibrons: An approach based on the nonequilibrium Green function technique and the self-consistent Born approximation *Phys. Rev. B-Condens Matter* **73** 045420

[259] Feynman R P and Hibbs A R 1965 *Quantum Mechanics and Path Integrals* (New York: McGraw-Hill)

[260] Schulman L S 1981 *Techniques and Applications of Path Integration* (New York: Wiley)

[261] Sebastian K L and Doyen G 1993 Dynamical image interaction in scanning tunneling microscopy *Physical Review B* **47** 7634

[262] Klipa N and Šunjić M 1995 Dynamical effective potentials in electron tunneling: Path-integral study *Physical Review B* **52** 12408

[263] Ness H and Fisher A J 1998 Dynamical effective potential for tunneling: an exact matrix method and a path-integral technique *Applied Physics A: Materials Science & Processing* **V66** S919-S23

[264] Mozyrsky D and Martin I 2005 Effects of strong correlations in single electron traps in field-effect transistors *Nanotechnology, IEEE Transactions on* **4** 90-5

[265] Martin I and Mozyrsky D 2005 Charge dynamics and Kondo effect in single-electron traps in field-effect transistors *Physical Review B (Condensed Matter and Materials Physics)* **71** 165115-7

[266] Sheng L, Xing D Y and Sheng D N 2004 Theory of the zero-bias anomaly in magnetic tunnel junctions: Inelastic tunneling via impurities *Phys. Rev. B-Condens Matter* **70** 094416

[267] Hastings M B, Martin I and Mozyrsky D 2003 Quantum dynamics in nonequilibrium strongly correlated environments *Physical Review B (Condensed Matter and Materials Physics)* **68** 035101-5

[268] Mozyrsky D, Martin I and Hastings. M B 2004 Quantum-Limited Sensitivity of Single-Electron-Transistor-Based Displacement Detectors *Phys. Rev. Lett.* **92** 018303

[269] Gogolin A O and Komnik A 2002 Multistable transport regimes and conformational changes in molecular quantum dots *condmat/ 0207513*

[270] Mitra A, Aleiner I and Millis A J 2005 Semiclassical Analysis of the Nonequilibrium Local Polaron *Physical Review Letters* **94** 076404

[271] Cornaglia P S, Ness H and Grempel D R 2004 Many-Body Effects on the Transport Properties of Single-Molecule Devices *Phys. Rev. Letters* **93** 147201

[272] Cornaglia P S, Grempel D R and Ness H 2005 Quantum transport through a deformable molecular transistor *Phys. Rev. B-Condens Matter* **71** 075320

[273] Paaske J and Flensberg K 2005 Vibrational Sidebands and the Kondo Effect in Molecular Transistors *Phys. Rev. Letters* **94** 176801

[274] Braig S and Flensberg K 2003 Vibrational sidebands and dissipative tunneling in molecular transistors. *Phys. Rev. B-Condens Matter* **68** 205324-1-10

[275] Smirnov A Y, Mourokh L G and Horing N J M 2003 Nonequilibrium fluctuations and decoherence in nanomechanical devices coupled to the tunnel junction *Phys. Rev. B-Condens Matter* **67** 115312

[276] Koch J and Oppen F v 2005 Franck-Condon Blockade and Giant Fano Factors in Transport through Single Molecules *Phys. Rev. Letters* **94** 206804

[277] Koch J and von Oppen F 2005 Effects of charge-dependent vibrational frequencies and anharmonicities in transport through molecules *Physical Review B (Condensed Matter and Materials Physics)* **72** 113308-4

[278] Koch J, Raikh M E and Oppen F v 2006 Pair Tunneling through Single Molecules *Phys. Rev. Letters* **96** 056803

[279] Ueba H, Mii T, Lorente N and Persson B N J 2005 Adsorbate motions induced by inelastic-tunneling current: Theoretical scenarios of two-electron processes *The Journal of Chemical Physics* **123** 084707

[280] Doiron C B, Belzig W and Bruder C 2006 Electrical transport through a single-electron transistor strongly coupled to an oscillator *Phys. Rev. B-Condens Matter* **74** 205336-10

[281] Harbola U, Maddox J and Mukamel S 2006 Nonequilibrium superoperator Green's function approach to inelastic resonances in STM currents *Phys. Rev. B-Condens Matter* **73** 205404

[282] Anariba F and McCreery R L 2002 Electronic conductance behavior of carbon-based molecular junctions with conjugated structures *Journal of Physical Chemistry B* **106** 10355-62

[283] Choi J W, Flood A H, Steuerman D W, Nygaard S, Braunschweig A B, Moonen N N P, Laursen B W, Luo Y, DeIonno E, Peters A J, Jeppesen J O, Xu K, Stoddart J F and Heath J R 2005 Ground-state equilibrium thermodynamics and switching kinetics of bistable
[2]rotaxanes switched in solution, polymer gels, and molecular electronic
devices *CHEMISTRY-A EUROPEAN JOURNAL* **12** 261-79

[284] Weiss E A, Tauber M J, Kelley R F, Ahrens M J, Ratner M A and Wasielewski M R 2005 Conformationally gated switching between superexchange and hopping within oligo-p-phenylene-based molecular wires *JOURNAL OF THE AMERICAN CHEMICAL SOCIETY* **127** 11842-50

[285] Lastapis M, Martin M, Riedel D, Hellner L, Comtet G and Dujardin G 2005 Picometer-Scale Electronic Control of Molecular Dynamics Inside a Single Molecule *Science* **308** 1000-3

[286] Persson B N J and Ueba H 2002 Theory of inelastic tunneling induced motion of adsorbates on metal surfaces *Surf. Sci.* **502-503** 12-7

[287] Komeda T, Kim Y, Kawai M, Persson B N J and Ueba H 2002 Lateral hopping of molecules induced by excitation of internal vibrational mode *Science* **295** 2055-8

[288] Lorente N, Rurali R and Tang H 2005 Single-molecule manipulation and chemistry with the STM *J. Phys.: Cond. Matter* **17** S1049

[289] Blanter Y M and Buttiker M 2000 Shot Noise In Mesoscopic Conductors *Physics Reports* **336** 1-166

[290] Nishiguchi N 2002 *Phys. Rev. Let.* **89** 066802

[291] Clerk A A and Girvin S M 2004 Shot noise of a tunnel junction displacement detector *Phys. Rev. B-Condens Matter* **70** 121303

[292] Novotny T, Donarini A, Flindt C and Jauho A 2004 Shot Noise of a Quantum Shuttle *Phys. Rev. Letters* **92** 248302

[293] Flindt C, Novotny T and Jauho A-P 2004 Current noise in a vibrating quantum dot array *Phys. Rev. B-Condens Matter* **70** 205334

[294] Wabnig J, Khomitsky D V, Rammer J and Shelankov A L 2005 Statistics of charge transfer in a tunnel junction coupled to an oscillator *Phys. Rev. B-Condens Matter* **72** 165347

[295] Camalet S, Kohler S and Hänggi P 2004 Shot-noise control in ac-driven nanoscale conductors *Phys. Rev. B-Condens Matter* **70** 155326

[296] Guyon R, Jonckheere T, Mujica V, Crepieux A and Martin T 2005 Current and noise in a model of an alternating current scanning tunneling microscope molecule-metal junction *The Journal of Chemical Physics* **122** 144703

[297] Shimizu A and Ueda. M 1992 Effects of dephasing and dissipation on quantum noise in conductors *Phys. Rev. Letters.* **69** 1403

[298] Bo O L and Galperin Y 1997 Low-frequency shot noise in phonon-assisted resonant magnetotunneling *Phys. Rev. B-Condens Matter* **55** 1696

[299] Dong B, Cui H L, Lei X L and Horing N J M 2005 Shot noise of inelastic tunneling through quantum dot systems *Phys. Rev. B-Condens Matter* **71** 045331

[300] Chen Y-C and Ventra M D 2005 Effect of Electron-Phonon Scattering on Shot Noise in Nanoscale Junctions *Phys. Rev. Letters* **95** 166802

[301] Galperin M, Nitzan A and Ratner M A 2006 Inelastic tunneling effects on noise properties of molecular junctions *Phys. Rev. B, in press; , cond-mat/0604029 (2006)*

[302] Djukic D and Ruitenbeek J M v 2006 Shot Noise Measurements on a Single Molecule *Nano Letters* **6** 789

[303] Cohen R E, Mehl M J and Papaconstantopoulos D A 1994 Tight-binding total-energy method for transition and noble metals *Physical Review B* **50** 14694

[304] Mehl M J and Papaconstantopoulos D A 1996 Applications of a tight-binding total-energy method for transition and noble metals: Elastic constants, vacancies, and surfaces of monatomic metals *Physical Review B* **54** 4519

[305] Papaconstantopoulos D A and Mehl M J 2003 The Slater–Koster tight-binding method: a computationally efficient and accurate approach *Journal of Physics: Condensed Matter* **15** R413-R40

[306] Di Ventra M and Pantelides S T 2000 Hellmann-Feynman theorem and the definition of forces in quantum time-dependent and transport problems *Phys. Rev. B-Condens Matter* **61** 16207–12

[307] Di Ventra M, Pantelides S T and Lang N D 2002 Current-induced forces in molecular wires *Phys. Rev. Letters* **88** 046801

[308] Di Ventra M, Chen Y C and Todorov T N 2004 Are Current-Induced Forces Conservative? *Phys. Rev. Letters* **92** 176803

[309] Yang Z and Di Ventra M 2003 Nonlinear current-induced forces in Si atomic wires *Phys. Rev. B-Condens Matter* **67** 161311

[310] Park H, Park J, Lim A K L, Anderson E H, Alivisatos A P and Mceuen P L 2000 Nanomechanical oscillations in a single-C60 transistor *Nature* **407** 57 - 60

[311] Reed M A, Zhou C, Muller C J, Burgin T P and Tour J M 1997 Conductance of a molecular junction *Science* **278** 252-4

[312] Pollard W T, Felts A K and Friesner R A 1996 The redfield equation in condensed phase quantum dynamics *Adv. Chem. Phys.* **93** 77-134

[313] Zurek W H 2003 Decoherence, einselection, and the quantum origins of the classical *Reviews of Modern Physics* **75** 715-75

[314] Weiss E A, Katz G, Goldsmith R H, Wasielewski M R, Ratner M A, Kosloff R and Nitzan A 2006 Electron transfer mechanism and the locality of the system-bath interaction: A comparison of local, semilocal, and pure dephasing models *The Journal of Chemical Physics* **124** 074501

[315] Skourtis S and Nitzan A 2003 **Effects of initial state preparation on the distance dependence of electron transfer through molecular bridges and wires** *The Journal of Chemical Physics* **119** 6271-6

[316] Devault D and Chance B 1966 *Biophys. J.* **6** 825

[317] Felts A K, Pollard W T and Friesner R A 1995 Multilevel Redfield treatment of bridge mediated long range electron transfer: A mechanism for anomalous distance dependence *J. Phys. Chem.* **99** 2929-40

[318] Skourtis S S and Mukamel S 1995 Superexchange versus sequential long range electron transfer; density matrix pathways in Liouville space *Chemical Physics* **197** 367-88

[319] Bixon M and Jortner J 1999 *Adv. Chem. Phys.: Electron transfer - from isolated molecules to biomolecules,* ed M Bixon and J Jortner (New York: Wiley) pp 35-202

[320] Mujica V, Nitzan A, Mao Y, Davis W, Kemp M, Roitberg A and Ratner M A 1999 *Adv. Chem. Phys: Electron Transfer-From Isoloted Moleciles to Biomolecules, Pt 2,* pp 403-29

[321] Wan C Z, Fiebig T, Kelley S O, Treadway C R, Barton J K and Zewail A H 1999 Femtosecond dynamics of DNA-mediated electron transfer *Proc. Natl. Acad. Sci. U. S. A.* **96** 6014-9

[322] Wolf E L 1985 *Principles of electron tunneling spectroscopy* vol 71 (New York: Oxford University Press)

[323] Hipps K W and Mazur U 1993 Inelastic Electron-Tunneling - an Alternative Molecular- Spectroscopy *J. Phys. Chem.* **97** 7803-14

[324] Bayman A, Hansma P and Kaska W C 1981 *Phys. Rev. B-Condens Matter* **24** 2449-56

[325] Stipe B C, Rezaei M A and Ho W 1999 Localization of Inelastic Tunneling and the Determination of Atomic-Scale Structure with Chemical Specificity *Phys. Rev. Lett.* **82** 1724-7

[326] Hahn J R, Lee H J and Ho W 2000 Electronic resonance and symmetry in single-molecule inelastic electron tunneling *Phys. Rev. Letters* **85** 1914-7

[327] Pascual J I, Jackiw J J, Song Z, Weiss P S, Conrad H and Rust H-P 2001 Adsorbate-Substrate Vibrational Modes of Benzene on Ag(110) Resolved with Scanning Tunneling Spectroscopy *Phys. Rev. Lett.* **86** 1050–3

[328] Wang W, Lee T and Reed M A 2004 Elastic and Inelastic Electron Tunneling in Alkane Self-Assembled Monolayers *J. Phys. Chem. B* **108** 18398 -407

[329] Kushmerick J, Allara D, Mallouk T and Mayer T 2004 Electrical and spectroscopic characterization of molecular junctions *MRS Bulletin* **29** 396-402

[330] Yu L H and Natelson D 2004 The Kondo Effect in C60 Single-Molecule Transistors *Nano Letters* **4** 79-83

[331] Nazin G V, Qiu X H and Ho W 2005 Vibrational spectroscopy of individual doping centers in a monolayer organic crystal *J. Chem Phys.* **122** 181105

[332] Grein C H, Runge E and Ehrenreich H 1993 Phonon-assisted transport in double-barrier resonant-tunneling structures *Phys. Rev. B-Condens Matter* **47** 12590

[333] Yamamoto T, Watanabe K and Watanabe S 2005 Electronic transport in fullerene C-20 bridge assisted by molecular vibrations *PHYSICAL REVIEW LETTERS* **95** 065501

[334] Ryndyk D A and Keller J 2005 Inelastic resonant tunneling through single molecules and quantum dots: Spectrum modification due to nonequilibrium effects *Phys. Rev. B-Condens Matter* **71** 073305

[335] Tikhodeev S G and Ueba H 2005 Theory of inelastic tunneling and its relation to vibrational excitation in ladder climbing processes of single adsorbates *SURFACE SCIENCE* **587** 25-33

[336] Tikhodeev S G and Ueba H 2004 Relation between inelastic electron tunneling and vibrational excitation of single adsorbates on metal surfaces *Phys. Rev. B-Condens Matter* **70** 125414-7

[337] Fano U 1961 Effects of Configuration Interaction on Intensities and Phase Shifts *Physical Review* **124** 1866

[338] Baratoff A and Persson B N J 1988 Theory of the local tunneling spectrum of a vibrating adsorbate *J. Vac. Sci. & Tech. A* **6** 331-5

[339] Lang I G and Firsov Y A 1963 *Sov. Phys. JETP* **16** 1301

[340] Koch J, von Oppen F and Andreev A V 2006 Theory of the Franck-Condon blockade regime *Phys. Rev. B-Condens Matter* **74** 205438-19

[341] Repp J, Meyer G, S. P, Olsson F E and Persson M 2005 Scanning tunneling spectroscopy of Cl vacancies in NaCl films: Strong electron-phonon coupling in double-barrier tunneling junctions *Phys. Rev. Letters* **95** 225503

[342] Ness H 2006 Quantum inelastic electron–vibration scattering in molecular wires: Landauer-like versus Green's function approaches and temperature effects *J. Phys.: Condens. Matter* **18** 6307–28

[343] M. Galperin, unpublished

[344] Aviram A and ratner M A 1974 *Chem. Phys. Lett.* **29** 277

[345] Martin A S and Sambles J R 1996 Molecular rectification, photodiodes and symmetry *Nanotechnology* **7** 401-5

[346] Metzger R M 2001 Rectification by a single molecule *Synthetic Metals* **124** 107-12

[347] Stokbro K, Taylor J and Brandbyge M 2003 *J. Am. Chem. Soc.* **125** 3674

[348] Galperin M, Nitzan A, Ratner M A and Stewart D R 2005 Molecular Transport Junctions: Asymmetry in Inelastic Tunneling Processes *J. Phys. Chem. B* **109** 8519-22

[349] Djukic D, Thygesen K S, Untiedt C, Smit R H M, Jacobsen K W and van Ruitenbeek J M 2005 Stretching dependence of the vibration modes of a single-

molecule Pt-H[sub 2]-Pt bridge *Physical Review B (Condensed Matter and Materials Physics)* **71** 161402-4

[350] Agrait N, Untiedt C, Rubio-Bollinger G and Vieira S 2002 Onset of Energy Dissipation in Ballistic Atomic Wires *Phys. Rev. Lett.* **88** 216803
[351] Davis L C 1970 *Phys. Rev.B* **2** 1714-32
[352] Persson B N J 1988 Inelastic vacuum tunneling *Physica Scripta* **38** 282-90
[353] Persson B N J and Baratoff A 1987 Inelastic electron tunneling from a metal tip: The contribution from resonance processes *Phys. Rev. Lett.* **59** 339-42
[354] Frederiksen T, Brandbyge M, Jauho A P and Lorente N 2004 Modeling of inelastic transport in one-dimensional metallic atomic wires *Journal of Computational Electronics* **3** 423-7
[355] Agrait N, Untiedt C, Rubio-Bollinger G and Vieira S 2002 Electron transport and phonons in atomic wires *Chemical Physics* **281** 231-4
[356] Kulik I O, Omel'yanchuk A N and Shekhter R I 1977 ELECTRICAL CONDUCTIVITY OF POINT MICROBRIDGES AND PHONON AND IMPURITY SPECTROSCOPY IN NORMAL METALS *Soviet Journal of Low Temperature Physics (English Translation of Fizika Nizkikh Temperatur)* **3** 740-8
[357] van Gelder A P, Jansen A G M and Wyder P 1980 Temperature dependence of point-contact spectroscopy in copper *Physical Review B* **22** 1515
[358] Thijssen W H A, Djukic D, Otte A F, Bremmer R H and van Ruitenbeek J M 2006 Vibrationally Induced Two-Level Systems in Single-Molecule Junctions *Physical Review Letters* **97** 226806-4
[359] Vitali L, Schneider M A, Kern K, Wirtz L and Rubio A 2004 Phonon and plasmon excitation in inelastic electron tunneling spectroscopy of graphite *Phys. Rev. B-Condens Matter* **69** 121414
[360] Rousseau R, Renzi V D, Mazzarello R, Marchetto D, Biagi R, Scandolo S and Pennino U d 2006 Interfacial Electrostatics of Self-Assembled Monolayers of Alkane Thiolates on Au(111): Work Function Modification and Molecular Level Alignments *The Journal of Physical Chemistry B* **110** 10862
[361] Levin C S, Janesko B G, Bardhan R, Scuseria G E, Hartgerink J D and Halas N J 2006 Chain-Length-Dependent Vibrational Resonances in Alkanethiol Self-Assembled Monolayers Observed on Plasmonic Nanoparticle Substrates *Nano Lett.*
[362] Kluth G J, Carraro C and Maboudian R 1999 Direct observation of sulfur dimers in alkanethiol self-assembled monolayers on Au(111) *Physical Review B* **59** R10449-R52
[363] Hayashi T, Morikawa Y and Nozoye H 2001 Adsorption state of dimethyl disulfide on Au(111): Evidence for adsorption as thiolate at the bridge site *The Journal of Chemical Physics* **114** 7615-21
[364] Kato H S, Noh J, Hara M and Kawai M 2002 An HREELS Study of Alkanethiol Self-Assembled Monolayers on Au(111) *J. Phys. Chem. B* **106** 9655-8
[365] Evers F, Weigend F and Koentopp M 2004 Conductance of molecular wires and transport calculations based on density-functional theory *Phys. Rev. B-Condens Matter* **69** 235411
[366] Gaudoin R and Burke K 2004 Lack of Hohenberg-Kohn Theorem for Excited States *Physical Review Letters* **93** 173001-4
[367] Toher C, Filippetti A, Sanvito S and Burke K 2005 Self-Interaction Errors in Density-Functional Calculations of Electronic Transport *Physical Review Letters* **95** 146402-4

[368] Stipe B C, Rezaei M A and Ho W 1998 Single-molecule vibrational spectroscopy and microscopy *Science* **280** 1732-5

[369] Stipe B C, Rezaei M A and Ho W 1998 Coupling of vibrational excitation to the rotational motion of a single adsorbed molecule *Phys. Rev. Letters* **81** 1263-6

[370] Emberly E and Kirczenow G 2000 Landauer theory, inelastic scattering, and electron transport in molecular wires *Phys. Rev. B-Condens Matter* **61** 5740-50

[371] Jiang J, Kula M and Luo Y 2006 A generalized quantum chemical approach for elastic and inelastic electron transports in molecular electronics devices *JOURNAL OF CHEMICAL PHYSICS* **124** 034708

[372] Park J, Pasupathy A N, Goldsmith J I, Soldatov A V, Chang C, Yaish Y, Sethna J P, Abruna H D, Ralph D C and McEuen P L 2003 Wiring up single molecules *Thin Solid Films* **438-439** 457-61

[373] Zhitenev N B, Meng H and Bao Z 2002 Conductance of small molecular junctions *Phys. Rev. Letters* **88** 226801

[374] van der Zant H S J, Kervennic Y V, Poot M, O'Neill K, de Groot Z, Thijssen J M, Heersche H B, Stuhr-Hansen N, Bjornholm T, Vanmaekelbergh D, van Walree C A and Jenneskens L W 2006 Molecular three-terminal devices: fabrication and measurements *FARADAY DISCUSSIONS* **131** 347-56

[375] Clay R T and Hardikar R P 2005 Intermediate Phase of the One Dimensional Half-Filled Hubbard-Holstein Model *Physical Review Letters* **95** 096401-4

[376] Chatterjee A and Takada Y 2004 The Hubbard–Holstein Model with Anharmonic Phonons in One Dimension *J. Phys. Soc. Japan* **73** 964-9

[377] Beenakker C W J 1991 Theory of Coulomb-blockade oscillations in the conductance of a quantum dot *Phys. Rev. B-Condens Matter* **44** 1646

[378] Meir Y, Wingreen N S and Lee P A 1991 Transport through a strongly interacting electron system: Theory of periodic conductance oscillations *Physical Review Letters* **66** 3048

[379] Craco L and Kang K 1999 Perturbation treatment for transport through a quantum dot *Physical Review B* **59** 12244

[380] Gurvitz S A, Mozyrsky D and Berman G P 2005 Coherent effects in magnetotransport through Zeeman-split levels *Phys. Rev. B-Condens Matter* **72** 205341-10

[381] Gurvitz S A, Mozyrsky D and Berman G P 2005 Publisher's Note: Coherent effects in magnetotransport through Zeeman-split levels [Phys. Rev. B [bold 72], 205341 (2005)] *Physical Review B (Condensed Matter and Materials Physics)* **72** 249902-1

[382] Muralidharan B, Ghosh A W and Datta S 2006 Probing electronic excitations in molecular conduction *Physical Review B* **73**

[383] Muralidharan B, Ghosh A W, Pati S K and Datta S 2006 Theory of high bias Coulomb Blockade in ultrashort molecules *cond-mat/0610244*

[384] Petrov E G and Hanggi P 2001 Nonlinear Electron Current through a Short Molecular Wire *Phys. Rev. Letters.* **86** 2862-5

[385] Petrov E G, May V and Hanggi P 2002 Controlling electron transfer processes through short molecular wires *Chemical Physics* **281** 211-24

[386] Petrov E G, May V and Hanggi P 2005 Kinetic theory for electron transmission through a molecular wire *CHEMICAL PHYSICS* **319** 380-408

[387] Kaiser F J, Strass M, Kohler S and Hanggi P 2006 Coherent charge transport through molecular wires: Influence of strong Coulomb repulsion *CHEMICAL PHYSICS* **322** 193-9

[388] Petrov E G, May V and Hanggi P 2006 Kinetic rectification of charge transmission through a single molecule *Phys. Rev. B-Condens Matter* **73** 045408

[389] Petrov E G, Zelinskyy Y R, May V and Hanggi P 2006 Kinetic control of the current through a single molecule *Chemical Physics* **328** 173-82

[390] Langreth D C and Nordlander P 1991 Derivation of a master equation for charge-transfer processes in atom-surface collisions *Physical Review B* **43** 2541

[391] Shao H, Langreth D C and Nordlander P 1994 Many-body theory for charge transfer in atom-surface collisions *Physical Review B* **49** 13929

[392] Nordlander P, Wingreen N S, Meir Y and Langreth D C 2000 Kondo physics in the single-electron transistor with ac driving *Physical Review B* **61** 2146

[393] Wingreen N S and Meir Y 1994 Anderson model out of equilibrium: Noncrossing approximation approach to transport through a quantum dot *Phys. Rev. B* **49** 11040-52

[394] Krawiec M and Wysoki„ski K I 2002 Nonequilibrium Kondo effect in asymmetrically coupled quantum dots *Physical Review B* **66** 165408

[395] Guclu A D, Sun Q F and Guo H 2003 Kondo resonance in a quantum dot molecule *Physical Review B (Condensed Matter and Materials Physics)* **68** 245323-5

[396] Ng T-K 1996 ac Response in the Nonequilibrium Anderson Impurity Model *Physical Review Letters* **76** 487

[397] Sun Q-f and Lin T-h 1997 Time-dependent electron tunnelling through a quantum dot with Coulomb interactions *Journal of Physics: Condensed Matter* **9** 4875-86

[398] Niu C, Lin D L and Lin T H 1999 Equation of motion for nonequilibrium Green functions *Journal of Physics: Condensed Matter* **11** 1511-21

[399] Krawiec M and Wysokinski K I 2006 Thermoelectric effects in strongly interacting quantum dot coupled to ferromagnetic leads *Phys. Rev. Letters* **submitted**

[400] Swirkowicz R, Barnas J and Wilczynski M 2003 Nonequilibrium Kondo effect in quantum dots *Physical Review B (Condensed Matter and Materials Physics)* **68** 195318-10

[401] Swirkowicz R, Wilczynski M and Barnas J 2006 Spin-polarized transport through a single-level quantum dot in the Kondo regime *Journal of Physics: Condensed Matter* **18** 2291-304

[402] Yeyati A L, Mart„n-Rodero A and Flores F 1993 Electron correlation resonances in the transport through a single quantum level *Physical Review Letters* **71** 2991

[403] Rosch A, Paaske J, Kroha J and Wolfle P 2003 Nonequilibrium Transport through a Kondo Dot in a Magnetic Field: Perturbation Theory and Poor Man's Scaling *Physical Review Letters* **90** 076804

[404] Rosch A, Paaske J, Kroha J and Wölfle P 2005 The Kondo Effect in Non-Equilibrium Quantum Dots: Perturbative Renormalization Group *J. Phys. Soc. Japan* **74** 118-26

[405] Paaske J, Rosch A, Kroha J and Wolfle P 2004 Nonequilibrium transport through a Kondo dot: Decoherence effects *Physical Review B* **70** 155301

[406] Kaminski A, Nazarov Y V and Glazman L I 1999 Suppression of the Kondo Effect in a Quantum Dot by External Irradiation *Phys. Rev. Lett.* **83** 384

[407] Kaminski A, Nazarov Y V and Glazman L I 2000 Universality of the Kondo effect in a quantum dot out of equilibrium *Phys. Rev. B-Condens Matter* **62** 8154

[408] Fujii T and Ueda K 2003 Perturbative approach to the nonequilibrium Kondo effect in a quantum dot *Phys. Rev. B-Condens Matter* **68** 155310-5

[409] Fujii T and Ueda K 2004 Theory of the nonequilibrium Kondo effect in a quantum dot *Physica E* **22** 498-501

[410] Komnik A and Gogolin A O 2004 Mean-field results on the Anderson impurity model out of equilibrium *Phys. Rev. B-Condens Matter* **69** 153102

[411] König J, Schoeller H and Schön G 1998 Cotunneling and renormalization effects for the single-electron transistor *Phys. Rev. B-Condens Matter* **58** 7882

[412] Hamasaki M 2004 Effect of electron correlation on current and current noise for the single- and the two-impurity Anderson model *Phys. Rev. B-Condens Matter* **69** 115313-9

[413] Al-Hassanieh K A, Busser C A, Martins G B and Dagotto E 2005 Electron Transport through a Molecular Conductor with Center-of-Mass Motion *Phys. Rev. Lett.* **95** 256807

[414] Braig S and Flensberg K 2004 Dissipative tunneling and orthogonality catastrophe in molecular transistors *Phys. Rev. B-Condens Matter* **70** 085317

[415] Siddiqui L, Ghosh A W and Datta S 2006 Phonon runaway in nanotube quantum dots *cond- mat/060273*

[416] Meir Y, Wingreen N S and Lee P A 1993 Low-temperature transport through a quantum dot: The Anderson model out of equilibrium *Physical Review Letters* **70** 2601

[417] Hanke U, Galperin Y, Chao K A, Gisselfält M, Jonson M and Shekhter R I 1995 Static and dynamic transport in parity-sensitive systems *Physical Review B* **51** 9084

[418] Bo O L and Galperin Y 1996 Low-frequency shot noise in double-barrier resonant-tunnelling GaAs/AlxGa1-xAs structures in a strong magnetic field *Journal of Physics-Condensed Matter* **8** 3033-45

[419] Chen J, Reed M A, Rawlett A M and Tour J M 1999 Large On-Off Ratios and Negative Differential Resistance in a Molecular Electronic Device *Science* **286** 1550-2

[420] Rawlett A, Hopson T J, Nagahara L A, Tsui R K, Ramachandran G K and Lindsay S M 2002 Electrical measurements of a dithiolated electronic molecule via conducting atomic force microscopy *Applied Physics Letters* **81** 3043-5

[421] Li C, Zhang D, Liu X, Han S, Tang T, Zhou C, Fan W, Koehne J, Han J, Meyyappan M, Rawlett A M, Price D W and Tour J M 2003 Fabrication approach for molecular memory arrays *Appl. Phys. Letters* **82** 645-7

[422] Chen F, He J, Nuckolls C, Roberts T, Klare J E and Lindsay S 2005 A Molecular Switch Based on Potential-Induced Changes of Oxidation State *Nano Letters* **5** 503-6

[423] Xiao X Y, Nagahara L A, Rawlett A M and Tao N J 2005 Electrochemical gate-controlled conductance of single oligo(phenylene ethynylene)s *J. Am. Chem. Soc.* **127** 9235-40

[424] Xu B Q, Li X L, Xiao X Y, Sakaguchi H and Tao N J 2005 Electromechanical and Conductance Switching Properties of Single Oligothiophene Molecules *Nano Letters* **5** 1491-5

[425] Li Z, Han B, Meszaros G, Pobelov I, Wandlowski T, Baszczyk A and Mayor M 2006 Two-dimensional assembly and local redox-activity of molecular hybrid structures in an electrochemical environment *Faraday Discussions* **131** 121-43

[426] Hewson A C and Newns D M 1974 Effect of Image Force in Chemisorption *Japanese Journal of Applied Physics* **Suppl 2** 121-30

[427] Hewson A C and Newns D M 1979 Polaronic Effects in Mixed-Valence and Intermediate-Valence Compounds *Journal of Physics C-Solid State Physics* **12** 1665-83

[428] Rego L G C and Kirczenow G 1998 Quantized Thermal Conductance of Dielectric Quantum Wires *Phys. Rev. Letters* **81** 232-5

[429] Ozpineci A and Ciraci S 2001 Quantum effects of thermal conductance through atomic chains *Phys. Rev. B-Condens Matter* **63** 125415/1-5

[430] Segal D, Nitzan A and Hanggi P 2003 Thermal conductance through molecular wires *J. Chem. Phys.* **119** 6840-55

[431] van den Brom H E, Yanson A I and van Ruitenbeek J M 1998 Characterization of individual conductance steps in metallic quantum point contacts *Physica B: Condensed Matter* **252** 69-75

[432] Huang Z F, Xu B Q, Chen Y C, Di Ventra M and Tao N J 2006 Measurement of current-induced local heating in a single molecule junction *Nano Letters* **6** 1240-4

[433] Yang Z, Chshiev M, Zwolak M, Chen Y C and Di Ventra M 2005 Role of heating and current-induced forces in the stability of atomic wires *Phys. Rev. B-Condens Matter* **71** 041402

[434] Peierls R E 1929 *Ann. Phys. (Liepzig)* **3** 1055

[435] Rieder Z, Lebowitz J L and Lieb E 1967 Properties of a harmonic crystal in a stationary non-equilibrium state *J. Chem. Phys.* **8** 1073-8

[436] Zürcher U and Talkner P 1990 Quantum mechanical harmonic chain attached to heat baths II. Nonequilibrium properties *Phys. Rev. A* **42** 3278-90

[437] Grassberger P, Nadler W and Yang L 2002 Heat Conduction and Entropy Production in a One-Dimensional Hard-Particle Gas *Phys. Rev. Lett.* **89** 180601

[438] Casati G and Prosen T 2003 Anomalous heat conduction in a one-dimensional ideal gas *Phys. Rev. E* **67** 015203

[439] Casher A and Lebowitz J L 1971 Heat flow in regular and disordered harmonic chains *J. Math. Phys.* **12** 1701-11

[440] O'Connor A J and Lebowitz J L 1974 Heat conduction and sound transmission in isotopically disordered harmonic chains *J. Math. Phys.* **15** 692-703

[441] Leitner D M and Wolynes P G 2000 Heat flow through an insulating nanocrystal *Phys. Rev. E* **61** 2902-8

[442] Lepri S, Livi R and Politi A 1997 Heat Conduction in Chains of Nonlinear Oscillators *Phys. Rev. Lett.* **78** 1896-99

[443] Mokross F and Buttner H 1983 *J.Phys. C* **16** 4539

[444] Hu B, Li B and Zhao H 1998 Heat conduction in one- dimensional chains *Phys. Rev. E* **57** 2992-5

[445] Terraneo M, Peyrard M and Casati G 2002 Controlling the Energy Flow in Nonlinear Lattices: A Model for a Thermal Rectifier *Phys. Rev. Lett.* **88** 094302

[446] Li B, Wang L and Casati G 2004 Thermal Diode: Rectification of Heat Flux *Physical Review Letters* **93** 184301

[447] Segal D and Nitzan A 2005 Spin-Boson Thermal Rectifier *Phys. Rev. Lettters* **94** 034301

[448] Segal D and Nitzan A 2005 Heat rectification in molecular junctions *J. Chem Phys.* **122** 194704

[449] Li B, Lan J and Wang L 2005 Interface Thermal Resistance between Dissimilar Anharmonic Lattices *Phys. Rev. Lett.* **95** 104302

[450] Saito K 2006 Asymmetric Heat Flow in Mesoscopic Magnetic System *Journal of the Physical Society of Japan* **75** 034603

[451] Kindermann M and Pilgram S 2004 Statistics of heat transfer in mesoscopic circuits *Phys. Rev. B-Condens Matter* **69** 155334

[452] Cho S Y and McKenzie R H 2005 Thermal and electrical currents in nanoscale electronic interferometers *Physical Review B (Condensed Matter and Materials Physics)* **71** 045317

[453] Cahill D, Goodson K and Majumdar A 2002 Thermometry and thermal transport in micro/nanoscale solid-state devices and structures *Journal of Heat Transfer* **124** 223-41

[454] Shi L and Majumdar A 2002 Thermal Transport Mechanisms at Nanoscale Point Contacts *Journal of Heat Transfer* **124** 329-37

[455] Kim P, Shi L, Majumdar A and McEuen P L 2001 Thermal transport measurements of individual multiwalled nanotubes *Phys. Rev. Letters* **87** 215502

[456] Kim P, Shi L, Majumdar A and McEuen P L 2002 Mesoscopic thermal transport and energy dissipation in carbon nanotubes *Physica B* **323** 67-70

[457] Small J P, Shi L and Kim P 2003 Mesoscopic thermal and thermoelectric measurements of individual carbon nanotubes *Solid State Communications* **127** 181-6

[458] Lin-Chung P and Rajagopal A K 2002 Green's function theory of electrical and thermal transport in single-wall carbon nanotubes *Phys. Rev. B-Condens Matter* **65** 1-4

[459] Zheng Q, Su G, Wang J and Guo H 2002 Thermal conductance for single wall carbon nanotubes *Eur. Phys. J. B* **25** 233-8

[460] Yao Z, Wang J-S, Li B and Liu G-R 2005 Thermal conduction of carbon nanotubes using molecular dynamics *Physical Review B (Condensed Matter and Materials Physics)* **71** 085417

[461] Wang J A and Wang J S 2006 Carbon nanotube thermal transport: Ballistic to diffusive *Applied Physics Letters* **88**

[462] Schwab K, Henriksen E A, Worlock J M and Roukes M L 2000 Measurement of the quantum of thermal conductance *Nature* **404** 974-7

[463] Cahill D, Ford W K, Goodson K E, Mahan G D, Majumdar A, Maris H J, Merlin R and Phillpot S R 2003 Nanoscale thermal transport *Journal of Applied Physics* **93** 793-818

[464] Schwarzer D, Kutne P, Schroder C and Troe J 2004 Intramolecular vibrational energy redistribution in bridged azulene-anthracene compounds: Ballistic energy transport through molecular chains *The Journal of Chemical Physics* **121** 1754-64

[465] Wang R Y, Segalman R A and Majumdar A 2006 Room temperature thermal conductance of alkanedithiol self-assembled monolayers *Applied Physics Letters* **89**. Note, however, that in this experiment one cannot rule out the possibility that the observed heat conduction is dominated by boundary effects.

[466] Buldum A and Ciraci S 1996 Controlled lateral and perpendicular motion of atoms on metal surfaces *Phys. Rev. B-Condens Matter* **54** 2175-83

[467] Stipe B C, Rezaei M A, Ho W, Gao S, Persson M and Lundqvist B I 1997 Single-molecule dissociation by tunneling electrons *Phys. Rev. Letters* **78** 4410-3

[468] Persson B N J and Avouris P 1997 Local bond breaking via STM-induced excitations: the role of temperature *Surface Science* **390** 45-54

[469] Foley E T, Kam A F, Lyding J W and Avouris P 1998 Cryogenic UHV-STM study of hysrogen and deuterium desorption from Si(100) *Phys. Rev. Lett.* **80** 1336-9

[470] Stipe B C, Rezaei M A and Ho W 1998 Inducing and Viewing the Rotational Motion of a Single Molecule *Science* **279** 1907-9

[471] Hla S-W, Meyer G and Rieder K-H 2001 Inducing Single-Molecule Chemical Reactions with a UHV-STM: A New Dimension for Nano-Science and Technology *CHEMPHYSCHEM* **2** 361 - 6

[472] Seideman T 2003 Current-driven dynamics in quantum electronics *Journal of Modern Optics* **50** 2393-410

[473] Seideman T and Guo H 2003 Quantum transport and current-triggered dynamics in molecular tunnel junctions *J. Theor. Comp. Chem.* **2** 439-58

[474] Pascual J I 2005 Single molecule vibrationally mediated chemistry: Towards state-specific strategies for molecular handling *Eur. Phys. J. D* **35** 327-40